\documentclass{article}
\usepackage{amsmath}
\usepackage{amssymb}
\usepackage[cmtip,arrow]{xy}
\usepackage{pb-diagram,pb-xy}

\input{tcilatex}
\begin{document}

\title{\textbf{Dirac method and symplectic submanifolds in the cotangent
bundle of a factorizable Lie group}\\
\smallskip\ }
\author{\textbf{S. Capriotti$^{\dag }$ \& H. Montani$^{\ddag }$\textbf{\ } }
\and   \\
\dag\ \textit{Departamento de Matem\'{a}tica, Universidad Nacional del
Sur,\smallskip }\ \\
\ ~\textit{Av. Alem 1253, 8000} - \textit{Bah\'{\i}a Blanca, Buenos Aires,
Argentina. }\\ \\
\ddag\ \textit{Instituto Balseiro - Centro At\'{o}mico Bariloche,\smallskip }\ \\
~\textit{8400 - S. C. de Bariloche, Rio Negro, Argentina.}\\
}
\maketitle

\begin{abstract}
In this work we study some symplectic submanifolds in the cotangent bundle
of a factorizable Lie group defined by second class constraints. By applying
the Dirac method, we study many issues of these spaces as fundamental Dirac
brackets, symmetries, and collective dynamics. This last item allows to
study integrability as inherited from a system on the whole cotangent
bundle, leading in a natural way to the AKS theory for integrable systems.
\end{abstract}

\tableofcontents

\section{Introduction}

Cotangent bundles of Lie groups are one of the most important symplectic
models for the phase spaces of dynamical systems. They correspond to systems
with configuration spaces being Lie groups, as it is the case for the rigid
body and its generalizations (as the main finite dimensional examples \cite%
{Arnold}) and with the sigma and WZNW models in field theory as some of the
infinite dimensional examples. They are symplectic manifolds and enjoy many
nice properties related to the symmetry issues \cite{Abr-Mars},\cite%
{Mars-Ratiu}, and they have close relationship with integrable systems, in
particular for those groups where some kind of factorization holds \cite%
{RSTS 0}.

Many interesting phase spaces arise as constrained submanifolds in a larger
phase space. The Dirac method \cite{Dirac} is a successful algorithm for
dealing with second class constraints, producing a Poisson bracket for a
constrained submanifold, and thus allowing a consistent quantization of this
kind of systems. It provides a way to obtain Lie derivatives of functions on
the whole phase space along the projection of the hamiltonian vector fields
on the tangent space of the constrained submanifold, and so it is a
representation of these vectors fields in terms of the geometrical data in
the total phase space.

In this work, we combine both the above ingredients to study a family of
symplectic submanifolds in the cotangent bundle of a factorizable Lie group $%
G=G_{+}\times G_{-}$, defined as the level set of a map which \emph{projects}
the cotangent bundle $T^{\ast }G$ onto the cotangent bundle of one of its
factors. To deal with this kind of constraints we develop, in a general
framework, a geometric approach to Dirac's brackets addressed to describe
immersed symplectic submanifold $\mathcal{N}$, $\imath :\mathcal{N}%
\hookrightarrow \mathcal{M}$, as the level sets of some submersive map from $%
\mathcal{M}$ into another manifold $\mathcal{P}$. This approach allows for a
straightforward application to the \emph{projection maps }$\Psi :G\times 
\mathfrak{g}^{\ast }\longrightarrow G_{-}\times \mathfrak{g}_{-}^{\ast }$
and $\Upsilon :G\times \mathfrak{g}^{\ast }\longrightarrow G_{+}\times 
\mathfrak{g}_{+}^{\ast }$, from where the symplectic submanifolds of $%
T^{\ast }G$ are obtained, equipped with the corresponding Poisson-Dirac
structure, turning them in \emph{symplectic} \emph{fibrations}. This
framework allows to study many aspects of theses spaces and related
dynamical systems such as symmetries and integrability. In fact, as by
product, we work out the application of the AKS ideas \cite{AKS} to some
collective models on these spaces exploiting the tools constructed
previously, showing how to make contact with the AKS results from the Dirac
bracket approach and offering an insight on the geometric aspects underlying
these issues.

To put in practice these constructions, we apply them on the group $SL(2,%
\mathbb{C)}$ which factorize as $SU(2)\times B$, where $B$ is the solvable
group of $2\times 2$ complex upper triangular matrices with real positive
diagonal elements and determinant 1, coming to generalize the results of 
\cite{CapMon}. We work out a collective hamiltonian models on a generic
fiber of $\Psi :SL(2,\mathbb{C)}\times \mathfrak{sl}_{2}^{\ast
}\longrightarrow B\times \mathfrak{b}^{\ast }$, and construct the
corresponding Lagrangian version.

This work is organized as follows: in the second Section we carry out a
brief review of the Dirac procedure addressed to deal with level set of
submersive maps from the phase space to another manifolds; in the third
Sections we define, by means of the Dirac procedure developed in the
previous Section, a class of phase spaces in the cotangent bundle of a
factorizable Lie group, working out many issues as fundamental brackets,
symmetries and Hamilton equations; in the fourth Section, the connection
with integrable models and factorization problem is studied from the Dirac
procedure and symplectic reduction points of view. Finally, in the fifth
Section we present the example built on $SL(2,\mathbb{C)}=SU(2)\times B$,
and in the sixth Section some conclusions are summarized.

\section{A brief review of Dirac procedure}

Let $\left( \mathcal{M},\omega \right) $ be a symplectic manifold and $%
\left( \mathcal{N},\imath ^{\ast }\omega \right) $ a symplectic submanifold
with the immersion $\imath :\mathcal{N}\hookrightarrow \mathcal{M}$.

\begin{description}
\item[Lemma:] \textit{The tangent space at a point }$\imath \left( n\right)
\in \mathcal{M}$ \textit{can be decomposed in a direct sum as }%
\begin{equation}
T_{\imath \left( n\right) }\mathcal{M}=\imath _{\ast }\left( T_{n}\mathcal{N}%
\right) \oplus \left[ \imath _{\ast }\left( T_{n}\mathcal{N}\right) \right]
^{\omega \bot }.  \label{gs-1}
\end{equation}
\end{description}

\textbf{Proof: }Here $\left[ V\right] ^{\omega \bot }$ means the symplectic
orthogonal of $V$ by the $2$-form $\omega $. This result stems from the non
degeneracy of $\omega $. $\blacksquare $

\begin{description}
\item[Lemma:] \textit{Let} \textit{be }$F\in \mathbb{C}^{\infty }\left( 
\mathcal{M}\right) $, $V_{F}\in \mathfrak{X}\left( \mathcal{M}\right) $ 
\textit{its Hamiltonian vector field and }$\left. \pi _{\mathcal{N}%
}\right\vert _{\imath \left( n\right) }:T_{\imath \left( n\right) }\mathcal{M%
}\longrightarrow \imath _{\ast }\left( T_{n}\mathcal{N}\right) $ \textit{the
projection}. \textit{Then}%
\begin{equation}
\left. \imath _{\ast }V_{\imath ^{\ast }F}\right\vert _{\imath \left(
n\right) }=\left. \pi _{\mathcal{N}}\left( V_{F}\right) \right\vert _{\imath
\left( n\right) }~~~.  \label{gs-2}
\end{equation}
\end{description}

The \emph{Dirac procedure} arises from considering a submanifold defined as
the level surfaces of some subrmersive map $\Psi $ from $\mathcal{M}$ into
another manifold $\mathcal{P}$. So, for~some~regular~value~$p_{\circ }\in 
\mathcal{P}$, we get a submanifold $\mathcal{N}$ defined as%
\begin{equation*}
\imath \left( \mathcal{N}\right) =\Psi ^{-1}\left( p_{\circ }\right)
\end{equation*}%
The following theorem is in truly realm of Dirac's idea.

\begin{description}
\item[Theorem:] \textit{Let }$\Psi :\mathcal{M}\twoheadrightarrow \mathcal{P}
$ \textit{a surjective map, and }$p_{\circ }\in \mathcal{P}$ \textit{a
regular value of }$\Psi .$\textit{Then, the cotangent space at a point }$%
\imath \left( n\right) \in \mathcal{M}$ \textit{is decomposed as }%
\begin{equation}
T_{\imath \left( n\right) }^{\ast }\mathcal{M}=\omega \left( T_{n}\mathcal{N}%
\right) \oplus \Psi ^{\ast }T_{p_{\circ }}^{\ast }\mathcal{P}.  \label{gs-3}
\end{equation}
\end{description}

\textbf{Proof: }For $n\in \Psi ^{-1}\left( p_{\circ }\right) $, we have the
exact short sequence 
\begin{equation*}
0\longrightarrow T_{n}\mathcal{N}\overset{\imath _{\ast }}{\longrightarrow }%
T_{\imath \left( n\right) }\mathcal{M}\overset{\Psi _{\ast \imath \left(
n\right) }}{\longrightarrow }T_{p_{\circ }}\mathcal{P}\longrightarrow 0
\end{equation*}%
and by duality%
\begin{equation*}
0\longrightarrow T_{p_{\circ }}^{\ast }\mathcal{P}\overset{\Psi _{\imath
\left( n\right) }^{\ast }}{\longrightarrow }T_{\imath \left( n\right)
}^{\ast }\mathcal{M}\overset{\imath ^{\ast }}{\longrightarrow }T_{n}^{\ast }%
\mathcal{N}\longrightarrow 0
\end{equation*}%
so that%
\begin{equation*}
T_{n}^{\ast }\mathcal{N}=\imath ^{\ast }\left[ \frac{T_{\imath \left(
n\right) }^{\ast }\mathcal{M}}{\Psi _{\imath \left( n\right) }^{\ast }\left(
T_{p_{\circ }}^{\ast }\mathcal{P}\right) }\right]
\end{equation*}%
from where we conclude that%
\begin{equation}
\left( \imath _{\ast }\left( T_{n}\mathcal{N}\right) \right) ^{\circ }=\Psi
_{\imath \left( n\right) }^{\ast }\left( T_{p_{\circ }}^{\ast }\mathcal{P}%
\right)  \label{gs-4a}
\end{equation}

Now, since $\left. \omega \right\vert \imath \left( \mathcal{N}\right) $ is
nondegenerate, $\omega \left( \imath _{\ast }\left( T_{n}\mathcal{N}\right)
\right) $ is complementary to $\left( \imath _{\ast }\left( T_{n}\mathcal{N}%
\right) \right) ^{\circ }$ in $T_{\imath \left( n\right) }^{\ast }\mathcal{M}
$, then 
\begin{equation*}
T_{\imath \left( n\right) }^{\ast }\mathcal{M}=\omega \left( \imath _{\ast
}\left( T_{n}\mathcal{N}\right) \right) \oplus \Psi _{\imath \left( n\right)
}^{\ast }\left( T_{p_{\circ }}^{\ast }\mathcal{P}\right)
\end{equation*}%
$\blacksquare $.

\begin{description}
\item[Proposition:] \textit{The induced bijection }$\omega :T\mathcal{M}%
\longrightarrow T^{\ast }\mathcal{M}$ \textit{provides the isomorphism }%
\begin{equation}
\Psi _{\imath \left( n\right) }^{\ast }\left( T_{p_{\circ }}^{\ast }\mathcal{%
P}\right) \overset{\omega }{\simeq }\left[ \imath _{\ast }\left( T_{n}%
\mathcal{N}\right) \right] ^{\omega \bot },  \label{gs-4b}
\end{equation}%
\textit{for all }$n\in \mathcal{N}$.
\end{description}

\textbf{Proof: }First, observe that $\left( \Psi ^{\ast }\alpha \right)
_{\imath \left( n\right) }\in \left[ \Psi ^{\ast }\left( T_{p_{\circ
}}^{\ast }\mathcal{P}\right) \right] _{\imath \left( n\right) }$, $\alpha
\in T_{p_{\circ }}^{\ast }\mathcal{P}$, the bijection $\omega $ assigns a $%
V_{\alpha }\in T_{\imath \left( n\right) }\mathcal{M}$ such that $%
i_{V_{\alpha }}\omega =d\Psi ^{\ast }\alpha $. Then, $\forall W\in \imath
_{\ast }\left( T_{n}\mathcal{N}\right) =\ker \Psi _{\ast \imath \left(
n\right) }$ we have%
\begin{equation*}
\left\langle \omega ,V_{\alpha }\otimes W\right\rangle =\left\langle d\Psi
^{\ast }\alpha ,W\right\rangle =\left\langle d\alpha ,\Psi _{\ast
}W\right\rangle =0
\end{equation*}%
then $V_{\alpha }\in \left[ \imath _{\ast }\left( T_{n}\mathcal{N}\right) %
\right] ^{\omega \bot }$.

Reciprocally, $\forall W\in \left[ \imath _{\ast }\left( T_{n}\mathcal{N}%
\right) \right] ^{\omega \bot },~\exists \alpha _{W}T_{\imath \left(
n\right) }^{\ast }\mathcal{M~}/i_{W}\omega =\alpha _{W}$. Moreover, $\forall
V\in \imath _{\ast }\left( T_{n}\mathcal{N}\right) =\ker \Psi _{\ast \imath
\left( n\right) }$,%
\begin{equation*}
\left\langle \alpha _{W},V\right\rangle =\left\langle \omega ,W\otimes
V\right\rangle =0
\end{equation*}%
then $\alpha _{W}\in \left( \imath _{\ast }\left( T_{n}\mathcal{N}\right)
\right) ^{o}=\Psi _{\imath \left( n\right) }^{\ast }\left( T_{p_{\circ
}}^{\ast }\mathcal{P}\right) $. $\blacksquare $

Before to going on, it is worth to distinguish the two following situations
which applies into the most important physical systems:

\begin{enumerate}
\item \emph{First class}$\Longleftrightarrow \left[ \imath _{\ast }\left(
T_{n}\mathcal{N}\right) \right] ^{\omega \bot }\subset \imath _{\ast }\left(
T_{n}\mathcal{N}\right) \Longleftrightarrow \mathcal{N}$ \emph{is a
coisotropic submanifold.}

\item \emph{Second class}$\Longleftrightarrow \left[ \imath _{\ast }\left(
T_{n}\mathcal{N}\right) \right] ^{\omega \bot }\cap \imath _{\ast }\left(
T_{n}\mathcal{N}\right) =\left\{ 0\right\} $
\end{enumerate}

From now on, we shall be concerned with the \emph{second class}\textit{\ }%
case.

\begin{description}
\item[Theorem:] $\mathcal{N}=\Psi ^{-1}\left( p_{\circ }\right) $ \textit{is
symplectic if and only if }%
\begin{equation*}
\omega \left( \ker \Psi _{\ast \imath \left( n\right) }\right) \cap \left.
\Psi ^{\ast }\left( T_{p_{\circ }}^{\ast }\mathcal{P}\right) \right\vert
_{\imath \left( n\right) }=\left\{ 0\right\}
\end{equation*}
\end{description}

\textbf{Proof: }Since $p_{\circ }$ is a regular value of $\Psi $ and $%
\mathcal{N}$ is a level surface of $\Psi $, $\ker \Psi _{\ast \imath \left(
n\right) }=\imath _{\ast }\left( T_{n}\mathcal{N}\right) .$Then, because $%
\omega :T_{n}\mathcal{N}\longrightarrow T_{n}^{\ast }\mathcal{N}$ is a
bijection, there is a one to one correspondence between tangent vectors to $%
\mathcal{N}$ and linear forms. So that seeking for null vectors of $\imath
^{\ast }\omega $ in $T_{n}\mathcal{N}$ is equivalent to look for $\alpha \in
\omega \left( \ker \Psi _{\ast \imath \left( n\right) }\right) $ vanishing
on $T_{n}\mathcal{N}$, that is $\alpha \in \omega \left( \ker \Psi _{\ast
\imath \left( n\right) }\right) \cap \left( \imath _{\ast }\left( T_{n}%
\mathcal{N}\right) \right) ^{o}$. Regularity of $\imath ^{\ast }\omega $
implies no nontrivial such an $\alpha $ does exist, so $\omega \left( \ker
\Psi _{\ast \imath \left( n\right) }\right) \cap \left. \Psi ^{\ast }\left(
T_{p_{\circ }}^{\ast }\mathcal{P}\right) \right\vert _{\imath \left(
n\right) }=\left\{ 0\right\} $. $\blacksquare $

\begin{description}
\item[Corollary:] $\Psi $ \textit{is a second class constraint if and only
if }%
\begin{equation*}
\omega \left( \ker \Psi _{\ast \imath \left( n\right) }\right) \cap \left.
\Psi ^{\ast }\left( T_{p_{\circ }}^{\ast }\mathcal{P}\right) \right\vert
_{\imath \left( n\right) }=\left\{ 0\right\}
\end{equation*}
\end{description}

\textbf{Proof:}$~\Psi $ is a second class if and only if $\left[ \imath
_{\ast }\left( T_{n}\mathcal{N}\right) \right] ^{\omega \bot }\cap \imath
_{\ast }\left( T_{n}\mathcal{N}\right) =\left\{ 0\right\} $, that is
equivalent to $\left( \imath _{\ast }\left( T_{n}\mathcal{N}\right) \right)
^{\circ }\cap \ker \Psi _{\ast \imath \left( n\right) }=\left\{ 0\right\} $.
And it happens if and only if $\omega \left( \ker \Psi _{\ast \imath \left(
n\right) }\right) \cap \left. \Psi ^{\ast }\left( T_{p_{\circ }}^{\ast }%
\mathcal{P}\right) \right\vert _{\imath \left( n\right) }=\left\{ 0\right\} $%
. $\blacksquare $

Let us now introduce a suitable set of functions $\left\{ f_{1},\cdots
,f_{r}\right\} \subset \mathbb{C}^{\infty }\left( \mathcal{P}\right) $, $%
r=\dim \mathcal{P}$, such that $\left\{ \left. df_{1}\right\vert _{p_{\circ
}},\cdots ,\left. df_{r}\right\vert _{p_{\circ }}\right\} $ is a basis of $%
T_{p_{\circ }}^{\ast }\mathcal{P}$. Then, $\left\{ f_{1},\cdots
,f_{r}\right\} $ is local coordinate system for $\mathcal{P}$. Let us
consider then the set of local hamiltonian vector fields associated with the
pullback of these functions%
\begin{equation*}
\Psi ^{\ast }f_{j}\longrightarrow V_{\Psi ^{\ast }f_{j}}~/~\imath _{V_{\Psi
^{\ast }f_{j}}}\omega =d\left( \Psi ^{\ast }f_{j}\right)
\end{equation*}

\begin{description}
\item[Proposition:] \textit{Let be} $\Psi $\textit{\ a second class
constraint, then the set }$\left\{ V_{\Psi ^{\ast }f_{j}}\right\} _{i=1}^{r}$
\textit{is a basis for }$\left[ \imath _{\ast }\left( T_{n}\mathcal{N}%
\right) \right] ^{\omega \bot }$.
\end{description}

\textbf{Proof:} Let be $W\in \imath _{\ast }\left( T_{n}\mathcal{N}\right)
=\ker \Psi _{\ast \imath \left( n\right) }$, then%
\begin{equation*}
\left\langle \omega ,V_{\Psi ^{\ast }f_{j}}\otimes W\right\rangle _{\imath
\left( n\right) }=\left\langle d\left( \Psi ^{\ast }f_{j}\right)
,W\right\rangle _{\imath \left( n\right) }=\left\langle df_{j},\Psi _{\ast
}W\right\rangle _{\Psi \left( n\right) }=0
\end{equation*}%
from where we conclude that each $V_{\Psi ^{\ast }f_{j}}\in \left[ \imath
_{\ast }\left( T_{n}\mathcal{N}\right) \right] ^{\omega \bot }$

One may see that the functions $\Psi ^{\ast }f_{1},\cdots ,\Psi ^{\ast
}f_{r} $ are \emph{constant} on $\mathcal{N}=\Psi ^{-1}\left( p_{\circ
}\right) $. Also, we know that $\left\{ \left. df_{1}\right\vert _{p_{\circ
}},\cdots ,\left. df_{r}\right\vert _{p_{\circ }}\right\} $ is a linearly
independent set, then so is $\left\{ \left. d\Phi ^{\ast }f_{1}\right\vert
_{\imath \left( n\right) },\cdots ,\left. d\Phi ^{\ast }f_{r}\right\vert
_{\imath \left( n\right) }\right\} $ because $\Psi $ is a surjective map and 
$\Psi ^{\ast }$ is an injective one. Then, the set $\left\{ V_{\Psi ^{\ast
}f_{j}}\right\} _{i=1}^{r}$ is linearly independent too.

From theorem $\left( \ref{gs-3}\right) $ we know that $T_{\imath \left(
n\right) }^{\ast }\mathcal{M}=\omega \left( \imath _{\ast }\left( T_{n}%
\mathcal{N}\right) \right) \oplus \Psi _{\imath \left( n\right) }^{\ast
}\left( T_{p_{\circ }}^{\ast }\mathcal{P}\right) $, and because $\omega $ is
nondegenerate on $\imath _{\ast }\left( T_{n}\mathcal{N}\right) $ we can
apply $\omega ^{-1}$ on this direct sum to map it to the tangent space 
\begin{equation*}
T_{\imath \left( n\right) }\mathcal{M}=\imath _{\ast }\left( T_{n}\mathcal{N}%
\right) \oplus \omega ^{-1}\left( \Psi _{\imath \left( n\right) }^{\ast
}\left( T_{p_{\circ }}^{\ast }\mathcal{P}\right) \right)
\end{equation*}%
and this decomposition is symplectically orthogonal. This shows that%
\begin{equation*}
\omega ^{-1}\left( \Psi _{\imath \left( n\right) }^{\ast }\left( T_{p_{\circ
}}^{\ast }\mathcal{P}\right) \right) =\left[ \imath _{\ast }\left( T_{n}%
\mathcal{N}\right) \right] ^{\omega \bot }
\end{equation*}%
This finish the proof.$\blacksquare $

\begin{description}
\item[Corollary:] \textit{The set}%
\begin{equation*}
\left\{ \left. d\Psi ^{\ast }f_{1}\right\vert _{\imath \left( n\right)
},\cdots ,\left. d\Psi ^{\ast }f_{r}\right\vert _{\imath \left( n\right)
}\right\} \subset T_{\imath \left( n\right) }^{\ast }\mathcal{M}
\end{equation*}
\textit{is a basis of }$\left[ \imath _{\ast }\left( T_{n}\mathcal{N}\right) %
\right] ^{\circ }=\Psi _{\imath \left( n\right) }^{\ast }\left( T_{p_{\circ
}}^{\ast }\mathcal{P}\right) $.
\end{description}

Then, from the last four assertions, we conclude that in the \emph{second
class case} any vector $V\in T_{\imath \left( n\right) }\mathcal{M}$ can be
written as 
\begin{equation}
V_{\imath \left( n\right) }=\left( \imath _{\ast }v\right) _{\imath \left(
n\right) }+\sum_{i=1}^{r}a^{i}\left( n\right) V_{\Psi ^{\ast }f_{i}}
\label{gs-5}
\end{equation}%
where $v\in T_{n}\mathcal{N}$. Let us introduce the \emph{Dirac matrix} 
\begin{equation}
C_{jk}\left( \imath \left( n\right) \right) =\left\langle d\Psi ^{\ast
}f_{j},V_{\Psi ^{\ast }f_{k}}\right\rangle _{\imath \left( n\right)
}=\left\{ \Psi ^{\ast }f_{j},\Psi ^{\ast }f_{k}\right\} \left( \imath \left(
n\right) \right)  \label{gs-6}
\end{equation}%
Observe that, in the case of $\Psi $ being second class, $\omega \left(
\imath _{\ast }\left( T_{n}\mathcal{N}\right) \right) \cap \left. \Psi
^{\ast }\left( T_{p_{\circ }}^{\ast }\mathcal{P}\right) \right\vert _{\imath
\left( n\right) }=\left\{ 0\right\} $, there are no hamiltonian forms
associated to vector of $\imath _{\ast }\left( T_{n}\mathcal{N}\right) $
contained in $\left. \Psi ^{\ast }\left( T_{p_{\circ }}^{\ast }\mathcal{P}%
\right) \right\vert _{\imath \left( n\right) }$. Then, in this case the
vector space $\Psi _{\imath \left( n\right) }^{\ast }\left( T_{p_{\circ
}}^{\ast }\mathcal{P}\right) =\left[ \imath _{\ast }\left( T_{n}\mathcal{N}%
\right) \right] ^{\circ }$ is the dual of $\left[ \imath _{\ast }\left( T_{n}%
\mathcal{N}\right) \right] ^{\omega \bot }$%
\begin{equation*}
\Psi _{\imath \left( n\right) }^{\ast }\left( T_{p_{\circ }}^{\ast }\mathcal{%
P}\right) =\left( \left[ \imath _{\ast }\left( T_{n}\mathcal{N}\right) %
\right] ^{\omega \bot }\right) ^{\ast }
\end{equation*}%
so that the matrix obtained from the contraction of both the basis $\left\{
\left. d\Psi ^{\ast }f_{j}\right\vert _{\imath \left( n\right) }\right\} $
and $\left\{ V_{\Psi ^{\ast }f_{j}}\right\} _{i=1}^{r}$, namely $%
C_{jk}\left( \imath \left( n\right) \right) $, is an invertible one.

Contracting the vector $V_{\imath \left( n\right) }$ with $d\Psi ^{\ast
}f_{j}$, and because $\left\langle d\Psi ^{\ast }f_{k},\imath _{\ast
}v\right\rangle _{\imath \left( n\right) }=\left\langle df_{k},\Psi _{\ast
}\imath _{\ast }v\right\rangle _{p_{\circ }}=0$ (remember that $\imath
_{\ast }\left( T_{n}\mathcal{N}\right) =\ker \Psi _{\ast \imath \left(
n\right) }$), we get%
\begin{equation*}
a^{j}\left( n\right) =\sum_{k=1}^{r}C^{jk}\left( \imath \left( n\right)
\right) \left\langle d\Psi ^{\ast }f_{k},V\right\rangle _{\imath \left(
n\right) }
\end{equation*}%
Writing the vector $V,W$ in terms of the expression $\left( \ref{gs-5}%
\right) $, including the coefficients given in the last equation, we get
from the contraction $\left\langle \omega ,V\otimes W\right\rangle _{\imath
\left( n\right) }$ the following relation 
\begin{equation*}
\left\langle \imath ^{\ast }\omega ,v\otimes w\right\rangle
_{n}=\left\langle \omega ,V\otimes W\right\rangle _{\imath \left( n\right)
}+\left\langle \sum_{k,l=1}^{r}C^{lk}\left( \imath \left( n\right) \right)
~\Psi ^{\ast }\left( df_{l}\wedge df_{k}\right) ,V\otimes W\right\rangle
_{\imath \left( n\right) }
\end{equation*}

\begin{description}
\item[Lemma:] \textit{Let} \textit{be }$\alpha \in \Omega ^{1}\left( 
\mathcal{M}\right) $ \textit{and let be} $V_{\alpha }\in \mathfrak{X}\left( 
\mathcal{M}\right) $\textit{\ the associated hamiltonian vector field}. 
\textit{Let the surjection }$\Psi :\mathcal{M}\longrightarrow \mathcal{P}$ 
\textit{be a second class constraint, and }$\left. \pi _{\mathcal{N}%
}\right\vert _{n}:T_{\imath \left( n\right) }\mathcal{M}\longrightarrow T_{n}%
\mathcal{N}$ \textit{the projection on the first factor in the direct sum in 
}$\left( \ref{gs-5}\right) $\textit{. Then}%
\begin{equation*}
\imath _{\ast }v_{\imath ^{\ast }\alpha }=\pi _{\mathcal{N}}V_{\alpha }~~~%
\text{~.}
\end{equation*}
\end{description}

So, let us now specialize the expression of $\left\langle \imath ^{\ast
}\omega ,v\otimes w\right\rangle _{n}$ given above to the hamiltonian vector
fields like in the last Lemma, and defining as usual 
\begin{equation*}
\left\{ f,g\right\} ^{\mathcal{N}}\left( n\right) =\left\langle \imath
^{\ast }\omega ,v_{f}\otimes v_{g}\right\rangle _{n}
\end{equation*}%
$\forall $ $f,g\in \mathbb{C}^{\infty }\left( \mathcal{N}\right) $, we
obtain the celebrated Dirac formula relating the Poisson bracket on $%
\mathcal{N}$ with the one defined on $\mathcal{M}:$%
\begin{eqnarray}
\left\{ \imath ^{\ast }F,\imath ^{\ast }H\right\} ^{\mathcal{N}}\left(
n\right) &=&\left\{ F,H\right\} ^{\mathcal{M}}\left( \imath \left( n\right)
\right)  \label{gs-9} \\
&&-\sum_{l,k=1}^{r}~\left\{ F,\Psi ^{\ast }f_{l}\right\} ^{\mathcal{M}%
}\imath \left( n\right) ~C^{lk}\left( \imath \left( n\right) \right)
~\left\{ \Psi ^{\ast }f_{k},H\right\} ^{\mathcal{M}}\left( \imath \left(
n\right) \right)  \notag
\end{eqnarray}

\section{Symplectic submanifolds in cotangent bundle of double Lie group as
constrained system}

In this section we shall consider tangent and cotanget bundles of Lie groups
trivilized by left translation, a procedure equivalent to take global \emph{%
body coordinates}.

Given three Lie groups $\left( G,G_{+},G_{-}\right) $\ where $G_{+}$ and $%
G_{-}$ are both closed Lie subgroups of $G$, we say they form a\ \emph{%
double Lie group} if\ there exist a diffeomorphism $\alpha :G_{+}\times
G_{-}\longrightarrow G$\ defined as $\left( g_{+},g_{-}\right) \rightarrow
g_{+}g_{-}~$\cite{Lu-We}. In this case, the Lie algebras $\left( \mathfrak{g}%
,\mathfrak{g}_{+},\mathfrak{g}_{-}\right) $ form a\ \emph{double Lie algebra}
if $\mathfrak{g}_{+}$ and $\mathfrak{g}_{-}$ are Lie subalgebras of $%
\mathfrak{g}$, and $\mathfrak{g}=\mathfrak{g}_{+}\oplus \mathfrak{g}_{-}$ as
vector spaces. By duality, this decomposition gives rise also to the
factorization of $\mathfrak{g}^{\ast }=\mathfrak{g}_{-}^{\circ }\oplus 
\mathfrak{g}_{+}^{\circ }$, where%
\begin{equation*}
\mathfrak{g}_{\pm }^{\circ }=\left\{ \eta \in \mathfrak{g}^{\ast
}/\left\langle \eta ,X_{\pm }\right\rangle =0,~\forall X_{\pm }\in \mathfrak{%
g}_{\pm }^{\circ }\right\}
\end{equation*}%
that allows to make the identifications $\mathfrak{g}_{\pm }^{\ast }\cong 
\mathfrak{g}_{\mp }^{\circ }$. In particular, if $\mathfrak{g}$ is supplied
with a $Ad^{G}$-invariant nondegenerate symmetric bilinear form turning $%
\mathfrak{g}_{+}^{\circ },\mathfrak{g}_{-}^{\circ }$ into isotropic
subspaces, the previous identifications implies $\mathfrak{g}_{\pm }^{\ast
}\cong \mathfrak{g}_{\mp }$. Let us denote the corresponding projectors $\Pi
_{G_{\pm }}:G\longrightarrow G_{\pm }$, $\Pi _{\mathfrak{g}_{\pm }^{\circ }}:%
\mathfrak{g}\longrightarrow \mathfrak{g}_{\pm }^{\circ }$ and $\Pi _{%
\mathfrak{g}_{\pm }^{\ast }}:\mathfrak{g}^{\ast }\longrightarrow \mathfrak{g}%
_{\pm }^{\ast }$. We also denote $\eta _{\pm }=\Pi _{\mathfrak{g}_{\pm
}^{\ast }}\eta $.

Writing out every element $g\in G$ as $g=g_{+}g_{-}$, with $g_{+}\in G_+$
and $g_{-}\in G_{-}$, the product $g_{-}g_{+}$ in $G$ can be expressed as $%
g_{-}g_{+}=g_{+}^{g_{-}}g_{-}^{g_{+}}$, with $g_{+}^{g_{-}}\in G_{+}$ and $%
g_{-}^{g_{+}}\in G_{-}$ \cite{STS},\cite{Lu-We}. The \emph{dressing action}
of $G_{-}$ on $G_{+}$ is defined to be 
\begin{equation}
\mathsf{Dr}:G_{-}\times G_{+}\longrightarrow G_{+}\qquad /\qquad \mathsf{Dr}%
\left( h_{-},g_{+}\right) :=\Pi _{G_{+}}h_{-}g_{+}=g_{+}^{h_{-}}
\label{dress-G*}
\end{equation}%
The infinitesimal generator of this action at the point $g_{+}\in G_{+}$ is,
for $X_{-}\in \mathfrak{g}_{-}$, 
\begin{equation*}
X_{-}\longrightarrow g_{+}^{X_{-}}=-\left. \dfrac{d}{dt}\mathsf{Dr}\left(
e^{tX_{-}},g_{+}\right) \right\vert _{t=0}
\end{equation*}%
such that, for $X_{-},Y_{-}\in \mathfrak{g}_{-}$, we have $\left[
g_{+}^{X_{-}},g_{+}^{Y_{-}}\right] =-g_{+}^{\left[ X_{-},Y_{-}\right] _{%
\mathfrak{g}_{-}}}$. It satisfies the relation 
\begin{equation}
Ad_{g_{+}^{-1}}^{G}X_{-}=g_{+}^{-1}g_{+}^{X_{-}}+Ad_{g_{+}}^{\ast }X_{-}
\label{Ad=dress+}
\end{equation}%
where $Ad_{g_{+}^{-1}}^{G}X_{-}\in \mathfrak{g}$ is the adjoint action of $G$%
, and $Ad_{g_{+}}^{\ast }X_{-}\in \mathfrak{g}_{-}$ is the coadjoint action
of $G_{+}$ on the dual of its Lie algebra $\mathfrak{g}_{+}^{\ast }\cong 
\mathfrak{g}_{-}^{\circ }$. Then, we may write $g_{+}^{X_{-}}=g_{+}\Pi _{%
\mathfrak{g}_{+}}Ad_{g_{+}^{-1}}^{G}X_{-}$ $.$

Let us introduce the fibrations 
\begin{equation}
\begin{array}{l}
\begin{array}{c}
\Psi :G\times \mathfrak{g}^{\ast }\longrightarrow G_{-}\times \mathfrak{g}%
_{-}^{\ast } \\ 
\\ 
\left( g,\eta \right) \longmapsto \left( \Pi _{G_{-}}\left( g\right) ,\Pi _{%
\mathfrak{g}_{-}^{\ast }}\left( \eta \right) \right)%
\end{array}
\\ 
\\ 
\begin{array}{c}
\Upsilon :G\times \mathfrak{g}^{\ast }\longrightarrow G_{+}\times \mathfrak{g%
}_{+}^{\ast } \\ 
\\ 
\left( g,\eta \right) \longmapsto \left( \Pi _{G_{+}}\left( g\right) ,\Pi _{%
\mathfrak{g}_{+}^{\ast }}\left( \eta \right) \right)%
\end{array}%
\end{array}
\label{proyeccion-1}
\end{equation}%
with fibers%
\begin{eqnarray*}
\mathcal{N}\left( g_{-},\eta _{-}\right) &=&\left\{ \left( g,\eta \right)
\in G\times \mathfrak{g}^{\ast }/\Psi \left( g,\eta \right) =\left(
g_{-},\eta _{-}\right) \right\} \\
&& \\
\mathcal{M}\left( g_{+},\eta _{+}\right) &=&\left\{ \left( g,\eta \right)
\in G\times \mathfrak{g}^{\ast }/\Upsilon \left( g,\eta \right) =\left(
g_{+},\eta _{+}\right) \right\}
\end{eqnarray*}%
Observe that the corresponding fibers on $\left( e,0\right) \in G_{-}\times 
\mathfrak{g}_{-}^{\ast }$ reduce to the trivialized cotangent bundles of the
factors $G_{+}$, $G_{-}$, namely $\mathcal{N}\left( e,0\right) =G_{+}\times 
\mathfrak{g}_{+}^{\ast }\cong T^{\ast }G_{+}$ and $\mathcal{M}\left(
e,0\right) =G_{-}\times \mathfrak{g}_{-}^{\ast }\cong $ $T^{\ast }G_{-}$.

The differential of the maps $\Psi $ and $\Upsilon $ involve the following
digression: a vector $v\in T_{g}G$ with $g=g_{+}g_{-}$ can be written as 
\begin{equation*}
v=v_{+}g_{-}+g_{+}v_{-}
\end{equation*}%
that in terms of left translated vectors on $G_{+}\times \mathfrak{g}%
_{+}^{\circ }$ and $G_{-}\times \mathfrak{g}_{-}^{\circ }$ reads%
\begin{equation*}
\dot{g}=g_{+}\left( g_{+}^{-1}v_{+}\right) g_{-}+g_{+}g_{-}\left(
g_{-}^{-1}v_{-}\right)
\end{equation*}%
Hence, the relation with the left trivialization of $TG\cong G\times 
\mathfrak{g}$ arises from 
\begin{equation*}
g^{-1}v=Ad_{g_{-}^{-1}}^{G}\left( g_{+}^{-1}v_{+}\right) +\left(
g_{-}^{-1}v_{-}\right)
\end{equation*}%
from where we define 
\begin{eqnarray}
X_{+} &=&\Pi _{\mathfrak{g}_{+}^{\circ }}g^{-1}v=Ad_{g_{-}}^{\ast }\left(
g_{+}^{-1}v_{+}\right)  \notag \\
&&  \label{tangent decomp 1} \\
X_{-} &=&\Pi _{\mathfrak{g}_{-}^{\circ }}g^{-1}v=\left( g_{-}^{-1}\right)
^{g_{+}^{-1}v_{+}}g_{-}+g_{-}^{-1}v_{-}  \notag
\end{eqnarray}%
that are equivalent to%
\begin{equation}
\left\{ 
\begin{array}{l}
g_{+}^{-1}\dot{g}_{+}=Ad_{g_{-}^{-1}}^{\ast }X_{+} \\ 
\\ 
g_{-}^{-1}\dot{g}_{-}=X_{-}+g_{-}^{-1}g_{-}^{X_{+}}%
\end{array}%
\right.  \label{tangent decomp 2}
\end{equation}%
We shall use these results in the next sections where $\mathcal{N}\left(
g_{-},\eta _{-}\right) $ and $\mathcal{M}\left( g_{+},\eta _{+}\right) $
shall be regarded as phase spaces.

In fact, since the cotangent bundle $T^{\ast }G\cong G\times \mathfrak{g}%
^{\ast }$ is a symplectic manifold with the canonical 2-form $\omega _{\circ
}$, which has associated the nondegenerate Poisson bracket%
\begin{equation*}
\left\{ \mathcal{F},\mathcal{G}\right\} \left( g,\eta \right) =\left\langle g%
\mathbf{d}\mathcal{F},\delta \mathcal{G}\right\rangle _{\left( g,\eta
\right) }-\left\langle g\mathbf{d}\mathcal{G},\delta \mathcal{F}%
\right\rangle _{\left( g,\eta \right) }-\left\langle \eta ,\left[ \left. d%
\mathcal{F}\right\vert _{\left( g,\eta \right) },\left. d\mathcal{G}%
\right\vert _{\left( g,\eta \right) }\right] \right\rangle
\end{equation*}%
where we wrote $\left. d\mathcal{F}\right\vert _{\left( g,\eta \right)
}=\left. \left( \mathbf{d}\mathcal{F},\delta \mathcal{G}\right) \right\vert
_{\left( g,\eta \right) }\in T_{g}^{\ast }G\oplus \mathfrak{g}^{\ast }$, we
shall study how to apply the Dirac method to supply the submanifolds $%
\mathcal{N}\left( g_{-},\eta _{-}\right) $ and $\mathcal{M}\left( g_{+},\eta
_{+}\right) $ with a Poisson-Dirac structure.

\subsection{The phase spaces $\mathcal{N}\left( g_{-},\protect\eta %
_{-}\right) $}

Let us consider the restriction of the canonical Poisson bracket of $T^{\ast
}G\cong G\times \mathfrak{g}^{\ast }$ to the fibers $\mathcal{N}\left(
g_{-},\eta _{-}\right) :=\Psi ^{-1}\left( g_{-},\eta _{-}\right) $. Then the
differential map $\Psi _{\ast }:T\left( G\times \mathfrak{g}^{\ast }\right)
\longrightarrow T\left( G_{-}\times \mathfrak{g}_{-}^{\ast }\right) $ is%
\begin{equation}
\left. \Psi _{\ast }\left( gX,\xi \right) \right\vert _{\left( g,\eta
\right) }=\left. \frac{d}{dt}\Pi _{G_{-}}\left( ge^{tX}\right) \right\vert
_{t=0}=\left( g_{-}^{X_{+}}+g_{-}X_{-},\xi _{-}\right) _{\left( g_{-},\eta
_{-}\right) }  \label{proyeccion-2}
\end{equation}%
Therefore, the kernel of $\Psi _{\ast }$, that coincides with $T\mathcal{N}%
\left( g_{-},\eta _{-}\right) $, is%
\begin{equation}
\left. \ker \Psi _{\ast }\right\vert _{\left( g,\eta \right) }=\left\{
\left( g_{+}\left( Ad_{g_{-}^{-1}}^{\ast }X_{+}\right) g_{-},\xi _{+}\right)
~/~\left( X_{+},\xi _{+}\right) \in \mathfrak{g}_{+}^{\circ }\oplus 
\mathfrak{g}_{+}^{\ast }\right\}  \label{kernel psi}
\end{equation}

At each point $\left( g,\eta \right) \in \mathcal{N}\left( g_{-},\eta
_{-}\right) $, the intersection of $T_{\left( g,\eta \right) }\mathcal{N}%
\left( g_{-},\eta _{-}\right) $ with its symplectic orthogonal is $\left\{
0\right\} $, so we have the following result.

\begin{description}
\item[Proposition:] $\left( \mathcal{N}\left( g_{-},\eta _{-}\right) ,\tilde{%
\omega}_{\circ }\right) $\textit{, where }$\tilde{\omega}_{\circ }$ \textit{%
is the restriction to }$\mathcal{N}\left( g_{-},\eta _{-}\right) $\textit{\
of the canonical symplectic form }$\omega _{\circ }$ \textit{on }$G\times 
\mathfrak{g}^{\ast }$\textit{, is a symplectic manifold.}
\end{description}

\textbf{Proof:} The restriction of the symplectic form to $T\mathcal{N}%
\left( g_{-},\eta _{-}\right) $ reduces to%
\begin{equation}
\begin{array}{l}
\left\langle \omega _{\circ },\left( -g_{+}\left( Ad_{g_{-}^{-1}}^{\ast
}X_{+}\right) g_{-},\xi _{+}\right) \otimes \left( -g_{+}\left(
Ad_{g_{-}^{-1}}^{\ast }Y_{+}\right) g_{-},\lambda _{+}\right) \right\rangle
_{\left( g,\eta \right) } \\ 
=-\left\langle \xi _{+},Y_{+}\right\rangle +\left\langle \lambda
_{+},X_{+}\right\rangle +\left\langle Ad_{g_{-}}\eta _{+},\left[
Ad_{g_{-}^{-1}}^{\ast }X_{+},Ad_{g_{-}^{-1}}^{\ast }Y_{+}\right]
\right\rangle \\ 
~~~+\left\langle \eta _{-},\left( g_{-}^{-1}\right) ^{\left[
Ad_{g_{-}^{-1}}^{\ast }X_{+},Ad_{g_{-}^{-1}}^{\ast }Y_{+}\right]
}g_{-}\right\rangle%
\end{array}
\label{restricted sympl form}
\end{equation}%
Here we can see that there are no null vectors of $\omega _{\circ }$ on $%
T\Psi ^{-1}\left( g_{-},\eta _{-}\right) =\left. \ker \Psi _{\ast
}\right\vert _{\left( g,\eta \right) }$.$\blacksquare $

\begin{description}
\item[Corollary:] $\left( T_{\left( g,\eta \right) }\mathcal{N}\left(
g_{-},\eta _{-}\right) \right) ^{\bot }\cap T_{\left( g,\eta \right) }%
\mathcal{N}\left( g_{-},\eta _{-}\right) =\left\{ 0\right\} $\textit{, then }%
$\mathcal{N}\left( g_{-},\eta _{-}\right) $ \textit{is a }\emph{second class
constraint.}
\end{description}

\subsubsection{Dirac brackets on $\mathcal{N}\left( g_{-},\protect\eta %
_{-}\right) $}

In order to built up the Dirac brackets, we choose a basis for $T_{\left(
g_{-},\eta _{-}\right) }^{\ast }\left( G_{-}\times \mathfrak{g}_{-}^{\ast
}\right) $ $\cong \mathfrak{g}_{-}^{\ast }\oplus \mathfrak{g}_{-}^{\circ }$.
In doing so, we introduce the basis $\left\{ T_{a}\right\} _{a=1}^{n}$ of $%
\mathfrak{g}_{+}^{\circ }\cong \mathfrak{g}_{-}^{\ast }$ and the basis $%
\left\{ T^{a}\right\} _{a=1}^{n}$ of $\mathfrak{g}_{-}^{\circ }=\mathfrak{g}%
_{+}^{\ast }$, which provide a set of linearly independent 1-forms on $%
G_{-}\times \mathfrak{g}_{-}^{\ast }$: 
\begin{eqnarray*}
\alpha _{a} &=&\left( L_{g_{-}^{-1}}^{\ast }T_{a},0\right) \in T_{\left(
g_{-},\eta _{-}\right) }^{\ast }G_{-}\times \mathfrak{g}_{-}^{\ast } \\
\beta _{a} &=&\left( 0,T^{a}\right) \in T_{\left( g_{-},\eta _{-}\right)
}^{\ast }G_{-}\times \mathfrak{g}_{-}^{\ast }
\end{eqnarray*}%
\bigskip for $a=1,2,...,n$. Their pullback amount a set $\left\{ \Psi ^{\ast
}\alpha _{a}\right\} _{a=1}^{n}\cup \left\{ \Psi ^{\ast }\beta ^{a}\right\}
_{a=1}^{n}$ of linearly independent 1-forms on $G\times \mathfrak{g}^{\ast }$%
, whose null distribution is precisely the tangent space of $\mathcal{N}%
\left( g_{-},\eta _{-}\right) $.

Hence, the hamiltonian vector fields associated with these forms through $%
\omega _{\circ }$ are%
\begin{eqnarray}
V_{\Psi ^{\ast }\alpha _{a}} &=&\left(
0,g_{-}^{-1}g_{-}^{T_{a}}-T_{a}\right) _{\left( g,\eta \right) }  \notag \\
&&  \label{Vham alpha - beta} \\
V_{\Psi ^{\ast }\beta ^{a}} &=&\left( gT^{a},ad_{T^{a}}^{\mathfrak{g}\ast
}\eta \right)  \notag
\end{eqnarray}%
So, we calculate the entries of the Dirac matrix:

\begin{eqnarray*}
C_{\Psi ^{\ast }\alpha _{a},\Psi ^{\ast }\alpha _{b}}\left( g,\eta \right)
&=&\left\langle \Psi ^{\ast }\alpha _{a},V_{\Psi ^{\ast }\alpha
_{b}}\right\rangle _{\left( g,\eta \right) }=0 \\
&& \\
C_{\Psi ^{\ast }\beta ^{a},\Psi ^{\ast }\beta ^{b}}\left( g,\eta \right)
&=&\left\langle \Psi ^{\ast }\beta ^{a},V_{\Psi ^{\ast }\beta
^{b}}\right\rangle _{\left( g,\eta \right) }=-\left\langle \eta _{-},\left[
T^{a},T^{b}\right] \right\rangle \\
&& \\
C_{\Psi ^{\ast }\alpha _{a},\Psi ^{\ast }\beta ^{b}}\left( g,\eta \right)
&=&\left\langle \Psi ^{\ast }\alpha _{a},V_{\Psi ^{\ast }\beta
^{b}}\right\rangle _{\left( g,\eta \right) }=\delta _{a}^{b}
\end{eqnarray*}%
that finally produces the matrix%
\begin{equation}
C\left( g,\eta \right) =\left[ 
\begin{array}{cc}
0_{n\times n} & -I_{n\times n} \\ 
I_{n\times n} & \Omega \left( \eta \right)%
\end{array}%
\right]  \label{Dirac matrix}
\end{equation}%
where $\Omega \left( \eta \right) $ stands for the $n\times n$ matrix of
entries 
\begin{equation}
\Omega _{ab}\left( \eta \right) =-\left\langle \eta _{-},\left[ T^{a},T^{b}%
\right] \right\rangle  \label{omega}
\end{equation}

Now, we are ready to introduce the Dirac brackets: carrying these results in
the expression $\left( \ref{gs-9}\right) $ for any couple of function $%
\mathcal{F},\mathcal{G}\in C^{\infty }\left( G\times \mathfrak{g}^{\ast
}\right) $, the Dirac bracket gives the restriction of Poisson bracket on $%
G\times \mathfrak{g}^{\ast }$ to constraint submanifold $\mathcal{N}\left(
g_{-},\eta _{-}\right) $ and it is defined as 
\begin{eqnarray*}
\left\{ \mathcal{F},\mathcal{G}\right\} ^{\mathcal{N}}\left( g,\eta \right)
&=&\left\{ \mathcal{F},\mathcal{G}\right\} \left( g,\eta \right) -\left\{ 
\mathcal{F},\alpha _{a}\right\} \left( g,\eta \right) \Omega _{ab}\left(
\eta \right) \left\{ \alpha _{b},\mathcal{G}\right\} \left( g,\eta \right) \\
&&+\left\{ \mathcal{F},\alpha _{a}\right\} \left( g,\eta \right) \left\{
\beta ^{a},\mathcal{G}\right\} \left( g,\eta \right) -\left\{ \mathcal{F}%
,\beta ^{a}\right\} \left( g,\eta \right) \left\{ \alpha _{a},\mathcal{G}%
\right\} \left( g,\eta \right)
\end{eqnarray*}%
that has the explicit form 
\begin{eqnarray}
\left\{ \mathcal{F},\mathcal{G}\right\} ^{\mathcal{N}}\left( g,\eta \right)
&=&\left\langle g\mathbf{d}\mathcal{F},Ad_{g_{-}^{-1}}^{G}\Pi _{\mathfrak{g}%
_{+}^{\circ }}Ad_{g_{-}}^{G}\delta \mathcal{G}\right\rangle -\left\langle g%
\mathbf{d}\mathcal{G},Ad_{g_{-}^{-1}}^{G}\Pi _{\mathfrak{g}_{+}^{\circ
}}Ad_{g_{-}}^{G}\delta \mathcal{F}\right\rangle  \notag \\
&&-\left\langle \eta ,\left[ Ad_{g_{-}^{-1}}^{G}\Pi _{\mathfrak{g}%
_{+}^{\circ }}Ad_{g_{-}}^{G}\delta \mathcal{F},Ad_{g_{-}^{-1}}^{G}\Pi _{%
\mathfrak{g}_{+}^{\circ }}Ad_{g_{-}}^{G}\delta \mathcal{G}\right]
\right\rangle  \label{Dirac bracket on N -1}
\end{eqnarray}%
Observe that, for $\mathcal{N}\left( e,0\right) $, it turns in%
\begin{equation*}
\left\{ \mathcal{F},\mathcal{G}\right\} ^{\mathcal{N}}\left( g_{+},\eta
_{+}\right) =\left\langle g\mathbf{d}\mathcal{F},\delta \mathcal{G}%
_{+}\right\rangle -\left\langle g\mathbf{d}\mathcal{G},\delta \mathcal{F}%
_{+}\right\rangle -\left\langle \eta _{+},\left[ \delta \mathcal{F}%
_{+},\delta \mathcal{G}_{+}\right] \right\rangle
\end{equation*}%
giving the usual canonical Poisson structure on $T^{\ast }G_{+}=G_{+}\times 
\mathfrak{g}_{+}^{\ast }$.

\subsubsection{The fundamental brackets}

Let $\mathrm{T}:G\longrightarrow GL\left( n,\mathbb{C}\right) $ a
representation of the group $G$, with the associated set of functions $%
\mathrm{T}_{i}^{j}:G\longrightarrow \mathbb{C}$ such that $\mathrm{T}%
=E_{i}^{j}\mathrm{T}_{i}^{j}$, being $E_{i}^{j}$, $i,j=1,...,n$ , the
elementary $n\times n$ matrices with entries $\left[ E_{i}^{j}\right]
_{k}^{l}=\delta _{ik}\delta ^{kl}$. For $g\in G$ 
\begin{equation*}
\mathrm{T}_{i}^{j}\left( g\right) =g_{i}^{j}\Longrightarrow \mathrm{T}\left(
g\right) =\sum_{i,j=1}^{n}g_{i}^{j}E_{i}^{j}
\end{equation*}%
Also, we consider the coordinates $\{\xi _{A}\}_{A=1}^{N}$ for $\mathfrak{g}%
^{\ast }$, associated with the basis $\{\mathbb{T}_{A}\}_{A=1}^{N}$ of $%
\mathfrak{g}$, such that 
\begin{equation*}
\xi _{A}:\mathfrak{g}^{\ast }\longrightarrow \mathbb{C~}/~\xi
_{A}=\left\langle \xi ,\mathbb{T}_{A}\right\rangle
\end{equation*}%
The representation map $\mathrm{T}$ induces the map $d\mathrm{T}:\mathfrak{g}%
\longrightarrow gl\left( n,\mathbb{C}\right) $ such that $\left( d\mathrm{T}%
\right) _{e}X=\sum_{i,j=1}^{n}X_{i}^{j}E_{i}^{j}$.

The corresponding fundamental Poisson bracket on $G\times \mathfrak{g}^{\ast
}$, associated with the canonical Poisson bracket, are:%
\begin{eqnarray*}
\{\mathrm{T}_{i}^{j},\mathrm{T}_{k}^{l}\}(g,\xi ) &=&0 \\
&& \\
\{\xi _{A},\mathrm{T}_{i}^{j}\}(g,\xi ) &=&-\sum_{k=1}^{n}g_{i}^{k}\left[
T_{A}\right] _{k}^{j}
\end{eqnarray*}

Let us now calculate the Dirac brackets for these functions. First observe
that%
\begin{equation*}
\begin{array}{ccc}
\delta \mathrm{T}_{i}^{j}=0 & , & \mathbf{d}\xi _{A}=0%
\end{array}%
\end{equation*}%
hence, carrying these differential into the Dirac bracket $\left( \ref{Dirac
bracket on N -1}\right) $, we get the fundamental brackets on $\mathcal{N}%
\left( g_{-},\eta _{-}\right) $. The differential $\delta \xi _{A}$
coincides with the generator $T_{A}$ of the Lie algebra $\mathfrak{g}$,
being $T_{A}\in \left\{ T_{a},T^{a}\right\} _{a=1}^{n}$. These relations can
be written in terms of the coordinates for $\eta _{+}=\left\langle \eta
_{+},T_{a}\right\rangle \mathbf{t}_{a}=\xi _{a}\left( \eta \right) \mathbf{t}%
_{a}$ and $\eta _{-}=\left\langle \eta _{-},T^{a}\right\rangle \mathbf{t}%
^{a}=\xi ^{a}\left( \eta \right) \mathbf{t}^{a}$.

The fundamental Dirac brackets on the submanifold\textit{\ }$\Psi
^{-1}\left( g_{-},\eta _{-}\right) \subset G\times \mathfrak{g}^{\ast }$ are%
\begin{equation*}
\begin{array}{l}
\left\{ \mathrm{T}_{i}^{j},\mathrm{T}_{k}^{l}\right\} ^{D}\left( g,\eta
\right) =0 \\ 
\left\{ \xi _{a},\mathrm{T}_{i}^{j}\right\} ^{D}\left( g,\eta \right)
=\left\langle g\mathbf{d}T_{i}^{j},g_{-}^{-1}g_{-}^{T_{a}}-T_{a}\right\rangle
\\ 
\left\{ \xi ^{a},\mathrm{T}_{i}^{j}\right\} ^{D}\left( g,\eta \right) =0 \\ 
\left\{ \xi _{a},\xi _{b}\right\} ^{D}\left( g,\eta \right) =-\left\langle
\eta ,\left[ T_{a}-g_{-}^{-1}g_{-}^{T_{a}},T_{b}-g_{-}^{-1}g_{-}^{T_{b}}%
\right] \right\rangle \\ 
\left\{ \xi _{a},\xi ^{b}\right\} ^{D}\left( g,\eta \right) =0 \\ 
\left\{ \xi ^{a},\xi ^{b}\right\} ^{D}\left( g,\eta \right) =0%
\end{array}%
\end{equation*}

Since 
\begin{equation*}
\left\langle g\mathbf{d}T_{i}^{j},g_{-}^{-1}g_{-}^{T_{a}}-T_{a}\right\rangle
=\mathrm{T}_{i}^{k}\left( g_{+}\right) \left[ g_{-}^{T_{a}}\right] _{k}^{j}-%
\mathrm{T}_{i}^{k}\left( g\right) \left[ T_{a}\right] _{k}^{j}
\end{equation*}%
and from the definition of the Lie bracket in $\mathfrak{g}=\mathfrak{g}%
_{+}^{\circ }\oplus \mathfrak{g}_{-}^{\circ }$,%
\begin{eqnarray*}
\left[ T_{a}-g_{-}^{-1}g_{-}^{T_{a}},T_{b}-g_{-}^{-1}g_{-}^{T_{b}}\right] &=&%
\left[ T_{a},T_{b}\right] +ad_{g_{-}^{-1}g_{-}^{T_{a}}}^{\ast
}T_{b}-ad_{g_{-}^{-1}g_{-}^{T_{b}}}^{\ast }T_{a} \\
&&+\left[ g_{-}^{-1}g_{-}^{T_{a}},g_{-}^{-1}g_{-}^{T_{b}}\right]
+\,ad_{T_{a}}^{\ast }g_{-}^{-1}g_{-}^{T_{b}}-ad_{T_{b}}^{\ast
}g_{-}^{-1}g_{-}^{T_{a}}
\end{eqnarray*}%
the fundamental Dirac brackets can be written as 
\begin{equation}
\begin{array}{l}
\left\{ \mathrm{T}_{i}^{j},\mathrm{T}_{k}^{l}\right\} ^{D}\left( g,\eta
\right) =0 \\ 
\left\{ \xi _{a},\mathrm{T}_{i}^{j}\right\} ^{D}\left( g,\eta \right) =%
\mathrm{T}_{i}^{k}\left( g_{+}\right) \left[ g_{-}^{T_{a}}\right] _{k}^{j}-%
\mathrm{T}_{i}^{k}\left( g\right) \left[ T_{a}\right] _{k}^{j} \\ 
\left\{ \xi ^{a},\mathrm{T}_{i}^{j}\right\} ^{D}\left( g,\eta \right) =0 \\ 
\left\{ \xi _{a},\xi _{b}\right\} ^{D}\left( g,\eta \right) =-f_{ab}^{c}\xi
_{c}\left( \eta \right) +m_{ab}^{c}\left( g_{-}\right) \xi _{c}\left( \eta
\right) +n_{ab}^{c}\left( g_{-}\right) \xi ^{c}\left( \eta \right) \\ 
\left\{ \xi _{a},\xi ^{b}\right\} ^{D}\left( g,\eta \right) =0 \\ 
\left\{ \xi ^{a},\xi ^{b}\right\} ^{D}\left( g,\eta \right) =0%
\end{array}
\label{fund Dirac bracket}
\end{equation}%
where 
\begin{eqnarray*}
m_{ab}^{c}\left( g_{-}\right) &=&\left\langle \left[
g_{-}^{-1}g_{-}^{T_{b}},T^{c}\right] ,T_{a}\right\rangle -\left\langle \left[
g_{-}^{-1}g_{-}^{T_{a}},T^{c}\right] ,T_{b}\right\rangle \\
&& \\
n_{ab}^{c}\left( g_{-}\right) &=&\left\langle \left[ T_{b},T_{c}\right]
,g_{-}^{-1}g_{-}^{T_{a}}\right\rangle -\left\langle \left[ T_{a},T_{c}\right]
,\,g_{-}^{-1}g_{-}^{T_{b}}\right\rangle -\left\langle T_{c},\left[
g_{-}^{-1}g_{-}^{T_{a}},g_{-}^{-1}g_{-}^{T_{b}}\right] \right\rangle
\end{eqnarray*}%
are constant coefficients on each submanifold $\mathcal{N}\left( g_{-},\eta
_{-}\right) $ and, in particular, they vanish for $g_{-}=e$, $%
m_{ab}^{c}\left( e\right) =n_{ab}^{c}\left( e\right) =0$.

\subsubsection{The action of $G$ on $\mathcal{N}\left( g_{-},\protect\eta %
_{-}\right) \label{left action}$}

The group $G$ acts on itself by left translations, $L:G\times
G\longrightarrow G$ as 
\begin{equation*}
L_{a}g=ag=a_{+}a_{-}g_{+}g_{-}
\end{equation*}%
for $a_{+},g_{+}\in G_{+}$ and $a_{-},g_{-}\in G_{-}$. It can be easily
lifted to the cotangent bundle $T^{\ast }G\cong G\times \mathfrak{g}^{\ast }$%
, in body coordinates, as the hamiltonian action

\begin{equation*}
\begin{array}{ccc}
\rho :G\times \left( G\times \mathfrak{g}^{\ast }\right) \longrightarrow
G\times \mathfrak{g}^{\ast } & / & \rho _{h}(g,\eta )=(hg,\eta )%
\end{array}%
\end{equation*}%
with associated $Ad$-equivariant momentum map $\Phi ^{L}:G\times \mathfrak{g}%
^{\ast }\longrightarrow \mathfrak{g}^{\ast }$%
\begin{equation}
\Phi ^{L}(g,\eta )=Ad_{g^{-1}}^{G\ast }\eta  \label{left trans mom map}
\end{equation}%
So, the map 
\begin{equation*}
\phi _{X}(g,\eta )=\left\langle \eta ,Ad_{g^{-1}}^{G}X\right\rangle
\end{equation*}%
is the hamiltonian function associated with the infinitesimal generator $%
X_{G\times \mathfrak{g}^{\ast }}$ corresponding to $X\in \mathfrak{g}$,
namely $\imath _{X_{G\times \mathfrak{g}^{\ast }}}\omega _{\circ }=d\phi
_{X} $.

We now use the Dirac bracket $\left( \ref{Dirac bracket on N -1}\right) $ to
get the hamiltonian vector field $X_{\mathcal{N}}$ on $\mathcal{N}\left(
g_{-},\eta _{-}\right) $, associated with $X\in \mathfrak{g}$, such that%
\begin{equation*}
\left\langle d\mathcal{F},X_{\mathcal{N}}\right\rangle _{\left( g,\eta
\right) }:=\left\{ \mathcal{F},\phi _{X}\right\} ^{D}\left( g,\eta \right)
\end{equation*}%
Writing $d\phi _{X}=\left( \mathbf{d}\phi _{X},\delta \phi _{X}\right) \in
T^{\ast }G\oplus T^{\ast }\mathfrak{g}^{\ast }$, 
\begin{equation*}
\left( g\mathbf{d}\phi _{X},\delta \phi _{X}\right) =\left( \left[ \eta
,Ad_{g^{-1}}^{G}X\right] ,Ad_{g^{-1}}^{G}X\right)
\end{equation*}%
the Poisson-Dirac bracket turns into%
\begin{eqnarray*}
\left\{ \mathcal{F},\phi _{X}\right\} ^{D}\left( g,\eta \right) =
&&\left\langle g\mathbf{d}\mathcal{F},Ad_{g_{-}^{-1}}^{G}\Pi _{\mathfrak{g}%
_{+}^{\circ }}Ad_{g_{+}^{-1}}^{G}X\right\rangle \\
&&-\left\langle \delta \mathcal{F},\left[ \eta ,Ad_{g_{-}^{-1}}^{G}\Pi _{%
\mathfrak{g}_{-}^{\circ }}Ad_{g_{+}^{-1}}^{G}X\right] \right\rangle \\
&&+\left\langle \delta \mathcal{F},Ad_{g_{-}^{-1}}^{G}\Pi _{\mathfrak{g}%
_{+}^{\circ }}Ad_{g_{-}}^{G}\left[ \eta ,Ad_{g_{-}^{-1}}^{G}\Pi _{\mathfrak{g%
}_{-}^{\circ }}Ad_{g_{+}^{-1}}^{G}X\right] \right\rangle
\end{eqnarray*}%
and from it we get the hamiltonian vector field associated with $\phi _{X}$, 
\begin{eqnarray}
X_{\mathcal{N}}\left( g,\eta \right) &=&V_{\phi _{X}}^{\mathcal{N}\left(
g_{-},\eta _{-}\right) }\left( g,\eta \right)  \label{ham vec field 0} \\
&=&\left( gAd_{g_{-}^{-1}}^{G}\Pi _{\mathfrak{g}_{+}^{\circ
}}Ad_{g_{+}^{-1}}^{G}X,Ad_{g_{-}^{-1}}^{G}\Pi _{\mathfrak{g}_{-}^{\circ }}%
\left[ \Pi _{\mathfrak{g}_{-}^{\circ
}}Ad_{g_{+}^{-1}}^{G}X,Ad_{g_{-}}^{G}\eta \right] \right)  \notag
\end{eqnarray}%
These vector fields are the projection on $T\mathcal{N}\left( g_{-},\eta
_{-}\right) $ of the \emph{infinitesimal generators} $X_{G\times \mathfrak{g}%
^{\ast }}$ of the action of $G$ on $G\times \mathfrak{g}^{\ast }$ associated
with $X\in \mathfrak{g}$. However, it is not clear at this point whether
these vector fields are infinitesimal generators for an action of $G$ on $%
\mathcal{N}\left( g_{-},\eta _{-}\right) $. This question is addressed in
the following proposition.

\begin{description}
\item[Proposition:] \textit{The assignment }$X\in \mathfrak{g}%
\longrightarrow X_{\mathcal{N}}\in \mathfrak{X}\left( \mathcal{N}\left(
g_{-},\eta _{-}\right) \right) $ \textit{defines a hamiltonian action of the
Lie algebra }$\mathfrak{g}$ \textit{on }$\mathcal{N}\left( g_{-},\eta
_{-}\right) $ \textit{provided }$\eta _{-}$ \textit{is a character of }$%
\mathfrak{g}_{-}^{\circ }$\textit{.}
\end{description}

\textbf{Proof: }The Dirac bracket of the hamiltonian functions $\phi
_{X},\phi _{Y}$ is 
\begin{equation*}
\left\{ \phi _{X},\phi _{Y}\right\} ^{D}\left( g,\eta \right) =\phi _{\left[
X,Y\right] }\left( g,\eta \right) -\left\langle Ad_{g_{-}^{-1}}^{\ast }\eta
_{-},\left[ \Pi _{\mathfrak{g}_{-}^{\circ }}Ad_{g_{+}^{-1}}^{G}X,\Pi _{%
\mathfrak{g}_{-}^{\circ }}Ad_{g_{+}^{-1}}^{G}Y\right] \right\rangle
\end{equation*}%
where it is obvious that the second term in the rhs vanish for every $X$ and 
$Y$ in $\mathfrak{g}$ only if $\eta _{-}$ \emph{is character of} $\mathfrak{g%
}_{-}^{\circ }$. In this case, it is easy to see using the Jacobi identity
that, for an arbitrary function $f$ on $G\times \mathfrak{g}^{\ast }$, 
\begin{eqnarray*}
\left( \mathbf{L}_{Y_{\mathcal{N}}}\mathbf{L}_{X_{\mathcal{N}}}-\mathbf{L}%
_{X_{\mathcal{N}}}\mathbf{L}_{Y_{\mathcal{N}}}\right) f\left( g,\eta \right)
&=&\left\{ \left\{ f,\phi _{X}\right\} ^{D},\phi _{Y}\right\} ^{D}\left(
g,\eta \right) \\
&&+\left\{ \left\{ \phi _{Y},f\right\} ^{D},\phi _{X}\right\} ^{D}\left(
g,\eta \right) \\
&=&\mathbf{L}_{\left[ X,Y\right] _{\mathcal{N}}}f\left( g,\eta \right)
\end{eqnarray*}%
that is equivalent to say that the assignment $X\in \mathfrak{g}%
\longrightarrow X_{\mathcal{N}}\in \mathfrak{X}\left( \mathcal{N}\left(
g_{-},\eta _{-}\right) \right) $ is an antihomomorphism of Lie algebras 
\begin{equation*}
\left[ X_{\mathcal{N}},Y_{\mathcal{N}}\right] =-\left[ X,Y\right] _{\mathcal{%
N}}
\end{equation*}%
Therefore, it defines a left action of $\mathfrak{g}$ on\textit{\ }$\mathcal{%
N}\left( g_{-},\eta _{-}\right) $.$\blacksquare $

In particular, at $\eta _{-}=0$ and $g_{-}=e$, we get%
\begin{equation*}
X_{\mathcal{N}}\left( g_{+},\eta _{+}\right) =\left( g_{+}\left( \Pi _{%
\mathfrak{g}_{+}^{\circ }}Ad_{g_{+}^{-1}}^{G}X\right) ,-\Pi _{\mathfrak{g}%
_{-}^{\circ }}\left[ \eta ,\Pi _{\mathfrak{g}_{-}^{\circ }}Ad_{g^{-1}}^{G}X%
\right] \right)
\end{equation*}

Observe that, for $X\in \left\{ T^{a}\right\} _{a=1}^{n}\subset \mathfrak{g}%
_{-}^{\circ }$, the momentum functions $\phi ^{a}:=\phi _{T^{a}}$ generate
the infinitesimal \emph{dressing action} of $\mathfrak{g}_{-}^{\circ }$ on $%
\mathcal{N}\left( g_{-},\eta _{-}\right) $.

This infinitesimal action of $\mathfrak{g}$ on\textit{\ }$\mathcal{N}\left(
g_{-},\eta _{-}\right) $ corresponds to the following action of $G$ on $%
\mathcal{N}\left( g_{-},\eta _{-}\right) $.

\begin{description}
\item[Proposition:] \textit{The vector field }$X_{\mathcal{N}}\in \mathfrak{X%
}\left( \mathcal{N}\left( g_{-},\eta _{-}\right) \right) $\textit{,} \textit{%
for }$X\in \mathfrak{g}$\textit{\ and }$\eta _{-}$ \textit{a character of }$%
\mathfrak{g}_{-}^{\circ }$\textit{, is the infinitesimal generator
associated with the action }$G\times \mathcal{N}\left( g_{-},\eta
_{-}\right) \longrightarrow \mathcal{N}\left( g_{-},\eta _{-}\right) $ 
\textit{defined as}%
\begin{equation}
\mathsf{d}\left( h,\left( g,\eta \right) \right) =\left(
gAd_{g_{-}^{-1}}^{G}\Pi _{\mathfrak{g}_{+}^{\circ }}\left(
g_{+}^{-1}hg_{+}\right) ,Ad_{g_{-}^{-1}}^{G}\Pi _{\mathfrak{g}_{-}^{\circ
}}Ad_{\Pi _{G_{-}}\left( g_{+}^{-1}hg_{+}\right) }^{G}Ad_{g_{-}}^{G}\eta
\right)  \label{SymplecticInducedAction}
\end{equation}%
$\forall $ $\left( g,\eta \right) =\left( g_{+}g_{-},\eta _{+}+\eta
_{-}\right) \in \mathcal{N}\left( g_{-},\eta _{-}\right) $.
\end{description}

\textbf{Proof: }It follows by straightforward calculation of the
differential of this map.$\blacksquare $

Observe that it can be written as 
\begin{equation*}
\begin{array}{l}
\mathsf{d}\left( h,\left( g_{+}g_{-},\eta _{+}+\eta _{-}\right) \right) \\ 
=\left( h_{+}g_{+}^{h_{-}}g_{-},Ad_{g_{-}^{-1}}^{G}\left[ \left(
h_{-}^{g_{+}}\right) ^{\eta _{-}^{h_{-}}}\left( h_{-}^{g_{+}}\right)
^{-1}+Ad_{h_{-}^{g_{+}}}^{G}g_{-}^{\eta _{-}}g_{-}^{-1}\right] \right. + \\ 
\qquad \qquad \qquad \qquad \qquad \qquad \qquad \qquad \qquad \left.
+Ad_{g_{-}^{-1}h_{-}^{g_{+}}g_{-}}\eta _{+}+\eta _{-}\right)%
\end{array}%
\end{equation*}%
so that for $g_{-}=e$ and $\eta _{-}=0$ it turns into%
\begin{equation*}
\mathsf{d}\left( h,\left( g_{+},\eta _{+}\right) \right) =\left(
h_{+}g_{+}^{h_{-}},Ad_{h_{-}^{g_{+}}}\eta _{+}\right)
\end{equation*}%
This action was introduced in \cite{CM} as the fundamental ingredient
underlying the Poisson Lie $T$-duality scheme.

\begin{description}
\item[Note:] The submanifolds we refer above are particular members of a
bigger family of symplectic submanifolds in $G\times \mathfrak{g}^{\ast }$
which are $G$-spaces and can be constructed by means of reduction theory 
\cite{MW}. In this framework, the procedure shows in particular that $%
\mathcal{N}\left( g_{-},\eta _{-}\right) $ and $\mathcal{M}\left( g_{+},\eta
_{+}\right) $ are symplectic submanifolds of $G\times \mathfrak{g}^{\ast }$
if $\eta _{\pm }$ are characters of the coadjoint action of $G$ on $%
\mathfrak{g}^{\ast }$, respectively. Additionally, it gives an
interpretation for the actions of the factorizable Lie group $G=G_{+}G_{-}$
on $\mathcal{N}\left( g_{-},\eta _{-}\right) $ and $\mathcal{M}\left(
g_{+},\eta _{+}\right) $, providing us an explanation to the symplecticity
of these actions. For instance, in the case of $\mathcal{N}\left( g_{-},\eta
_{-}\right) $ we can think of it as follows: the right action of $G_{-}$ on $%
G$ induces on $G\times \mathfrak{g}^{\ast }$ a symplectic action by lifting
and, by applying the Marsden-Weinstein reduction via its momentum map $%
J_{-}:G\times \mathfrak{g}^{\ast }\rightarrow \mathfrak{g}_{-}^{\ast }\simeq 
\mathfrak{g}_{+}^{\circ }$, for $\eta _{-}\in \mathfrak{g}_{+}^{\circ }$, we
get the quotient map 
\begin{equation*}
\pi _{\eta _{-}}:J_{-} ^{-1}\left( \eta _{-}\right) \rightarrow J_{-}
^{-1}\left( \eta _{-}\right) /\left( G_{-}\right) _{\eta _{-}}
\end{equation*}%
such that the symplectic form $\omega _{\eta _{-}}$ on $J_{-} ^{-1}\left(
\eta _{-}\right) /\left( G_{-}\right) _{\eta _{-}}$ is defined by the
condition 
\begin{equation*}
\pi _{\eta _{-}}^{\ast }\omega _{\eta _{-}}=\left. \omega _{\circ
}\right\vert \left( J_{-}\right) ^{-1}\left( \eta _{-}\right) ,
\end{equation*}%
where $\omega _{\circ }$ indicates the canonical $2$-form on $G\times 
\mathfrak{g}^{\ast }$. In this context we have the following remarkable
facts:

\begin{itemize}
\item for $\eta _{-}\in \mathfrak{g}_{+}^{\circ }$ a character, we have that 
$\left( G_{-}\right) _{\eta _{-}}=G_{-}$ for its isotropy group,

\item any submanifold $S\subset J_{-} ^{-1}\left( \eta _{-}\right) $
transverse to the $G_{-}$-orbits and such that $\dim S=\dim \left[ J_{-}
^{-1}\left( \eta _{-}\right) /G_{-}\right] $ (i.e. $S$ is a cross-section
for the $G_{-}$-action on $J_{-} ^{-1}\left( \eta _{-}\right) $), is a
symplectic submanifold of $G\times \mathfrak{g}^{\ast }$ via the restriction
of the canonical $2$-form, and

\item $\left. \pi _{\eta _{-}}\right\vert S$ is a symplectomorphism.
\end{itemize}

\item By viewing $\mathcal{N}\left( g_{-},\eta _{-}\right) $ ($\eta _{-}$ a
character!) as submanifolds of $J_{-} ^{-1}\left( \eta _{-}\right) $
transverse to the $G_{-}$-orbits, we find that they are symplectic
submanifolds of $G\times \mathfrak{g}^{\ast }$ and symplectomorphic to the
Marsden-Weinstein reduced space $J_{-} ^{-1}\left( \eta _{-}\right) /G_{-}$. 
\newline
On the other side, it can be proved that the Marsden-Weinstein reduced space
obtained above is symplectomorphic to the cotangent bundle of a (reductive)
homogeneous space $M$ of $G$, and this fact has two consequences:

\begin{itemize}
\item any slice in $J_{-} ^{-1}\left( \eta _{-}\right) $ is symplectomorphic
to $T^{\ast }M$, and

\item the lifted canonical action of $G$ on $T^{\ast }M$ induces a
symplectic action on these slices.
\end{itemize}

\item We see that the action of $G$ on $\mathcal{N}\left( g_{-},\eta
_{-}\right) $ defined above \eqref{SymplecticInducedAction} 
comes from this construction. \newline
Summarizing, we can fit all the relevant structures in the following
diagram: 
\begin{equation*}
\begin{diagram}
\node{\mathcal{N}\left(g_-,\eta_-\right)}\arrow{e,J}\arrow{s,..,l}{\left.\pi%
\right|\mathcal{N}\left(g_-,\eta_-\right)}\node{J_-^{-1}\left(\eta_-\right)}%
\arrow{s,r}{\pi}\\
\node{T^*M}\arrow{e,=,b}\node{J_-^{-1}\left(\eta_-\right)/G_-} \end{diagram}
\end{equation*}%
The dotted arrow can be viewed as a consequence of the rest of the
structures in the diagram, inducing the action while keeps it symplectic.
The particular case $\left( g_{-},\eta _{-}\right) =\left( e,0\right) $
gives the the $G$-action on $G_{+}$ considered in \cite{CM}.
\end{description}

\subsubsection{The Hamilton equations on $\mathcal{N}\left( g_{-},\protect%
\eta _{-}\right) $}

Let us consider a generic hamiltonian function $\mathcal{H}$ on $G\times 
\mathfrak{g}^{\ast }$, so the associated hamiltonian vector field by the
Poisson-Dirac structure $\left( \ref{Dirac bracket on N -1}\right) $ is%
\begin{equation*}
\begin{array}{l}
V_{\mathcal{H}}^{\mathcal{N}}\left( g,\eta \right) \\ 
=\left( g~Ad_{g_{-}^{-1}}^{G}\Pi _{\mathfrak{g}_{+}^{\circ
}}Ad_{g_{-}}^{G}\delta \mathcal{H},Ad_{g_{-}^{-1}}^{G}\Pi _{\mathfrak{g}%
_{-}^{\circ }}Ad_{g_{-}}^{G}\left( \left[ \eta ,Ad_{g_{-}^{-1}}^{G}\Pi _{%
\mathfrak{g}_{+}^{\circ }}Ad_{g_{-}}^{G}\delta \mathcal{H}\right] -g\mathbf{d%
}\mathcal{H}\right) \right)%
\end{array}%
\end{equation*}%
and the Hamilton equation are 
\begin{equation*}
\left\{ 
\begin{array}{l}
g^{-1}\dot{g}=\left( Ad_{g_{-}^{-1}}^{G}\Pi _{\mathfrak{g}_{+}^{\circ
}}Ad_{g_{-}}^{G}\right) \delta \mathcal{H} \\ 
\\ 
\dot{\eta}=\left( Ad_{g_{-}^{-1}}^{G}\Pi _{\mathfrak{g}_{-}^{\circ
}}Ad_{g_{-}}^{G}\right) \left( \left[ \eta ,Ad_{g_{-}^{-1}}^{G}\Pi _{%
\mathfrak{g}_{+}^{\circ }}Ad_{g_{-}}^{G}\delta \mathcal{H}\right] -g\mathbf{d%
}\mathcal{H}\right)%
\end{array}%
\right.
\end{equation*}%
In terms of the factors, $g=g_{+}g_{-}$ and $\eta =\eta _{+}+\eta _{-}$,
they are equivalent to%
\begin{equation*}
\begin{array}{l}
\left\{ 
\begin{array}{l}
g_{+}^{-1}\dot{g}_{+}=\Pi _{\mathfrak{g}_{+}^{\circ }}Ad_{g_{-}}^{G}\delta 
\mathcal{H} \\ 
\\ 
\dot{\eta}_{+}=Ad_{g_{-}^{-1}}^{G}\Pi _{\mathfrak{g}_{-}^{\circ
}}Ad_{g_{-}}^{G}\left( \left[ \eta ,Ad_{g_{-}^{-1}}^{G}\Pi _{\mathfrak{g}%
_{+}^{\circ }}Ad_{g_{-}}^{G}\delta \mathcal{H}\right] -g\mathbf{d}\mathcal{H}%
\right)%
\end{array}%
\right. \\ 
\\ 
\left\{ 
\begin{array}{l}
\dot{g}_{-}g_{-}^{-1}=0 \\ 
\\ 
\dot{\eta}_{-}=0%
\end{array}%
\right.%
\end{array}%
\end{equation*}

By using the fundamental coordinates, the Hamilton equation on $\mathcal{N}%
\left( g_{-},\eta _{-}\right) $ are%
\begin{equation*}
\left\{ 
\begin{array}{c}
\mathrm{\dot{T}}_{j}^{k}=\left\{ \mathrm{T}_{j}^{k},\mathcal{H}\right\} ^{D}
\\ 
\\ 
\dot{\xi}_{a}=\left\{ \xi _{a},\mathcal{H}\right\} ^{D}%
\end{array}%
\right.
\end{equation*}%
that turn out to be%
\begin{equation*}
\left\{ 
\begin{array}{l}
\mathrm{\dot{T}}_{j}^{k}=\dfrac{\partial \mathcal{H}}{\partial \xi _{b}}%
\left\{ \mathrm{T}_{j}^{k},\xi _{b}\right\} ^{D} \\ 
\\ 
\dot{\xi}_{a}=\dfrac{\partial \mathcal{H}}{\partial \mathrm{T}_{i}^{j}}%
\left\{ \xi _{a},\mathrm{T}_{i}^{j}\right\} ^{D}+\dfrac{\partial \mathcal{H}%
}{\partial \xi _{b}}\left\{ \xi _{a},\xi _{b}\right\} ^{D}%
\end{array}%
\right.
\end{equation*}

\subsection{The phase spaces $\mathcal{M}\left( g_{+},\protect\eta %
_{+}\right) $}

Let us consider the fibration $\Upsilon :G\times \mathfrak{g}^{\ast
}\longrightarrow G_{+}\times \mathfrak{g}_{+}^{\ast }$ with fiber $\mathcal{M%
}\left( g_{+},\eta _{-}\right) $ at $\left( g_{+},\eta _{-}\right) \in
G_{+}\times \mathfrak{g}_{+}^{\ast }$. The differential of the projection of
a vector $\left( v,\xi \right) \in G\times \mathfrak{g}^{\ast }$ on $%
G_{+}\times \mathfrak{g}_{+}^{\ast }$ is 
\begin{equation*}
\Pi _{+\ast }\left( v,\xi \right) =\left( g_{+}Ad_{g_{-}^{-1}}^{\ast
}X_{+},\xi _{+}\right) _{\left( g_{+},\eta _{+}\right) }
\end{equation*}%
where $X_{+}=\Pi _{\mathfrak{g}_{+}^{\circ }}g^{-1}v$ and $\xi =\xi _{+}+\xi
_{-}$. Hence, coming back to the expression for $\Upsilon _{\ast }$, we have%
\begin{equation*}
\left. \Upsilon _{\ast }\left( gX,\xi \right) \right\vert _{\left( g,\eta
\right) }=\left( g_{+}Ad_{g_{-}^{-1}}^{\ast }X_{+},\xi _{+}\right) _{\left(
g_{+},\eta _{+}\right) }
\end{equation*}

Let us find out $\left. \ker \Upsilon _{\ast }\right\vert _{\left( g,\eta
\right) }=T_{\left( g,\eta \right) }\mathcal{M}\left( g_{+},\eta _{+}\right) 
$, using $\left( \ref{tangent decomp 2}\right) $ we get%
\begin{equation*}
\left. \ker \Upsilon _{\ast }\right\vert _{\left( g,\eta \right) }=\left\{
\left( gX_{-},\xi _{-}\right) ~/~\left( X_{-},\xi _{-}\right) \in \mathfrak{g%
}_{-}^{\circ }\oplus \mathfrak{g}_{-}^{\ast }\right\}
\end{equation*}%
We may use this result to analyze the intersection of $\ker \Upsilon _{\ast
} $ with the symplectic orthogonal of $T\Upsilon ^{-1}\left( g_{+},\eta
_{+}\right) $. In order to do this we evaluate 
\begin{equation*}
\left\langle \omega ,\left( gX_{-},\xi _{-}\right) \otimes \left(
gY_{-},\lambda _{-}\right) \right\rangle _{\left( g,\eta \right)
}=-\left\langle \xi _{-},Y_{-}\right\rangle +\left\langle \lambda
_{-},X_{-}\right\rangle +\left\langle \eta _{-},\left[ X_{-},Y_{-}\right]
\right\rangle
\end{equation*}%
that is regular on $T\mathcal{M}\left( g_{+},\eta _{+}\right) $, so that $%
\left[ T_{\left( g,\eta \right) }\mathcal{M}\left( g_{+},\eta _{+}\right) %
\right] ^{\bot \omega }\cap \left. \ker \Upsilon _{\ast }\right\vert
_{\left( g,\eta \right) }=\left\{ 0\right\} $.

\begin{description}
\item[Proposition:] $\left( \mathcal{M}\left( g_{+},\eta _{+}\right) ,\hat{%
\omega}_{\circ }\right) $\textit{, where }$\hat{\omega}_{\circ }$ \textit{%
stands for the restriction of the canonical symplectic form }$\omega _{\circ
}$ \textit{on }$G\times \mathfrak{g}^{\ast }$ \textit{to }$\Upsilon
^{-1}\left( g_{+},\eta _{+}\right) $\textit{, is a symplectic manifold.
Moreover }$\hat{\omega}_{\circ }$ \textit{coincides with the canonical
symplectic form of} $G_{-}\times \mathfrak{g}_{-}^{\ast }$.
\end{description}

Consequently, the restriction to $\Upsilon ^{-1}\left( g_{+},\eta
_{+}\right) $ is a\textit{\ }\emph{second class constraint}, as expected.

In order to built up the Dirac brackets, we take a set of linearly
independent 1-forms on $G_{+}\times \mathfrak{g}_{+}^{\ast }$: 
\begin{eqnarray*}
\theta _{a} &=&\left( L_{g_{+}^{-1}}^{\ast }T^{a},0\right) \in T_{\left(
g_{+},\eta _{+}\right) }^{\ast }\left( G_{+}\times \mathfrak{g}_{+}^{\ast
}\right) \\
\gamma _{a} &=&\left( 0,T_{a}\right) \in T_{\left( g_{+},\eta _{+}\right)
}^{\ast }\left( G_{+}\times \mathfrak{g}_{+}^{\ast }\right)
\end{eqnarray*}%
Then their pullback $\left\{ \Upsilon ^{\ast }\theta ^{a},\Upsilon ^{\ast
}\gamma _{a}\right\} _{a=1}^{n}$ are a set of linearly independent 1-forms
on $G\times \mathfrak{g}^{\ast }$, such that $\left( \left\{ \Upsilon ^{\ast
}\theta ^{a},\Upsilon ^{\ast }\gamma _{a}\right\} _{a=1}^{n}\right) ^{\circ
}=T\mathcal{M}\left( g_{+},\eta _{+}\right) $. The hamiltonian vector fields
of these 1-forms on $G\times \mathfrak{g}^{\ast }$ are%
\begin{eqnarray*}
\left. V_{\Upsilon ^{\ast }\theta ^{a}}\right\vert _{\left( g,\eta \right) }
&=&\left( 0,-Ad_{g_{-}^{-1}}T^{a}\right) _{\left( g,\eta \right) } \\
&& \\
\left. V_{\Upsilon ^{\ast }\gamma _{a}}\right\vert _{\left( g,\eta \right) }
&=&\left( gT_{a},ad_{T_{a}}^{\mathfrak{g}\ast }\eta \right)
\end{eqnarray*}%
and the Dirac matrix 
\begin{equation*}
C=\left( 
\begin{array}{cc}
\left\langle \Upsilon ^{\ast }\theta ^{a},V_{\Upsilon ^{\ast }\theta
^{b}}\right\rangle _{\left( g,\eta \right) } & \left\langle \Upsilon ^{\ast
}\theta ^{a},V_{\Upsilon ^{\ast }\gamma _{b}}\right\rangle _{\left( g,\eta
\right) } \\ 
\left\langle \Upsilon ^{\ast }\gamma _{a},V_{\Upsilon ^{\ast }\theta
^{b}}\right\rangle _{\left( g,\eta \right) } & \left\langle \Upsilon ^{\ast
}\gamma _{a},V_{\Upsilon ^{\ast }\gamma _{b}}\right\rangle _{\left( g,\eta
\right) }%
\end{array}%
\right)
\end{equation*}%
is%
\begin{equation*}
C\left( g,\eta \right) =\left[ 
\begin{array}{cc}
0_{n\times n} & F\left( g_{-}\right) \\ 
-F^{\top }\left( g_{-}\right) & \Theta \left( \eta _{+}\right)%
\end{array}%
\right]
\end{equation*}%
where $\Theta \left( \eta _{+}\right) $ is a $n\times n$ matrix with entries 
\begin{equation*}
\Theta _{ab}\left( \eta _{+}\right) =\left\{ \Upsilon ^{\ast }\gamma
_{a},\Upsilon ^{\ast }\gamma _{b}\right\} \left( \eta _{+}\right)
=-\left\langle \eta _{+},\left[ T_{a},T_{b}\right] \right\rangle
\end{equation*}%
and 
\begin{equation*}
F_{b}^{a}\left( g_{-}\right) =\left\{ \Upsilon ^{\ast }\theta ^{a},\Upsilon
^{\ast }\gamma _{b}\right\} \left( g_{-}\right) =\left\langle
Ad_{g_{-}^{-1}}T^{a},T_{b}\right\rangle
\end{equation*}

The inverse of the Dirac matrix is 
\begin{equation*}
C^{-1}\left( g,\eta \right) =\left( 
\begin{array}{cc}
F_{a}^{c}\left( g_{-}^{-1}\right) \Theta _{cd}\left( \eta _{+}\right)
F_{b}^{d}\left( g_{-}^{-1}\right) & -F_{a}^{b}\left( g_{-}^{-1}\right) \\ 
F_{b}^{a}\left( g_{-}^{-1}\right) & 0_{n\times n}%
\end{array}%
\right)
\end{equation*}%
that reduces to 
\begin{equation*}
C^{-1}\left( g,\eta \right) =\left( 
\begin{array}{cc}
-\left\langle \eta _{+},\left[ Ad_{g_{-}}^{\ast }T_{a},Ad_{g_{-}}^{\ast
}T_{b}\right] \right\rangle & -\left\langle
Ad_{g_{-}}T^{b},T_{a}\right\rangle \\ 
\left\langle Ad_{g_{-}}T^{a},T_{b}\right\rangle & 0_{n\times n}%
\end{array}%
\right)
\end{equation*}

We are now ready to construct the Dirac bracket on $\mathcal{M}\left(
g_{+},\eta _{+}\right) $. Introducing this result into $\left( \ref{gs-9}%
\right) $, we get%
\begin{eqnarray}
\left\{ \mathcal{F},\mathcal{H}\right\} ^{\mathcal{M}}\left( g,\eta \right)
&=&\left\langle g\mathbf{d}\mathcal{F},\Pi _{\mathfrak{g}_{-}^{\circ
}}\delta \mathcal{H}\right\rangle -\left\langle g\mathbf{d}\mathcal{H},\Pi _{%
\mathfrak{g}_{-}^{\circ }}\delta \mathcal{F}\right\rangle
\label{Dirac bracket on M - 2} \\
&&-\left\langle \eta _{-},\left[ \Pi _{\mathfrak{g}_{-}^{\circ }}\delta 
\mathcal{F},\Pi _{\mathfrak{g}_{-}^{\circ }}\delta \mathcal{H}\right]
\right\rangle  \notag
\end{eqnarray}%
that coincides with the canonical Poisson bracket on $G_{-}\times \mathfrak{g%
}_{-}^{\ast }$.

\subsubsection{The left action of $G$ on $G\times \mathfrak{g}^{\mathfrak{%
\ast }}$ and its restriction to $\mathcal{M}\left( g_{+},\protect\eta %
_{+}\right) $}

As described at the beginning of the section $\left( \ref{left action}%
\right) $, we want to study how the left action of $G$ on its cotangent
bundle project on the fibers $\mathcal{M}\left( g_{+},\eta _{+}\right) $. In
doing so, we consider the Dirac bracket $\left( \ref{Dirac bracket on M - 2}%
\right) $ involving the momentum function $\phi _{X}$, which for an
arbitrary function $\mathcal{F}\in C^{\infty }\left( G\times \mathfrak{g}^{%
\mathfrak{\ast }}\right) $ gives%
\begin{eqnarray*}
\left\{ \mathcal{F},\phi _{X}\right\} ^{\mathcal{M}}\left( g,\eta \right)
&=&\left\langle g\mathbf{d}\mathcal{F},\Pi _{\mathfrak{g}_{-}^{\circ
}}Ad_{g^{-1}}^{G}X\right\rangle -\left\langle \left[ \eta ,Ad_{g^{-1}}^{G}X%
\right] ,\Pi _{\mathfrak{g}_{-}^{\circ }}\delta \mathcal{F}\right\rangle \\
&&-\left\langle \eta _{-},\left[ \Pi _{\mathfrak{g}_{-}^{\circ }}\delta 
\mathcal{F},\Pi _{\mathfrak{g}_{-}^{\circ }}Ad_{g^{-1}}^{G}X\right]
\right\rangle
\end{eqnarray*}%
therefore the hamiltonian vector field of $\phi _{X}$ is%
\begin{equation}
V_{\phi _{X}}^{\mathcal{M}\left( g_{+},\eta _{+}\right) }\left( g,\eta
\right) =\left( g\Pi _{\mathfrak{g}_{-}^{\circ }}Ad_{g^{-1}}^{G}X,\Pi _{%
\mathfrak{g}_{+}^{\circ }}\left[ \Pi _{\mathfrak{g}_{+}^{\circ
}}Ad_{g^{-1}}^{G}X,\eta \right] \right)  \label{G action ham vec field}
\end{equation}

On the other side, the Dirac bracket between two moment functions reduces to%
\begin{equation*}
\left\{ \phi _{X},\phi _{y}\right\} ^{\mathcal{M}}\left( g,\eta \right)
=\phi _{\left[ X,Y\right] }(g,\eta )-\left\langle \eta _{+},\left[ \Pi _{%
\mathfrak{g}_{+}^{\circ }}Ad_{g^{-1}}^{G}X,\Pi _{\mathfrak{g}_{+}^{\circ
}}Ad_{g^{-1}}^{G}Y\right] \right\rangle
\end{equation*}%
so, as it happens on $\mathcal{N}\left( g_{-},\eta _{-}\right) $, it closes
an algebra provided $\eta _{+}$ is a \emph{character }of $\mathfrak{g}%
_{+}^{\circ }$.

From the expression of the hamiltonian vector field $V_{\phi _{X}}^{\mathcal{%
M}\left( g_{+},\eta _{+}\right) }$ given in eq. $\left( \ref{G action ham
vec field}\right) $, we retrieve the action of $G$ on $\mathcal{M}\left(
g_{+},\eta _{+}\right) $, $\mathsf{b}:G\times \mathcal{M}\left( g_{+},\eta
_{+}\right) \longrightarrow \mathcal{M}\left( g_{+},\eta _{+}\right) $, in
the case $\eta _{+}$ is a character of $\mathfrak{g}_{+}^{\circ }$, 
\begin{equation*}
\mathsf{b}\left( h,\left( g_{+}g_{-},\eta _{+}+\eta _{-}\right) \right)
=\left( g\Pi _{-}\left( g^{-1}hg\right) ,\Pi _{\mathfrak{g}_{+}^{\circ
}}Ad_{\Pi _{-}\left( g^{-1}hg\right) }^{G}\eta \right)
\end{equation*}

\subsubsection{The Hamilton equations}

The Hamilton equations on $\mathcal{M}\left( g_{+},\eta _{+}\right) $ are
defined by the Dirac bracket $\left( \ref{Dirac bracket on M - 2}\right) $%
\begin{equation*}
\left\{ \mathcal{F},\mathcal{H}\right\} ^{\mathcal{M}}\left( g,\eta \right)
=\left\langle g\mathbf{d}\mathcal{F},\Pi _{\mathfrak{g}_{-}^{\circ }}\delta 
\mathcal{H}\right\rangle -\left\langle g\mathbf{d}\mathcal{H},\Pi _{%
\mathfrak{g}_{-}^{\circ }}\delta \mathcal{F}\right\rangle -\left\langle \eta
_{-},\left[ \Pi _{\mathfrak{g}_{-}^{\circ }}\delta \mathcal{F},\Pi _{%
\mathfrak{g}_{-}^{\circ }}\delta \mathcal{H}\right] \right\rangle
\end{equation*}%
from where we get the hamiltonian vector field for the Hamilton function $%
\mathcal{H}$:%
\begin{equation*}
V_{\mathcal{H}}^{\mathcal{M}}\left( g,\eta \right) =\left( g\Pi _{\mathfrak{g%
}_{-}^{\circ }}\delta \mathcal{H},\Pi _{\mathfrak{g}_{+}^{\circ }}\left( g%
\mathbf{d}\mathcal{H}-\left( \left[ \eta _{-},\Pi _{\mathfrak{g}_{-}^{\circ
}}\delta \mathcal{H}\right] \right) \right) \right)
\end{equation*}%
The reduced Hamilton equations are then 
\begin{equation*}
\left\{ 
\begin{array}{l}
g^{-1}\dot{g}=\Pi _{\mathfrak{g}_{-}^{\circ }}\delta \mathcal{H} \\ 
\\ 
\dot{\eta}=\Pi _{\mathfrak{g}_{+}^{\circ }}\left( g\mathbf{d}\mathcal{H}%
-\left( \left[ \eta _{-},\Pi _{\mathfrak{g}_{-}^{\circ }}\delta \mathcal{H}%
\right] \right) \right)%
\end{array}%
\right.
\end{equation*}%
that, when expressed in terms of the factors, $g=g_{+}g_{-}$ and $\eta =\eta
_{+}+\eta _{-}$, becomes in%
\begin{equation*}
\begin{array}{l}
\left\{ 
\begin{array}{l}
g_{-}^{-1}\dot{g}_{-}=\Pi _{\mathfrak{g}_{-}^{\circ }}\delta \mathcal{H} \\ 
\\ 
\dot{\eta}_{-}=\Pi _{\mathfrak{g}_{+}^{\circ }}\left( g\mathbf{d}\mathcal{H}%
-\left( \left[ \eta _{-},\Pi _{\mathfrak{g}_{-}^{\circ }}\delta \mathcal{H}%
\right] \right) \right)%
\end{array}%
\right. \\ 
\\ 
\left\{ 
\begin{array}{l}
g_{+}^{-1}\dot{g}_{+}=0 \\ 
\\ 
\dot{\eta}_{+}=0%
\end{array}%
\right.%
\end{array}%
\end{equation*}

\section{The Dirac method and integrable systems}

\subsection{Involutive function algebra in $\mathcal{N}\left( e,\protect\eta %
_{-}\right) $}

The vector space $\mathfrak{g}^{\ast }$ turns into a Poisson manifold
provided we equip the set $C^{\infty }\left( \mathfrak{g}^{\ast }\right) $
with one of the Kirillov-Kostant bracket $\left\{ ,\right\} _{\pm
}:C^{\infty }\left( \mathfrak{g}^{\ast }\right) \times C^{\infty }\left( 
\mathfrak{g}^{\ast }\right) \longrightarrow C^{\infty }\left( \mathfrak{g}%
^{\ast }\right) $ defined as 
\begin{equation*}
\left\{ \mathsf{f},\mathsf{g}\right\} _{-}\left( \eta \right) =-\left\langle
\eta ,\left[ \mathcal{L}_{\mathsf{f}}\left( \eta \right) ,\mathcal{L}_{%
\mathsf{g}}\left( \eta \right) \right] \right\rangle
\end{equation*}%
where $\mathcal{L}_{\mathsf{h}}:\mathfrak{g}^{\ast }\longrightarrow 
\mathfrak{g}$ stands for the Legendre transformation of a function $\mathsf{h%
}:\mathfrak{g}^{\ast }\longrightarrow \mathbb{R}$ such that for any $\xi \in 
\mathfrak{g}^{\ast }$%
\begin{equation*}
\mathbf{\langle }\xi ,\mathcal{L}_{\mathsf{h}}(\eta )\mathbf{\rangle }_{%
\mathfrak{g}}=\mathbf{\langle }\left. d\mathsf{h}\right\vert _{\eta },\xi 
\mathbf{\rangle }_{\mathfrak{g}^{\ast }}=\left. \frac{d\mathsf{h}(\eta +t\xi
)}{dt}\right\vert _{t=0}
\end{equation*}

The canonical Poisson bracket on $G\times \mathfrak{g}^{\ast }$ 
\begin{equation*}
\left\{ \mathcal{F},\mathcal{G}\right\} \left( g,\eta \right) =\left\langle 
\mathbf{d}\mathcal{F},g\delta \mathcal{G}\right\rangle -\left\langle \mathbf{%
d}\mathcal{G},g\delta \mathcal{F}\right\rangle -\left\langle \eta ,[\delta 
\mathcal{F},\delta \mathcal{G}]\right\rangle
\end{equation*}%
that, for $\mathcal{F},\mathcal{G}=\mathsf{f},\mathsf{g}$ $C^{\infty }\left( 
\mathfrak{g}^{\ast }\right) $, such that $\left. d\mathsf{f}\right\vert
_{\eta }=\left. \left( 0,\delta \mathsf{f}\right) \right\vert _{\eta
}=\left( 0,\mathcal{L}_{\mathsf{f}}(\eta )\right) $, it reduces to 
\begin{equation}
\left\{ \mathsf{f},\mathsf{g}\right\} \left( g,\eta \right) =-\left\langle
\eta ,[\mathcal{L}_{\mathsf{f}}(\eta ),\mathcal{L}_{\mathsf{g}}(\eta
)]\right\rangle  \label{KK bracket on g*}
\end{equation}%
A remarkable fact here is that the symplectic leaves of this Poisson
structure coincides with the orbits of the coadjoint action of $G$ on $%
\mathfrak{g}^{\ast }$.

Let us now study the constraint submanifold $\mathcal{N}\left( e,\eta
_{-}\right) $, equipped with Poisson-Dirac structure derived from the Dirac
bracket $\left( \ref{Poisson-Dirac bracket 1}\right) $%
\begin{equation*}
\left\{ \mathcal{F},\mathcal{G}\right\} ^{\mathcal{N}}\left( g_{+},\eta
\right) =\left\langle g\mathbf{d}\mathcal{F},\Pi _{\mathfrak{g}_{+}^{\circ
}}\delta \mathcal{G}\right\rangle -\left\langle g\mathbf{d}\mathcal{G},\Pi _{%
\mathfrak{g}_{+}^{\circ }}\delta \mathcal{F}\right\rangle -\left\langle \eta
,\left[ \Pi _{\mathfrak{g}_{+}^{\circ }}\delta \mathcal{F},\Pi _{\mathfrak{g}%
_{+}^{\circ }}\delta \mathcal{G}\right] \right\rangle
\end{equation*}%
which, when applied to $\mathsf{f},\mathsf{g}$ $C^{\infty }\left( \mathfrak{g%
}^{\ast }\right) $, gives 
\begin{equation*}
\left\{ \mathsf{f},\mathsf{g}\right\} ^{\mathcal{N}}\left( g_{+},\eta
\right) =-\left\langle \eta ,\left[ \Pi _{\mathfrak{g}_{+}^{\circ }}\mathcal{%
L}_{\mathsf{f}}(\eta ),\Pi _{\mathfrak{g}_{+}^{\circ }}\mathcal{L}_{\mathsf{g%
}}(\eta )\right] \right\rangle
\end{equation*}%
meaning that $\mathcal{N}\left( e,\eta _{-}\right) $ is equipped with a kind
of Kirillov-Kostant bracket just as $\mathfrak{g}_{+}^{\ast }$ where $\eta
_{-}$ enters as a parameter in the Legendre transform of the functions $%
\mathsf{f},\mathsf{g}$ $C^{\infty }\left( \mathfrak{g}^{\ast }\right) $.
Then, for $\mathsf{f},\mathsf{g}$ being $Ad^{G}$-invariant functions we have
the relation 
\begin{equation}
\mathcal{L}_{\mathsf{h}}(Ad_{g}^{\ast }\eta )=Ad_{g^{-1}}\mathcal{L}_{%
\mathsf{h}}(\eta )  \label{legendre and Ad-inv}
\end{equation}%
that infinitesimally is $\left[ \mathcal{L}_{\mathsf{f}}(\eta ),\eta \right]
=0$, therefore we may also write 
\begin{equation*}
\left\{ \mathsf{f},\mathsf{g}\right\} ^{\mathcal{N}}\left( g_{+},\eta
\right) =\left\langle \eta ,\left[ \Pi _{\mathfrak{g}_{-}^{\circ }}\mathcal{L%
}_{\mathsf{g}}(\eta ),\Pi _{\mathfrak{g}_{-}^{\circ }}\mathcal{L}_{\mathsf{f}%
}(\eta )\right] \right\rangle
\end{equation*}%
Hence, provided $\eta _{-}$ is a \emph{character} of $\mathfrak{g}%
_{-}^{\circ }$, we get 
\begin{equation*}
\left\{ \mathsf{f},\mathsf{g}\right\} ^{\mathcal{N}}\left( g_{+},\eta
\right) =0
\end{equation*}%
meaning that $\mathsf{f},\mathsf{g}$ \emph{are involutive in relation to the
Dirac bracket which restricts them to the submanifold }$\mathcal{N}\left(
e,\eta _{-}\right) $. For the special value $\eta _{-}=0$, it is just the 
\emph{AKS result }\cite{AKS}.

\subsection{Solving a system in $\mathcal{N}\left( g_{-},\protect\eta %
_{-}\right) $ by factorization}

Let us now consider the collective hamiltonian $\mathsf{h}\circ \Phi
^{L}:G\times \mathfrak{g}^{\ast }\longrightarrow \mathbb{R}$, with $\mathsf{h%
}$ being $Ad^{G}$-invariant as above, and $\Phi ^{L}$ the momentum map
associated with the left translation symmetry given in eq. $\left( \ref{left
trans mom map}\right) $. In this way, the hamiltonian function is \emph{%
bi-invariant}. Since%
\begin{equation*}
d\left( \mathsf{h}\circ \Phi ^{L}\right) =d\mathsf{h}\circ \Phi _{\ast
}^{L}=\delta \mathsf{h}\circ \Phi _{\ast }^{L}
\end{equation*}%
Let $\mathcal{L}_{\mathsf{h}}:\mathfrak{g}^{\ast }\longrightarrow \mathfrak{g%
}$ be the Legendre transformation of $\mathsf{h}$ so, the differential of
the Hamilton function $\mathsf{h}\circ \Phi ^{L}$ reduces to 
\begin{equation*}
\left. d\left( \mathsf{h}\circ \Phi ^{L}\right) \right\vert _{\left( g,\eta
\right) }=\left( 0,\mathcal{L}_{\mathsf{h}}(Ad_{g}^{\ast }\Phi ^{L}\left(
g,\eta \right) )\right) =\left( 0,\mathcal{L}_{\mathsf{h}}(\eta )\right)
\end{equation*}%
and the associated hamiltonian vector field on $G\times \mathfrak{g}^{\ast }$
is%
\begin{equation*}
V_{\mathsf{h}\circ \Phi ^{L}}=\left( g\mathcal{L}_{\mathsf{h}}(\eta
),0\right)
\end{equation*}

With this result we evaluate the hamiltonian vector field using the Dirac
bracket $\left\{ \mathcal{F},\mathsf{h}\circ \Phi ^{L}\right\} ^{D}\left(
g,\eta \right) $ defined in $\left( \ref{Dirac bracket G+xg+ I}\right) $,
obtaining

\begin{equation}
V_{\mathsf{h}\circ \Phi ^{L}}^{D}=\left( gAd_{g_{-}^{-1}}^{G}\Pi _{\mathfrak{%
g}_{+}^{\circ }}Ad_{g_{-}}^{G}\mathcal{L}_{\mathsf{h}}(\eta
),Ad_{g_{-}^{-1}}\Pi _{\mathfrak{g}_{-}^{\circ }}Ad_{g_{-}}^{G}\left[ \eta
,Ad_{g_{-}^{-1}}^{G}\Pi _{\mathfrak{g}_{+}^{\circ }}Ad_{g_{-}}^{G}\mathcal{L}%
_{\mathsf{h}}(\eta )\right] \right)  \label{Dirac ham vec field}
\end{equation}

Therefore, the Hamilton eqs. motion are%
\begin{equation}
\left\{ 
\begin{array}{l}
g^{-1}\dot{g}=Ad_{g_{-}^{-1}}^{G}\Pi _{\mathfrak{g}_{+}^{\circ
}}Ad_{g_{-}}^{G}\mathcal{L}_{\mathsf{h}}(\eta ) \\ 
\\ 
\dot{\eta}=Ad_{g_{-}^{-1}}\Pi _{\mathfrak{g}_{-}^{\circ }}Ad_{g_{-}}^{G}%
\left[ \eta ,Ad_{g_{-}^{-1}}^{G}\Pi _{\mathfrak{g}_{+}^{\circ
}}Ad_{g_{-}}^{G}\mathcal{L}_{\mathsf{h}}(\eta )\right]%
\end{array}%
\right.  \label{ham eq on N(g-,eta-) 0}
\end{equation}%
that in terms of $g=g_{+}g_{-}$ and $\eta =\eta _{+}+\eta _{-}$ gives rise
to the 
\begin{equation}
\begin{array}{l}
\left\{ 
\begin{array}{l}
g_{+}^{-1}\dot{g}_{+}=\Pi _{\mathfrak{g}_{+}^{\circ }}Ad_{g_{-}}^{G}\mathcal{%
L}_{\mathsf{h}}(\eta ) \\ 
\\ 
\dot{\eta}_{+}=-Ad_{g_{-}^{-1}}\Pi _{\mathfrak{g}_{-}^{\circ }}\left[ \Pi _{%
\mathfrak{g}_{+}^{\circ }}Ad_{g_{-}}^{G}\mathcal{L}_{\mathsf{h}}(\eta ),\Pi
_{\mathfrak{g}_{-}^{\circ }}Ad_{g_{-}}^{G}\eta \right]%
\end{array}%
\right. \\ 
\\ 
\left\{ 
\begin{array}{l}
\dot{g}_{-}g_{-}^{-1}=0 \\ 
\\ 
\dot{\eta}_{-}=0%
\end{array}%
\right.%
\end{array}
\label{ham eq on N(g-,eta-)}
\end{equation}%
Introducing $\lambda =Ad_{g_{-}}^{G}\eta $ so, if $\eta _{-}$ is a character
of $\mathfrak{g}_{-}^{\circ }$, 
\begin{equation*}
\lambda _{-}=\Pi _{\mathfrak{g}_{+}^{\circ }}Ad_{g_{-}}^{G}\eta _{+}+\Pi _{%
\mathfrak{g}_{+}^{\circ }}Ad_{g_{-}}^{G}\eta _{-}=Ad_{g_{-}^{-1}}^{\ast
}\eta _{-}=\eta _{-}
\end{equation*}
$\lambda _{-}$ is also a character of $\mathfrak{g}_{-}^{\circ }$, the first
couple of equation turns in%
\begin{equation*}
\left\{ 
\begin{array}{l}
g_{+}^{-1}\dot{g}_{+}=\Pi _{\mathfrak{g}_{+}^{\circ }}\mathcal{L}_{\mathsf{h}%
}(\lambda ) \\ 
\\ 
\dot{\lambda}_{+}=-\Pi _{\mathfrak{g}_{-}^{\circ }}\left[ \Pi _{\mathfrak{g}%
_{+}^{\circ }}\mathcal{L}_{\mathsf{h}}(\lambda ),\lambda \right]%
\end{array}%
\right.
\end{equation*}

Because the Ad-invariance of $\mathsf{h}$ and having in mind that $\lambda
_{-}$ is a character of $\mathfrak{g}_{-}^{\circ }$, the second equation is
equivalent to%
\begin{equation*}
\dot{\lambda}_{+}=ad_{\Pi _{\mathfrak{g}_{-}^{\circ }}\mathcal{L}_{\mathsf{h}%
}(\lambda )}^{\mathfrak{g}}\lambda
\end{equation*}%
and because $\lambda _{-}=\eta _{-}$ $=cte$, we may write%
\begin{equation}
\dot{\lambda}=ad_{\Pi _{\mathfrak{g}_{-}^{\circ }}\mathcal{L}_{\mathsf{h}%
}(\lambda )}^{\mathfrak{g}}\lambda  \label{lambda punto}
\end{equation}%
Introducing the curve $h_{-}\left( t\right) \subset G_{-}$ satisfying the
differential equation%
\begin{equation}
\dot{h}_{-}h_{-}^{-1}=\Pi _{\mathfrak{g}_{-}^{\circ }}\mathcal{L}_{\mathsf{h}%
}(\lambda )  \label{h- punto}
\end{equation}%
we see that%
\begin{equation*}
g_{+}^{-1}\dot{g}_{+}+\dot{h}_{-}h_{-}^{-1}=\mathcal{L}_{\mathsf{h}}(\lambda
)
\end{equation*}%
Let us now write $\xi =Ad_{h_{-}^{-1}}^{G}\lambda $ so, the last equation is 
\begin{equation*}
Ad_{g_{-}^{-1}}^{G}\left( g_{+}^{-1}\dot{g}_{+}+\dot{h}_{-}h_{-}^{-1}\right)
=\mathcal{L}_{\mathsf{h}}(Ad_{h_{-}}^{G}\xi )
\end{equation*}%
that can be written as%
\begin{equation}
k^{-1}\left( t\right) \dot{k}\left( t\right) =\mathcal{L}_{\mathsf{h}}(\xi
\left( t\right) )  \label{eq. k}
\end{equation}%
\newline
where $k\left( t\right) :=g_{+}\left( t\right) h_{-}\left( t\right) $.

Having in mind eqs $\left( \ref{h- punto}\right) $ and $\left( \ref{lambda
punto}\right) $, we get%
\begin{equation*}
\dot{\xi}=0
\end{equation*}%
Thus, the couple of equations%
\begin{equation*}
\left\{ 
\begin{array}{l}
k^{-1}\dot{k}=\mathcal{L}_{\mathsf{h}}(\xi ) \\ 
\\ 
\dot{\xi}=0%
\end{array}%
\right.
\end{equation*}%
are solved by the curves%
\begin{equation*}
\left\{ 
\begin{array}{l}
k\left( t\right) =e^{\mathcal{L}_{\mathsf{h}}(\xi )} \\ 
\\ 
\xi =\xi _{\circ }%
\end{array}%
\right.
\end{equation*}

Therefore, we have shown that the Hamilton equations $\left( \ref{ham eq on
N(g-,eta-)}\right) $ on $\mathcal{N}\left( g_{-},\eta _{-}\right) $ have the
solutions 
\begin{equation*}
\left\{ 
\begin{array}{l}
g_{+}\left( t\right) =\Pi _{G_{+}}e^{\mathcal{L}_{\mathsf{h}}(\xi )} \\ 
\\ 
\eta \left( t\right) =Ad_{g_{-}^{-1}\Pi _{G_{-}}e^{\mathcal{L}_{\mathsf{h}%
}(\xi )}}^{G}\xi _{\circ }%
\end{array}%
\right.
\end{equation*}%
encountering the Adler-Kostant-Symes\ result \cite{AKS},\cite{RSTS 1},
relating a system of differential equations on a coadjoint orbit with the
factorization problem of an exponential curve in $G$. This issue deserves a
deeper insight which is addressed in the next subsection.

\subsection{AKS theory}

\label{AKS}

In order to understand the above results, let us digress on the meaning of
the Adler-Kostant-Symes approach to integrability \cite{AKS}\cite{RSTS 1}.
An AKS systems can be characterized as a reduced spaces derived from a
dynamical systems defined on the cotangent bundle of a Lie group, so our
main concern in this section will be to provide a connection between this
kind of systems and the dynamical systems studied above on the phase spaces $%
\mathcal{N}\left( g_{-},\eta _{-}\right) $ (analogous considerations can be
made for the spaces $\mathcal{M}\left( g_{+},\eta _{+}\right) $).

Let us first stress the role played by some symmetries of a factorizable Lie
group. The space $G\times \mathfrak{g}^{\ast }$ can be considered as a $%
G_{+}\times G_{-}$-space, if as above $G\times \mathfrak{g}^{\ast }\simeq
T^{\ast }G$ via left trivialization and we lift the $G_{+}\times G_{-}$%
-action on $G$ given by 
\begin{equation*}
G_{+}\times G_{-}\times G\rightarrow G:\left( a_{+},a_{-};g\right) \mapsto
a_{+}ga_{-}^{-1}.
\end{equation*}%
By using the facts that the action is lifted and the symplectic form on $%
G\times \mathfrak{g}^{\ast }$ is exact, we can determine the momentum map
associated to this action; then we obtain that%
\begin{equation*}
\begin{array}{rcl}
J:G\times \mathfrak{g}^{\ast } & \rightarrow & \mathfrak{g}_{-}^{\circ
}\times \mathfrak{g}_{+}^{\circ } \\ 
&  &  \\ 
\left( g,\xi \right) & \mapsto & \left( \Pi _{\mathfrak{g}_{-}^{\circ
}}\left( Ad_{g}^{G}\xi \right) ,\Pi _{\mathfrak{g}_{+}^{\circ }}\left( \xi
\right) \right)%
\end{array}%
\end{equation*}
where $Ad^{G}$ indicates the coadjoint \emph{action} of $G$ on $\mathfrak{g}%
^{\ast }$. Let us now define the submanifold 
\begin{equation*}
\Lambda _{\eta _{+}\eta _{-}}:=\left\{ \left( g,\xi \right) \in G\times 
\mathfrak{g}^{\ast }/\Pi _{\mathfrak{g}_{-}^{\circ }}\left( Ad_{g}^{G}\xi
\right) =\eta _{+},~\Pi _{\mathfrak{g}_{+}^{\circ }}\left( \xi \right) =\eta
_{-}\right\}
\end{equation*}%
for each pair $\eta _{+}\in \mathfrak{g}_{-}^{\circ },\eta _{-}\in \mathfrak{%
g}_{+}^{\circ }$. We have the following lemma.

\begin{description}
\item[Lemma] \textit{Let }$\xi _{\pm }\in \mathfrak{g}_{\mp }^{\circ }$%
\textit{, }$a_{+}\in G_{+}$\textit{, }$a_{-}\in G_{-}$\textit{\ be arbitrary
elements. Then the formulas}%
\begin{equation*}
\begin{array}{rcl}
a_{+}\cdot \xi _{+} & := & \Pi _{\mathfrak{g}_{-}^{\circ }}\left(
Ad_{a_{+}}^{G}\xi _{+}\right) \\ 
&  &  \\ 
a_{-}\cdot \xi _{-} & := & \Pi _{\mathfrak{g}_{+}^{\circ }}\left(
Ad_{a_{-}}^{G}\xi _{-}\right)%
\end{array}%
\end{equation*}
\textit{\ defines an action of }$G_{\pm }$\textit{\ on }$\mathfrak{g}_{\mp
}^{\circ }$\textit{; in fact, under the identification }$\mathfrak{g}_{\pm
}^{\ast }\simeq \mathfrak{g}_{\mp }^{\circ }$\textit{\ induced by the
decomposition }$\mathfrak{g}=\mathfrak{g}_{+}^{\circ }\oplus \mathfrak{g}%
_{-}^{\circ }$\textit{\ these actions are just the coadjoint actions of each
factor on the dual of its Lie algebras.}
\end{description}

\begin{description}
\item[Note:] The symbol $\mathcal{O}_{\xi _{\pm }}^{G_{\pm }}$ will denotes
the orbit in $\mathfrak{g}_{\mp }^{\circ }$ under the actions defined in the
previous lemma. Additionally, for each $\xi \in \mathfrak{g}^{\ast }$, the
form $\xi ^{\flat }\in \mathfrak{g}^{\ast }$ is given by $\xi ^{\flat }:={B}%
\left( \xi ,\cdot \right) $, where ${B}\left( \cdot ,\cdot \right) $ is the
invariant bilinear form on $\mathfrak{g}^{\ast }$ induced by the Killing
form.
\end{description}

Therefore $\Lambda _{\eta _{+}\eta _{-}}=J^{-1}\left( \eta _{+},\eta
_{-}\right) $ and, taking into account the Marsden-Weinstein reduction (see 
\cite{Abr-Mars}), the projection of $\Lambda _{\eta _{+}\eta _{-}}$ on $%
\Lambda _{\eta _{+}\eta _{-}}/\left( G_{+}\right) _{\eta _{+}}\times \left(
G_{-}\right) _{\eta _{-}}$ is presymplectic, and the solution curves for the
dynamical system defined there by the invariant Hamiltonian $\mathcal{H}%
\left( g,\xi \right) :=\frac{1}{2}\xi \left( \xi ^{\flat }\right) $ are
closely related with the solution curves of the system induced in the
quotient. To work out these equations, let us introduce some convenient
coordinates. The map $L_{\eta _{+}\eta _{-}}:\Lambda _{\eta _{+}\eta
_{-}}\longrightarrow \mathcal{O}_{\eta _{+}}^{G_{+}}\times \mathcal{O}_{\eta
_{-}}^{G_{-}}$ defined as%
\begin{equation*}
L_{\eta _{+}\eta _{-}}\left( g,\xi \right) =\left( \Pi _{\mathfrak{g}%
_{-}^{\circ }}Ad_{g_{+}^{-1}}^{G}\eta _{+},\Pi _{\mathfrak{g}_{+}^{\circ
}}Ad_{g_{-}}^{G}\eta _{-}\right) =\left( \Pi _{\mathfrak{g}_{-}^{\circ
}}Ad_{g_{-}}^{G}\xi ,\Pi _{\mathfrak{g}_{+}^{\circ }}Ad_{g_{-}}^{G}\xi
\right)
\end{equation*}%
where $g=g_{+}g_{-}$, induces a diffeomorphism on $\Lambda _{\eta _{+}\eta
_{-}}/\left[ \left( G_{+}\right) _{\eta _{+}}\times \left( G_{-}\right)
_{\eta _{-}}\right] $. If $\left( g,\xi ;X,\lambda \right) $ is a tangent
vector to $G\times \mathfrak{g}^{\ast }$ at $\left( g,\xi \right) $ (all the
relevant bundles are left trivialized) then the derivative of $L_{\eta
_{+}\eta _{-}}$ is 
\begin{eqnarray*}
&&\left. \left( L_{\eta _{+}\eta _{-}}\right) _{\ast }\right\vert _{\left(
g,\xi \right) }\left( g,\xi ;X,\lambda \right) \\
&=&\left( -\Pi _{\mathfrak{g}_{-}^{\circ }}ad_{X_{+}}^{\mathfrak{g}%
}Ad_{g_{+}^{-1}}^{G}\eta _{+},\Pi _{\mathfrak{g}_{+}^{\circ }}ad_{X_{-}}^{%
\mathfrak{g}}Ad_{g_{-}}^{G}\eta _{-}\right) \\
&=&\left( \Pi _{\mathfrak{g}_{-}^{\circ }}ad_{X_{-}}^{\mathfrak{g}%
}Ad_{g_{-}}^{G}\xi +\Pi _{\mathfrak{g}_{-}^{\circ }}Ad_{g_{-}}^{G}\lambda
,\Pi _{\mathfrak{g}_{+}^{\circ }}ad_{X_{-}}^{\mathfrak{g}}Ad_{g_{-}}^{G}\xi
+\Pi _{\mathfrak{g}_{+}^{\circ }}Ad_{g_{-}}^{G}\lambda \right)
\end{eqnarray*}%
if and only if $g=g_{+}g_{-}$, $X_{\pm }=\Pi _{\mathfrak{g}_{\pm }}\left(
Ad_{g_{-}}^{G}X\right) $. So the following remarkable result is true.

\begin{description}
\item[Proposition] \textit{Let }$\mathcal{O}_{\eta _{+}}^{G_{+}}\times 
\mathcal{O}_{\eta _{-}}^{G_{-}}$\textit{\ be the phase space whose
symplectic structure is }$\omega _{\eta _{+}\eta _{-}}=\omega _{\eta
_{+}}-\omega _{\eta _{-}}$\textit{, where }$\omega _{\eta _{\pm }}$\textit{\
are the corresponding Kirillov-Kostant symplectic structures on each orbit.
If }$i_{\eta _{+}\eta _{-}}:\Lambda _{\eta _{+}\eta _{-}}\hookrightarrow
G\times \mathfrak{g}^{\ast }$\textit{\ is the inclusion map, then we have
that }%
\begin{equation*}
i_{\eta _{+}\eta _{-}}^{\ast }\omega =L_{\eta _{+}\eta _{-}}^{\ast }\omega
_{\eta _{+}\eta _{-}}.
\end{equation*}
\end{description}

\textbf{Proof: }Let $\left( \varsigma _{+},\varsigma _{-}\right) =\left( \Pi
_{\mathfrak{g}_{-}^{\circ }}\left( Ad_{a_{+}}^{G}\eta _{+}\right) ,\Pi _{%
\mathfrak{g}_{+}^{\circ }}\left( Ad_{a_{-}}^{G}\eta _{-}\right) \right) $ be
an arbitrary element of $\mathcal{O}_{\eta _{+}}^{G_{+}}\times \mathcal{O}%
_{\eta _{-}}^{G_{-}}$, then the tangent space at this point is given by 
\begin{equation*}
T_{\left( \varsigma _{+},\varsigma _{-}\right) }\left( \mathcal{O}_{\eta
_{+}}^{G_{+}}\times \mathcal{O}_{\eta _{-}}^{G_{-}}\right) =\left\{
\left(\Pi _{\mathfrak{g}_{-}^{\circ }} ad_{X_{+}}^{\mathfrak{g}}\varsigma
_{+} ,\Pi _{\mathfrak{g}_{+}^{\circ }} ad_{X_{-}}^{\mathfrak{g}}\varsigma
_{-}\right) ~/~X_{\pm }\in \mathfrak{g}_{\pm }\right\}
\end{equation*}%
The symplectic structure $\omega _{\eta _{+}\eta _{-}}$ is given in these
terms as 
\begin{eqnarray*}
&&\left\langle \omega _{\eta _{+}\eta _{-}},\left( \Pi _{\mathfrak{g}%
_{-}^{\circ }}ad_{X_{+}}^{\mathfrak{g}}\varsigma _{+},\Pi _{\mathfrak{g}%
_{+}^{\circ }}ad_{X_{-}}^{\mathfrak{g}}\varsigma _{-}\right) \otimes \left(
\Pi _{\mathfrak{g}_{-}^{\circ }}ad_{Y_{+}}^{\mathfrak{g}}\varsigma _{+},\Pi
_{\mathfrak{g}_{+}^{\circ }}ad_{Y_{-}}^{\mathfrak{g}}\varsigma _{-}\right)
\right\rangle _{\left( \varsigma _{+},\varsigma _{-}\right) } \\
&=&\left\langle \varsigma _{+},\left[ X_{+},Y_{+}\right] \right\rangle
-\left\langle \varsigma _{-},\left[ X_{-},Y_{-}\right] \right\rangle
\end{eqnarray*}

Let us take now an element $\left( g,\xi ;X,\lambda \right) $ tangent to $%
\Lambda _{\eta _{+}\eta _{-}}$ at $\left( g,\xi \right) $; then it is true
that 
\begin{equation}
\left\{ 
\begin{array}{l}
\Pi _{\mathfrak{g}_{-}^{\circ }}\left( Ad_{g}^{G}\left( ad_{X}^{\mathfrak{g}%
}\xi +\lambda \right) \right) =0 \\ 
\\ 
\Pi _{\mathfrak{g}_{+}^{\circ }}\left( \lambda \right) =0%
\end{array}%
\right.  \label{CondTangencyLambdamunu}
\end{equation}%
Because of $g=g_{+}g_{-}$, the first condition can be written as 
\begin{equation*}
\Pi _{\mathfrak{g}_{-}^{\circ }}\left( Ad_{g_{+}}^{G}\left(
ad_{Ad_{g_{-}}X}^{\mathfrak{g}}Ad_{g_{-}}^{G}\xi +Ad_{g_{-}}^{G}\lambda
\right) \right) =0
\end{equation*}%
and because of the nondegeneracy condition \footnote{%
That is, such that $Ad_{g}^{G}\mathfrak{g}_{+}^{\circ }\cap \mathfrak{g}%
_{-}^{\circ }=0$ for all $g\in G$.}, it is equivalent to 
\begin{equation}
\Pi _{\mathfrak{g}_{-}^{\circ }}\left( ad_{Ad_{g_{-}}X}^{\mathfrak{g}%
}Ad_{g_{-}}^{G}\xi +Ad_{g_{-}}^{G}\lambda \right) =0
\label{CondTangencyLambdamunu2}
\end{equation}%
Let $\left( g,\xi ;X,\lambda \right) ,\left( g,\xi ;Y,\mu \right) \in
T_{\left( g,\xi \right) }\Lambda _{\eta _{+}\eta _{-}}$; then by evaluating
on the canonical form we have that 
\begin{eqnarray*}
&&\left\langle \omega ,\left( g,\xi ;X,\lambda \right) \otimes \left( g,\xi
;Y,\mu \right) \right\rangle _{\left( g,\xi \right) } \\
&=&\left\langle \lambda ,Y\right\rangle -\left\langle \mu ,X\right\rangle
-\left\langle \xi ,\left[ X,Y\right] \right\rangle \\
&=&\left\langle \Pi _{\mathfrak{g}_{-}^{\circ }}Ad_{g_{-}}^{G}\lambda
,Y_{+}\right\rangle +\left\langle \Pi _{\mathfrak{g}_{+}^{\circ
}}Ad_{g_{-}}^{G}\lambda ,Y_{-}\right\rangle \\
&&-\left\langle \Pi _{\mathfrak{g}_{-}^{\circ }}Ad_{g_{-}}^{G}\mu
,X_{+}\right\rangle -\left\langle \Pi _{\mathfrak{g}_{+}^{\circ
}}Ad_{g_{-}}^{G}\mu ,X_{+}\right\rangle \\
&&-\left\langle Ad_{g_{-}}^{G}\xi ,\left[ X,Y_{+}\right] \right\rangle
-\left\langle Ad_{g_{-}}^{G}\xi ,\left[ X_{-},Y\right] \right\rangle
-\left\langle Ad_{g_{-}}^{G}\xi ,\left[ X_{+},Y_{-}\right] \right\rangle \\
&=&\left\langle \Pi _{\mathfrak{g}_{-}^{\circ }}\left( Ad_{g_{-}}^{G}\lambda
+ad_{X}^{\mathfrak{g}}Ad_{g_{-}}^{G}\xi \right) ,Y_{+}\right\rangle
+\left\langle \Pi _{\mathfrak{g}_{+}^{\circ }}Ad_{g_{-}}^{G}\lambda
,Y_{-}\right\rangle - \\
&&-\left\langle \Pi _{\mathfrak{g}_{-}^{\circ }}\left( Ad_{g_{-}}^{G}\mu
+ad_{Y_{-}}^{\mathfrak{g}}Ad_{g_{-}}^{G}\xi \right) ,X_{+}\right\rangle
-\left\langle \Pi _{\mathfrak{g}_{+}^{\circ }}Ad_{g_{-}}^{G}\mu
,X_{+}\right\rangle - \\
&&-\left\langle Ad_{g_{-}}^{G}\xi ,\left[ X_{-},Y_{-}\right] \right\rangle
\end{eqnarray*}%
where we have used the notation according to which $\left( X\right) _{\pm
}=\Pi _{\mathfrak{g}_{\pm }}\left( Ad_{g_{-}}^{G}X\right) $ and $\left(
Y\right) _{\pm }=\Pi _{\mathfrak{g}_{\pm }}\left( Ad_{g_{-}}^{G}Y\right) $.
The first term in this expression annihilates because of the eq. $\left( \ref%
{CondTangencyLambdamunu2}\right) $; additionally, the second and fourth
vanishes as a consequence of the second eq. in $\left( \ref%
{CondTangencyLambdamunu}\right) $, which implies that $\lambda ,\mu \in 
\mathfrak{g}_{-}^{\circ }$, and because this subspace is invariant for the $%
G_{-}$-action through the coadjoint action. Moreover, for the third term in
the second hand side, we use the eq. $\left( \ref{CondTangencyLambdamunu}%
\right) $ again, and therefore we can write 
\begin{equation*}
\Pi _{\mathfrak{g}_{-}^{\circ }}\left( Ad_{g_{-}}^{G}\mu +ad_{Y_{-}}^{%
\mathfrak{g}}Ad_{g_{-}}^{G}\xi \right) \left( X_{+}\right) =-\left\langle
Ad_{g_{-}}^{G}\xi ,\left[ X_{+},Y_{+}\right] \right\rangle
\end{equation*}%
and so 
\begin{eqnarray*}
&&\left\langle \omega ,\left( g,\xi ;X,\lambda \right) \otimes \left( g,\xi
;Y,\mu \right) \right\rangle _{\left( g,\xi \right) } \\
&=&\left\langle \Pi _{\mathfrak{g}_{-}^{\circ }}Ad_{g_{-}}^{G}\xi ,\left[
X_{+},Y_{+}\right] \right\rangle -\left\langle \Pi _{\mathfrak{g}_{+}^{\circ
}}Ad_{g_{-}}^{G}\xi ,\left[ X_{-},Y_{-}\right] \right\rangle
\end{eqnarray*}%
From eq. $\left( \ref{EstrSimplecticaSobreProductoOrbitas}\right) $ and by
using formula $\left( \ref{DerivadaLmunu}\right) $ for the derivative along
the map $L_{\eta _{+}\eta _{-}}$, we conclude the proof.$\blacksquare $

Let us now address the dynamical data related to an AKS system. We have
shown that the symplectic manifold $\left( \mathcal{O}_{\eta
_{+}}^{G_{+}}\times \mathcal{O}_{\eta _{-}}^{G_{-}},\omega _{\eta _{+}\eta
_{-}}\right) $ is symplectomorphic to the reduced space associated to the $%
G_{+}\times G_{-}$-action defined above on $G\times \mathfrak{g}^{\ast }$.
As it was pointed out above, the Hamilton function $\mathcal{H}\left( g,\xi
\right) =\frac{1}{2}\left\langle \xi ,\xi ^{\flat }\right\rangle $ is
invariant for this action, and it implies that the solutions of the
dynamical system defined by such a hamiltonian on $G\times \mathfrak{g}%
^{\ast }$ are in one-to-one correspondence with those of the dynamical
system induced on $\mathcal{O}_{\eta _{+}}^{G_{+}}\times \mathcal{O}_{\eta
_{-}}^{G_{-}}$ by the hamiltonian $\mathcal{H}_{\eta _{+}\eta _{-}}$ \cite%
{Abr-Mars} defined according to the formula 
\begin{equation*}
i_{\eta _{+}\eta _{-}}^{\ast }\mathcal{H}=L_{\eta _{+}\eta _{-}}^{\ast }%
\mathcal{H}_{\eta _{+}\eta _{-}}
\end{equation*}%
Let us note now that if $L_{\eta _{+}\eta _{-}}\left( g,\xi \right) =\left(
\varsigma _{+},\varsigma _{-}\right) $, then $Ad_{g_{-}}^{G}\xi =\varsigma
_{+}+\varsigma _{-}$, and so%
\begin{eqnarray*}
\mathcal{H}_{\eta _{+}\eta _{-}}\left( \varsigma _{+},\varsigma _{-}\right)
&=&\frac{1}{2}\left\langle Ad_{g_{-}^{-1}}^{G}\left( \varsigma
_{+}+\varsigma _{-}\right) ,\left( Ad_{g_{-}^{-1}}^{G}\left( \varsigma
_{+}+\varsigma _{-}\right) \right) ^{\flat }\right\rangle \\
&=&\frac{1}{2}\left\langle Ad_{g_{-}^{-1}}^{G}\left( \varsigma
_{+}+\varsigma _{-}\right) ,Ad_{g_{-}^{-1}}\left( \varsigma _{+}+\varsigma
_{-}\right) ^{\flat }\right\rangle \\
&=&\frac{1}{2}\left\langle \varsigma _{+},\varsigma _{+}^{\flat
}\right\rangle +\frac{1}{2}\left\langle \varsigma _{-},\varsigma _{-}^{\flat
}\right\rangle +\left\langle \varsigma _{+},\varsigma _{-}^{\flat
}\right\rangle
\end{eqnarray*}%
Therefore 
\begin{eqnarray*}
&&\left\langle d\mathcal{H}_{\eta _{+}\eta _{-}},\left( \Pi _{\mathfrak{g}%
_{-}^{\circ }}ad_{X}^{\mathfrak{g}}\varsigma _{+},\Pi _{\mathfrak{g}%
_{+}^{\circ }}ad_{Y}^{\mathfrak{g}}\varsigma _{-}\right) \right\rangle
_{\left( \varsigma _{+},\varsigma _{-}\right) } \\
&=&\left\langle \Pi _{\mathfrak{g}_{-}^{\circ }}ad_{X}^{\mathfrak{g}%
}\varsigma _{+},\varsigma _{+}^{\flat }+\varsigma _{-}^{\flat }\right\rangle
+\left\langle \Pi _{\mathfrak{g}_{+}^{\circ }}ad_{Y}^{\mathfrak{g}}\varsigma
_{-},\varsigma _{+}^{\flat }+\varsigma _{-}^{\flat }\right\rangle
\end{eqnarray*}%
and the hamiltonian vector field is then 
\begin{equation}
\left. V_{\mathcal{H}_{\eta _{+}\eta _{-}}}\right\vert _{\left( \varsigma
_{+},\varsigma _{-}\right) }=\left( \Pi _{\mathfrak{g}_{-}^{\circ }}ad_{\Pi
_{\mathfrak{g}_{+}}\left( \varsigma _{+}^{\flat }+\varsigma _{-}^{\flat
}\right) }^{\mathfrak{g}}\varsigma _{+},\Pi _{\mathfrak{g}_{+}^{\circ
}}ad_{\Pi _{\mathfrak{g}_{-}}\left( \varsigma _{+}^{\flat }+\varsigma
_{-}^{\flat }\right) }^{\mathfrak{g}}\varsigma _{-}\right)
\label{CampoVectHmunu}
\end{equation}%
In general, in dealing with a dynamical systems via reduction it is expected
the reduced system to be easier to solve than the original because it
involves less degrees of freedom, and the solution of the original system is
found by lifting the solution of the reduced system. In case of AKS systems,
we proceed in the reverse direction: to this end let us note that the
dynamical system on $G\times \mathfrak{g}^{\ast }$ determined by $\mathcal{H}
$ has hamiltonian vector field given by 
\begin{equation}
\left. V_{\mathcal{H}}\right\vert _{\left( g,\xi \right) }=\left( \xi
^{\flat },-ad_{\xi ^{\flat }}^{\mathfrak{g}}\xi \right) =\left( \xi ^{\flat
},0\right)  \label{CampoVectHmunuSimp}
\end{equation}%
because of the invariance condition $ad_{X}^{\mathfrak{g}}X^{\flat }=0$ for
all $X\in \mathfrak{g}$. Then the solution in the original phase space
passing through $\left( g,\xi \right) $ at the initial time is 
\begin{equation*}
t\mapsto \left( g\exp {t\xi ^{\flat }},\xi \right)
\end{equation*}%
and if this initial data verifies $\Pi _{\mathfrak{g}_{-}^{\circ }}\left(
Ad_{g}^{\sharp }\xi \right) =\eta _{+},\Pi _{\mathfrak{g}_{+}^{\circ
}}\left( \xi \right) =\eta _{-}$, then this curve belongs to $\Lambda _{\eta
_{+}\eta _{-}}$ for all $t$. Therefore the map 
\begin{equation*}
t\mapsto \left( \Pi _{\mathfrak{g}_{-}^{\circ }}Ad_{\left( g_{+}\left(
t\right) \right) ^{-1}}^{G}\eta _{+},\Pi _{\mathfrak{g}_{+}^{\circ
}}Ad_{g_{-}\left( t\right) }^{G}\eta _{-}\right)
\end{equation*}%
where $g_{\pm }:\mathbb{R}\rightarrow G_{\pm }$ are the curves defined by
the factorization problem $g_{+}\left( t\right) g_{-}\left( t\right) =g\exp {%
t\xi ^{\flat }}$, is solution for the dynamical system associated to the
vectorial field $\left( \ref{CampoVectHmunu}\right) $, yielding to a
differential equation which is harder to solve than the original system.

\subsection{AKS dynamics on the spaces $\mathcal{N}\left(g_-,\protect\eta%
_-\right)$}

Let $\eta _{-}\in \mathfrak{g}_{+}^{\circ }$ be a character. As it was shown
above, the spaces $\mathcal{N}\left( g_{-},\eta _{-}\right) $ are $G$%
-spaces, and in particular $G_{+}$-spaces; therefore there exists a momentum
map $J_{+}:\mathcal{N}\left( g_{-},\eta _{-}\right) \rightarrow \mathfrak{g}%
_{-}^{\circ }$. The connection with the dynamics on the phase spaces $%
\mathcal{N}\left( g_{-},\eta _{-}\right) $ is given by the following
theorem, which is a consequence of the hamiltonian reduction by stages \cite%
{marsden98:_sympl_reduc_semid_produc_centr_exten}$\clubsuit $.

\begin{description}
\item[Theorem] \emph{\ Let }${\eta }_{+}\in \mathfrak{g}_{-}^{\circ }$ \emph{%
be an arbitrary element. The reduced space } 
\begin{equation*}
J_{+}^{-1}\left( \eta _{+}\right) /\left( G_{+}\right) _{\eta_+ }
\end{equation*}%
\emph{is therefore symplectomorphic to the orbit space} $\mathcal{O}_{{%
\eta_{+}}}^{G_{+}}\times \mathcal{O}_{\eta _{-}}^{G_{-}}$.
\end{description}

\textbf{Proof: }Simply note that $G_{+}$ and $G_{-}$ commutes each other as
subgroups in $G_{+}\times G_{-}$, and the same is true for their actions on $%
G$; then by performing the reduction into $J^{-1}\left( \eta _{+},\eta
_{-}\right) /\left[ \left( G_{+}\right) _{\eta _{+}}\times \left(
G_{-}\right) _{\eta _{-}}\right] $ by stages, first using the right $G_{-}$%
-action and then the left $G_{+}$-action, we prove the theorem.$\blacksquare 
$

The following diagram can be useful to explain the contents of the previous
statement: 
\begin{equation*}
\begin{diagram}
\node{G\times\mathfrak{g}^*}\node{J_-^{-1}\left(\eta_-\right)}%
\arrow{w,t,L}{}\arrow{s,l,A}{\pi_{\eta_+\eta_-}}\arrow{e,t,A}{\pi_{\eta_-}}%
\node{\frac{J_-^{-1}\left(\eta_-\right)}{\left(G_-\right)_{\eta_-}}}%
\arrow{sw,r,A}{\pi_{\eta_+}}\\
\node[2]{\frac{J^{-1}\left(\eta_+,\eta_-\right)}{\left[\left(G_+\right)_{%
\eta_+}\times\left(G_-\right)_{\eta_-}\right]}}\arrow{e,t,=}{L_{\eta_+%
\eta_-}}\node{\mathcal{O}^{G_+}_{\eta_+}\times\mathcal{O}^{G_-}_{\eta_-}}
\end{diagram}
\end{equation*}%
where the $\pi $-maps denotes the canonical projections onto the
corresponding quotients. Accordingly, the picture can be described as
follows: the hamiltonian $\mathcal{H}\left( g,\xi \right) =\frac{1}{2}\xi
\left( \xi ^{\flat }\right) $ defines on $G\times \mathfrak{g}^{\ast }$ a
dynamical system whose solutions are easier to be found; they are given by
the formula 
\begin{equation*}
t\mapsto \left( g\exp {t\xi ^{\flat }},\xi \right)
\end{equation*}%
for any initial data $\left( g_{\circ },\xi _{\circ }\right) \in G\times 
\mathfrak{g}^{\ast }$. Since $G$ is a factorizable Lie group, $G=G_{+}G_{-}$
for some subgroups $G_{\pm }\subset G$ such that $G_{+}\cap G_{-}=\left\{
e\right\} $; on $G$ there exists a $G_{+}\times G_{-}$-action defined as 
\begin{equation*}
\left( a_{+},a_{-}\right) \cdot g:=a_{+}ga_{-}^{-1}
\end{equation*}%
for all $a_{+}\in G_{+},a_{-}\in G_{-}$ and $g\in G$. This action lift to a
symplectic action on $G\times \mathfrak{g}^{\ast }$ (left trivialized), and $%
\mathcal{H}$ turns out to be an invariant hamiltonian; therefore there
exists a hamiltonian system on the reduced space with solutions related to
the solutions of the system determined by $\mathcal{H}$ on $G\times 
\mathfrak{g}^{\ast }$. Because of the discussion in the previous subsection,
the dynamical system so induced on $J^{-1}\left( \eta _{+},\eta _{-}\right) /%
\left[ \left( G_{+}\right) _{\eta _{+}}\times \left( G_{-}\right) _{\eta
_{-}}\right] $ is an AKS system.\newline

Let us now suppose that $\eta _{-}\in \mathfrak{g}_{+}^{\circ }$ is a
character for the coadjoint action. It was shown above that the spaces $%
\mathcal{N}\left( g_{-},\eta _{-}\right) $ are slices for the right $G_{-}$%
-action on $G\times \left( \mathfrak{g}_{-}^{\circ }+\eta _{-}\right) $, and
therefore they are symplectic submanifolds of $G\times \mathfrak{g}^{\ast }$%
. Further, it was shown that these spaces are symplectomorphic to the space $%
G\times _{G_{-}}\left( \mathfrak{g}_{-}^{\circ }+\eta _{-}\right) $, the
reduced space for the right $G_{-}$-action on $G\times \mathfrak{g}^{\ast }$%
. Additionally, this reduced space is a $G$-space via the lift of the
canonical action of $G$ on its homogeneous space $M=G/G_{-}$; in particular,
it is a $G_{+}$-space, and we can apply the reduction scheme once more. The
reduction by stages theorem guarantees that the resulting space is
symplectomorphic to the AKS system as obtained in section \ref{AKS}, and the
solution curves for the system induced on $G\times _{G_{-}}\left( \mathfrak{g%
}_{-}^{\circ }+\eta _{-}\right) $ by $\mathcal{H}$ projects down onto the
solution curves for the AKS systems; we can think of them as an intermediate
lift of the dynamics of the AKS system.

Let us now address the lifting the dynamics to $\mathcal{N}\left( g_{-},\eta
_{-}\right) $. As before, here $\eta _{-}$ is a character for the $Ad^{\ast
} $-action and $g_{-}$ an arbitrary element in $G_{-}$. We want to describe
the dynamical system on $\mathcal{N}\left( g_{-},\eta _{-}\right) $ defined
through the reduction procedure described above. In order to achieve this,
let us recall that the map 
\begin{equation*}
\tilde{\pi}_{\eta _{-}}:=\left. \pi _{\eta _{-}}\right\vert \mathcal{N}%
\left( g_{-},\eta _{-}\right) :\mathcal{N}\left( g_{-},\eta _{-}\right)
\rightarrow J_{-}^{-1}\left( \eta _{-}\right) /G_{-}=:M_{\eta _{-}}^{-}
\end{equation*}%
between this symplectic submanifold and the reduced space $M_{\eta _{-}}^{-}$
is a symplectomorphism (i.e. the note at the end of section \ref{left action}%
$\clubsuit $). Let us denote by $\omega _{\mathcal{N}}$ the symplectic
structure on $\mathcal{N}\left( g_{-},\eta _{-}\right) $; now the reduction
procedure defines on $M_{\eta _{-}}^{-}$ both the symplectic structure $%
\omega _{\eta _{-}}$ and the hamiltonian function $\mathcal{H}_{\eta _{-}}$
through the formulas%
\begin{eqnarray*}
\pi _{\eta _{-}}^{\ast }\omega _{\eta _{-}} &=&\left. \omega _{\circ
}\right\vert J_{-}^{-1}\left( \eta _{-}\right) \\
&& \\
\pi _{\eta _{-}}^{\ast }H_{\eta _{-}} &=&\left. \mathcal{H}\right\vert
J_{-}^{-1}\left( \eta _{-}\right)
\end{eqnarray*}%
so that if $i_{\eta _{-}}:\mathcal{N}\left( g_{-},\eta _{-}\right)
\hookrightarrow G\times \mathfrak{g}^{\ast }$ is the inclusion, then%
\begin{eqnarray*}
i_{\eta _{-}}^{\ast }\omega _{\circ } &=&i_{\eta _{-}}^{\ast }\left[ \left.
\omega _{\circ }\right\vert J_{-}^{-1}\left( \eta _{-}\right) \right] \\
&=&i_{\eta _{-}}^{\ast }\pi _{\eta _{-}}^{\ast }\omega _{\eta _{-}} \\
&=&\left( \pi _{\eta _{-}}\circ i_{\eta _{-}}\right) ^{\ast }\omega _{\eta
_{-}} \\
&=&\tilde{\pi}_{\eta _{-}}^{\ast }\omega _{\eta _{-}} \\
&=&\omega _{\mathcal{N}}
\end{eqnarray*}%
and accordingly $\mathcal{H}_{\mathcal{N}}:=i_{\eta _{-}}^{\ast }\mathcal{H}$
verifies $\tilde{\pi}_{\eta _{-}}^{\ast }\mathcal{H}_{\eta _{-}}=\mathcal{H}%
_{\mathcal{N}}$. Therefore the dynamics defined on $\mathcal{N}\left(
g_{-},\eta _{-}\right) $ via the identification with the reduced space $%
M_{\eta _{-}}^{-}$ is equivalent to the dynamics defined on the same space
through the restriction of the dynamical data $\left( \omega _{\circ },%
\mathcal{H}\right) $. These considerations enables to lift the AKS dynamics
from the full reduced space $J^{-1}\left( \eta _{+},\eta _{-}\right) /\left[
\left( G_{+}\right) _{\eta _{+}}\times G_{-}\right] $ to $M_{\eta _{-}}^{-}$%
; that is, if $\left( h,\xi \right) $ is an arbitrary point at $%
J_{-}^{-1}\left( \eta _{-}\right) $, then the solution $\gamma :t\mapsto
\left( h\exp {t\xi ^{\flat }},{\xi }\right) $ for the dynamical system
defined on $G\times \mathfrak{g}^{\ast }$ by the data $\left( \omega _{\circ
},\mathcal{H}\right) $ remains there for all $t$, and $t\mapsto \pi _{\eta
_{-}}\left( \gamma \left( t\right) \right) $ is the solution for the
dynamical system defined on $M_{\eta _{-}}^{-}$ via $\left( \omega _{\eta
_{-}},\mathcal{H}_{\eta _{-}}\right) $ passing through $\pi _{\eta
_{-}}\left( h,\xi \right) $. Finally, $t\mapsto \left[ \tilde{\pi}_{\eta
_{-}}^{-1}\circ \pi _{\eta _{-}}\right] \left( \gamma \left( t\right)
\right) $ is the solution to the Hamilton equations on $\mathcal{N}\left(
g_{-},\eta _{-}\right) $ associated to the dynamical data $\left( \omega _{%
\mathcal{N}},\mathcal{H}_{\mathcal{N}}\right) $, with initial data $\gamma
\left( 0\right) =\tilde{\pi}_{\eta _{-}}^{-1}\circ \pi _{\eta _{-}}\left(
h,\xi \right) $. An explicit formula for this solution curve can be given,
because%
\begin{eqnarray*}
\left[ \left( \tilde{\pi}_{\eta _{-}}\right) ^{-1}\circ \pi _{\eta _{-}}%
\right] \left( h_{+}h_{-},\xi _{+}+\eta _{-}\right) &=&\tilde{\pi}_{\eta
_{-}}^{-1}\left( \left[ h_{+}h_{-},\xi _{+}+\eta _{-}\right] \right) \\
&=&\tilde{\pi}_{\eta _{-}}^{-1}\left( \left[
h_{+}g_{-},Ad_{g_{-}^{-1}h_{-}}^{G}\xi _{+}+\eta _{-}\right] \right) \\
&=&\left( h_{+}g_{-},Ad_{g_{-}^{-1}h_{-}}^{G}\xi _{+}+\eta _{-}\right)
\end{eqnarray*}%
This means that the solution curve has the following form 
\begin{equation*}
t\mapsto \left( h_{+}\left( t\right) g_{-},Ad_{g_{-}^{-1}h_{-}\left(
t\right) }^{G}\xi _{+}+\eta _{-}\right)
\end{equation*}%
where $h_{\pm }:\mathbb{R}\rightarrow G_{\pm }$ are the solutions of the
factorization problem $h_{+}\left( t\right) h_{-}\left( t\right) =h\exp {%
t\xi ^{\flat }}$.

In the same vein, the previous setting can be applied in order to find the
hamiltonian vector field associated to the hamiltonian $\mathcal{H}_{%
\mathcal{N}}$ on $\mathcal{N}\left( g_{-},\eta _{-}\right) $. First of all,
we need the following proposition.

\begin{description}
\item[Proposition:] \textit{By left trivializing}\textsl{\ }$T_{\left( h,\xi
\right) }\left( G\times \mathfrak{g}^{\ast }\right) =T_{h}G\times T_{\xi }%
\mathfrak{g}^{\ast }=\mathfrak{g}\times \mathfrak{g}^{\ast }$\textit{, the
derivative of the map }$\Xi _{\eta _{-}}:=\tilde{\pi}_{\eta _{-}}^{-1}\circ
\pi _{\nu }$\textit{\ is given by}%
\begin{equation*}
\begin{array}{l}
\left. \Xi _{\eta _{-}\ast }\right\vert _{\left( h,\xi \right) }\left(
X,\lambda _{+}\right) \\ 
=\left( Ad_{g_{-}^{-1}}\Pi _{\mathfrak{g}%
_{+}}Ad_{h_{-}}^{G}X,Ad_{g_{-}^{-1}}^{G}ad_{Ad_{h_{-}}X_{-}}^{\mathfrak{g}%
}Ad_{h_{-}}^{G}\xi _{+}+Ad_{g_{-}^{-1}h_{-}}^{G}\lambda _{+}\right)%
\end{array}%
\end{equation*}%
\textit{\ for all }$\left( X,\lambda _{+}\right) \in T_{\left( h,\xi \right)
}J_{-}^{-1}\left( \eta _{-}\right) \subset \mathfrak{g}\times \mathfrak{g}%
^{\ast }$\textit{. \ }
\end{description}

\textbf{Proof: }Let $h_{\pm }:\mathbb{R}\rightarrow G_{\pm }$ be the
solution curves for the factorization problem 
\begin{equation*}
h_{+}\left( t\right) h_{-}\left( t\right) =h\exp {tX}
\end{equation*}%
then if 
\begin{equation*}
X_{\pm }:=L_{g_{\pm }^{-1}\ast }\left\{ \left. \frac{d}{dt}\right\vert _{t=0}%
\left[ h_{\pm }\left( t\right) \right] \right\}
\end{equation*}%
we see that 
\begin{equation*}
X=Ad_{h_{-}^{-1}}^{G}X_{+}+X_{-}
\end{equation*}%
It is immediate from here to conclude that 
\begin{equation*}
X_{+}=\Pi _{\mathfrak{g}_{+}}\left( Ad_{h_{-}}^{G}X\right)
\end{equation*}%
and additionally 
\begin{equation*}
Ad_{h_{-}}X_{-}=\Pi _{\mathfrak{g}_{-}}Ad_{h_{-}}^{G}X
\end{equation*}%
By using the previous formula for map $\pi _{\eta _{-}}$ and the identity 
\begin{equation*}
\left. \frac{d}{dt}\right\vert _{t=0}\left[ Ad_{h_{-}\left( t\right)
}^{G}\xi _{+}\right] =ad_{Ad_{h_{-}}X_{-}}^{\mathfrak{g}}Ad_{h_{-}}^{G}\xi
_{+}
\end{equation*}%
we obtain the proposition.$\blacksquare $

Finally it remains to take into account that 
\begin{equation*}
\left. V_{\mathcal{H}_{\mathcal{N}}}\right\vert _{\left( g_{+}g_{-},\eta
_{+}+\eta _{-}\right) }=\left. \Xi _{\eta _{-}\ast }\right\vert _{\left(
h,\xi \right) }\left( \left. V_{\mathcal{H}}\right\vert _{\left( h,\xi
\right) }\right)
\end{equation*}%
for any $\left( h,\xi \right) \in \Xi _{\eta _{-}}^{-1}\left(
g_{+}g_{-},\eta _{+}+\eta _{-}\right) $. From the expression 
\begin{equation*}
\left. V_{\mathcal{H}}\right\vert _{\left( h,\xi \right) }=\left( \xi
^{\flat },0\right)
\end{equation*}%
in the left trivialization, and by realizing that $\left( h,\xi \right) \in
\Xi _{\eta_- }^{-1}\left( g_{+}g_{-},\eta _{+}+\eta _{-}\right) $ if and
only if 
\begin{equation*}
\begin{cases}
h_{+}=g_{+} \\ 
\\ 
\xi _{+}=Ad_{h_{-}^{-1}g_{-}}^{G}\eta _{+}%
\end{cases}%
\end{equation*}%
we can conclude that 
\begin{equation}
\left. V_{H_{\mathcal{N}}}\right\vert _{\left( g,\eta \right) }=\left(
Ad_{g_{-}^{-1}}^{G}\Pi _{\mathfrak{g}_{+}}Ad_{g_{-}}\eta _{+}^{\flat
},Ad_{g_{-}^{-1}}^{G}ad_{{\Pi }_{\mathfrak{g}_{-}}Ad_{g_{-}}\eta _{+}^{\flat
}}^{\mathfrak{g}}Ad_{g_{-}}^{G}\eta _{+}\right)  \label{VectorFieldAKS}
\end{equation}%
for $\left( g,\eta \right) =\left( g_{+}g_{-},\eta _{+}+\eta _{-}\right) $.
It was used here that the $Ad$-invariance of the Killing form implies the
equivariance formula 
\begin{equation*}
Ad_{h_{-}^{-1}}^{G}\xi _{+}^{\flat }=\left( Ad_{h_{-}}^{G}\xi _{+}\right)
^{\flat }
\end{equation*}%
for all $g_{-}\in G_{-},\xi _{+}\in \mathfrak{g}_{-}^{\circ }$.

It is interesting to relate this vector field with the vector field obtained
in eq. $\left( \ref{Dirac ham vec field}\right) $. If $\mathsf{h}\left( \xi
\right) :=\frac{1}{2}\left\langle \xi ,\xi ^{\flat }\right\rangle $, then $%
\mathcal{L}_{\mathsf{h}}\left( \eta \right) =\eta ^{\flat }$ and the vector
field $\left( \ref{Dirac ham vec field}\right) $ (evaluated at $\left(
h_{+}h_{-},\eta _{+}+\eta _{-}\right) $) can be written as 
\begin{eqnarray*}
&&V_{\mathsf{h}\circ \Phi ^{L}}^{D} \\
&=&\left( 
\begin{array}{l}
gAd_{h_{-}^{-1}}\Pi _{\mathfrak{g}_{+}^{\circ }}Ad_{h_{-}}(\eta _{+}+\eta
_{-})^{\flat }, \\ 
\qquad Ad_{h_{-}^{-1}}\Pi _{\mathfrak{g}_{-}^{\circ }}Ad_{h_{-}}\left[ \eta
_{+}+\eta _{-},Ad_{h_{-}^{-1}}\Pi _{\mathfrak{g}_{+}^{\circ
}}Ad_{h_{-}}(\eta _{+}+\eta _{-})^{\flat }\right]%
\end{array}%
\right) \\
&=&\left( gAd_{h_{-}^{-1}}\Pi _{\mathfrak{g}_{+}^{\circ }}Ad_{h_{-}}(\eta
_{+}),Ad_{h_{-}^{-1}}\Pi _{\mathfrak{g}_{-}^{\circ }}Ad_{h_{-}}\left[ \eta
_{+},Ad_{h_{-}^{-1}}\Pi _{\mathfrak{g}_{+}^{\circ }}Ad_{h_{-}}(\eta _{+})%
\right] \right)
\end{eqnarray*}%
where it was used that, under the performed identifications, $\eta _{\pm
}^{\flat }=\eta _{\pm }$ for all $\eta _{\pm }\in \mathfrak{g}_{\pm }^{\circ
}$, and moreover, that $\eta _{-}\in \mathfrak{g}_{-}^{\circ }$ is a
character. Finally 
\begin{eqnarray*}
V_{\mathsf{h}\circ \Phi ^{L}}^{D} &=&\left( gAd_{h_{-}^{-1}}\Pi _{\mathfrak{g%
}_{+}^{\circ }}Ad_{h_{-}}(\eta _{+}),Ad_{h_{-}^{-1}}\Pi _{\mathfrak{g}%
_{-}^{\circ }}\left[ Ad_{h_{-}}\eta _{+},\Pi _{\mathfrak{g}_{+}^{\circ
}}Ad_{h_{-}}(\eta _{+})\right] \right) \\
&=&\left( gAd_{h_{-}^{-1}}\Pi _{\mathfrak{g}_{+}^{\circ }}Ad_{h_{-}}(\eta
_{+}),-Ad_{h_{-}^{-1}}\Pi _{\mathfrak{g}_{-}^{\circ }}\left[ Ad_{h_{-}}\eta
_{+},\Pi _{\mathfrak{g}_{-}^{\circ }}Ad_{h_{-}}(\eta _{+})\right] \right) \\
&=&\left( gAd_{h_{-}^{-1}}\Pi _{\mathfrak{g}_{+}^{\circ }}Ad_{h_{-}}(\eta
_{+}),-Ad_{h_{-}^{-1}}\left[ Ad_{h_{-}}\eta _{+},\Pi _{\mathfrak{g}%
_{-}^{\circ }}Ad_{h_{-}}(\eta _{+})\right] \right)
\end{eqnarray*}%
gives us the desired expression, to be compared with eq.$\left( \ref{Dirac
ham vec field}\right) $.

\section{Example: $SL\left( 2,\mathbb{C}\right) =SU\left( 2\right) \times B$}

We now specialize the above abstract structure to $G=SL(2,\mathbb{C})$ and
its Iwasawa decomposition $SL(2,\mathbb{C})\cong SU\left( 2\right) \times B$%
, where $B$ is the group of $2\times 2$ complex upper triangular matrices,
with real diagonal and determinant equal to $1,$and we identify $%
G_{+}=SU\left( 2\right) $ and $G_{-}=B$.

The Killing form for $\mathfrak{sl}_{2}(\mathbb{C})$ is $\kappa (X,Y):=%
\mathsf{tr}\,{(ad}\left( {X}\right) {ad}\left( {Y}\right) {)}=4\mathsf{tr}\,{%
(XY)}$, the restrictions to $\mathfrak{su}_{2}$, $\mathfrak{a}$, and $%
\mathfrak{n}$ are negative defined; positive defined, and $0$, respectively.
We consider the non degenerate symmetric bilinear form on $\mathfrak{sl}_{2}(%
\mathbb{C})$ 
\begin{equation}
(X,Y)_{\mathfrak{sl}_{2}}=-\frac{1}{4}\mathrm{\func{Im}}\,\kappa (X,Y)
\label{prod-escal 0}
\end{equation}%
which turns $\mathfrak{b}$ and $\mathfrak{su}_{2}$ into isotropic subspaces.
Also, we take the basis 
\begin{equation*}
\begin{array}{ccccc}
T_{1}=\left[ 
\begin{array}{cc}
0 & i \\ 
i & 0%
\end{array}%
\right] &  & T_{2}=\left[ 
\begin{array}{cc}
0 & 1 \\ 
-1 & 0%
\end{array}%
\right] &  & T_{3}=\left[ 
\begin{array}{cc}
i & 0 \\ 
0 & -i%
\end{array}%
\right]%
\end{array}%
\end{equation*}%
for $\mathfrak{su}_{2}$, and%
\begin{equation*}
\begin{array}{ccccc}
T^{1}=\left( 
\begin{array}{cc}
0 & -1 \\ 
0 & 0%
\end{array}%
\right) ~, &  & T^{2}=\left( 
\begin{array}{cc}
0 & i \\ 
0 & 0%
\end{array}%
\right) ~, &  & T^{3}=-\frac{1}{2}\left( 
\begin{array}{cc}
1 & 0 \\ 
0 & -1%
\end{array}%
\right)%
\end{array}%
\end{equation*}%
in $\mathfrak{b}$. Then, the crossed product are 
\begin{equation}
\begin{array}{ccccc}
(T_{1},T^{1})=1 &  & (T_{2},T^{1})=0 &  & (T_{3},T^{1})=0 \\ 
(T_{1},T^{2})=0 &  & (T_{2},T^{2})=1 &  & (T_{3},T^{2})=0 \\ 
(T_{1},T^{3})=0 &  & (T_{1},T^{3})=0 &  & (T_{3},T^{3})=1%
\end{array}
\label{prod-escal}
\end{equation}%
allowing for the identification 
\begin{eqnarray}
\psi &:&\mathfrak{su}_{2}\rightarrow \mathfrak{b}^{\ast }  \label{psi} \\
&&  \notag \\
\psi (T_{1}) &=&\mathbf{t}^{1},\qquad \psi (T_{2})=\mathbf{t}^{2},\qquad
\psi (T_{3})=\mathbf{t}^{3}  \notag
\end{eqnarray}%
where $\left\{ \mathbf{t}^{1},\mathbf{t}^{2},\mathbf{t}^{3}\right\} \subset 
\mathfrak{b}^{\ast }$ is the dual basis to $\left\{
T^{1},T^{2},T^{3}\right\} \subset \mathfrak{b}$.

Let us come back to the expression $\left( \ref{Dirac bracket G+xg+ I}%
\right) $%
\begin{eqnarray*}
\left\{ \mathcal{F},\mathcal{G}\right\} ^{D}\left( g,\eta \right)
&=&\left\langle g\mathbf{d}\mathcal{F},Ad_{g_{-}^{-1}}^{G}\Pi _{\mathfrak{g}%
_{+}^{\circ }}Ad_{g_{-}}^{G}\delta \mathcal{G}\right\rangle \\
&&-\left\langle g\mathbf{d}\mathcal{G},Ad_{g_{-}^{-1}}^{G}\Pi _{\mathfrak{g}%
_{+}^{\circ }}Ad_{g_{-}}^{G}\delta \mathcal{F}\right\rangle \\
&&-\left\langle \eta ,\left[ Ad_{g_{-}^{-1}}^{G}\Pi _{\mathfrak{g}%
_{+}^{\circ }}Ad_{g_{-}}^{G}\delta \mathcal{F},Ad_{g_{-}^{-1}}^{G}\Pi _{%
\mathfrak{g}_{+}^{\circ }}Ad_{g_{-}}^{G}\delta \mathcal{G}\right]
\right\rangle
\end{eqnarray*}%
and apply it to the functions 
\begin{eqnarray*}
\mathrm{T}_{i}^{j} &:&G\longrightarrow \mathbb{C~}/~\mathrm{T}_{i}^{j}\left(
g\right) =g_{i}^{j} \\
\xi _{A} &:&\mathfrak{g}^{\ast }\longrightarrow \mathbb{C~}/~\xi
_{A}=\left\langle \xi ,\mathbb{T}_{A}\right\rangle
\end{eqnarray*}%
with%
\begin{eqnarray*}
\delta \mathrm{T}_{i}^{j} &=&0 \\
\mathbf{d}\xi _{A} &=&0
\end{eqnarray*}

So, it is necessary to calculate the expressions%
\begin{equation*}
Ad_{g_{-}^{-1}}^{G}\Pi _{\mathfrak{g}_{+}^{\circ }}Ad_{g_{-}}^{G}\delta \xi
_{A}
\end{equation*}%
The differential $\delta \xi _{A}$ coincides with the generator $\mathbb{T}%
_{A}$ of the Lie algebra $\mathfrak{g}$, being $\mathbb{T}_{A}\in \left\{
T_{a},T^{a}\right\} _{a=1}^{n}$. This relations can be written in terms of
the coordinates for $\eta _{+}=\left\langle \eta _{+},T_{a}\right\rangle 
\mathbf{t}_{a}=\xi _{a}\left( \eta \right) \mathbf{t}_{a}$ and $\eta
_{-}=\left\langle \eta _{-},T^{a}\right\rangle \mathbf{t}^{a}=\xi ^{a}\left(
\eta \right) \mathbf{t}^{a}$. Therefore, for $A$ running on the superindex $%
\delta \xi _{A}\equiv \delta \xi ^{a}=T^{a}$ we have $\delta \xi _{A}=\delta
\xi _{A-}$ and $\delta \xi _{A+}=0$. On the other side, for $A$ running on
the subindex $\delta \xi _{A}\equiv \delta \xi _{a}=T_{a}$, we have $\delta
\xi _{A}=\delta \xi _{A+}=\delta \xi _{a}$ and $\delta \xi _{A-}=0$.
Therefore, since the only non vanishing Dirac brackets just involve $\xi
_{a} $, we evaluate 
\begin{equation*}
Ad_{g_{-}^{-1}}^{G}\Pi _{\mathfrak{g}_{+}^{\circ }}Ad_{g_{-}}^{G}\delta \xi
_{a}=Ad_{g_{-}^{-1}}^{G}\Pi _{\mathfrak{g}_{+}^{\circ }}Ad_{g_{-}}^{G}T_{a}
\end{equation*}%
Writing a generic element 
\begin{equation*}
g_{+}=\left( 
\begin{array}{cc}
\alpha & \beta \\ 
-\bar{\beta} & \bar{\alpha}%
\end{array}%
\right) \in G_{+}\cong SU\left( 2\right)
\end{equation*}%
and 
\begin{equation*}
g_{-}=\left( 
\begin{array}{cc}
a & b+ic \\ 
0 & a^{-1}%
\end{array}%
\right) =\left( 
\begin{array}{cc}
a & z \\ 
0 & a^{-1}%
\end{array}%
\right) \in G_{-}\cong B
\end{equation*}%
with $a\in \mathbb{R}_{>0}$, $b,c\in \mathbb{R}$, we have that 
\begin{equation*}
\begin{array}{c}
Ad_{g_{-}^{-1}}^{G}\Pi _{\mathfrak{g}_{+}^{\circ
}}Ad_{g_{-}}^{G}T_{1}=T_{1}-\left( 1-\frac{b^{2}}{a^{2}}-\frac{c^{2}}{a^{2}}-%
\frac{1}{a^{4}}\right) T^{2}+2\frac{c}{a}T^{3} \\ 
\\ 
Ad_{g_{-}^{-1}}^{G}\Pi _{\mathfrak{g}_{+}^{\circ
}}Ad_{g_{-}}^{G}T_{2}=T_{2}+\left( 1-\frac{b^{2}}{a^{2}}-\frac{c^{2}}{a^{2}}-%
\frac{1}{a^{4}}\right) T^{1}+2\frac{b}{a}T^{3} \\ 
\\ 
Ad_{g_{-}^{-1}}^{G}\Pi _{\mathfrak{g}_{+}^{\circ
}}Ad_{g_{-}}^{G}T_{3}=T_{3}+2\frac{c}{a}T^{1}+2\frac{b}{a}T^{2}%
\end{array}%
\end{equation*}

Explicit calculation for the non trivial Dirac brackets gives 
\begin{equation}
\begin{array}{l}
\left\{ \xi _{1},\mathrm{T}_{i}^{j}\right\} ^{D}\left( g,\eta \right) =-%
\mathrm{T}_{i}^{k}\left( g\right) \left[ T_{1}-\left( 1-\dfrac{b^{2}}{a^{2}}-%
\dfrac{c^{2}}{a^{2}}-\dfrac{1}{a^{4}}\right) T^{2}+2\dfrac{c}{a}T^{3}\right]
_{k}^{j} \\ 
\\ 
\left\{ \xi _{2},\mathrm{T}_{i}^{j}\right\} ^{D}\left( g,\eta \right) =-%
\mathrm{T}_{i}^{k}\left( g\right) \left[ T_{2}+\left( 1-\dfrac{b^{2}}{a^{2}}-%
\dfrac{c^{2}}{a^{2}}-\dfrac{1}{a^{4}}\right) T^{1}+2\dfrac{b}{a}T^{3}\right]
_{k}^{j} \\ 
\\ 
\left\{ \xi _{3},\mathrm{T}_{i}^{j}\right\} ^{D}\left( g,\eta \right) =-%
\mathrm{T}_{i}^{k}\left( g\right) \left[ T_{3}+2\dfrac{c}{a}T^{1}+2\dfrac{b}{%
a}T^{2}\right] _{k}^{j}%
\end{array}
\label{fund Dirac Bracket 1}
\end{equation}

\begin{equation}
\begin{array}{l}
\left\{ \xi _{1},\xi _{2}\right\} ^{D}\left( g,\eta \right) =2\dfrac{b}{a}%
\xi _{1}\left( \eta \right) -2\dfrac{c}{a}\xi _{2}\left( \eta \right)
+2\left( \dfrac{b^{2}}{a^{2}}+\dfrac{c^{2}}{a^{2}}+\dfrac{1}{a^{4}}\right)
\xi _{3}\left( \eta \right) \\ 
~~~~~~~~~~~~~\text{~~~~~~~~~~~~~~}~~-2\dfrac{c}{a}\left( 1+\dfrac{b^{2}}{%
a^{2}}+\dfrac{c^{2}}{a^{2}}+\dfrac{1}{a^{4}}\right) \xi ^{1}\left( \eta
\right) \\ 
~~~~~~~~~~~~~\text{~~~~~~~~~~~~~~}~~-2\dfrac{b}{a}\left( 1+\dfrac{b^{2}}{%
a^{2}}+\dfrac{c^{2}}{a^{2}}+\dfrac{1}{a^{4}}\right) \xi ^{2}\left( \eta
\right) \\ 
\\ 
\left\{ \xi _{1},\xi _{3}\right\} ^{D}\left( g,\eta \right) =-2\xi
_{2}\left( \eta \right) -2\dfrac{c}{a}\xi _{3}\left( \eta \right) \\ 
~~~~~~\text{~~~~~~~}~~~~~~~~~~~~~~~~+2\left( \dfrac{b^{2}}{a^{2}}+3\dfrac{%
c^{2}}{a^{2}}+\dfrac{1}{a^{4}}-1\right) \xi ^{1}\left( \eta \right) \\ 
~~~~~~\text{~~~~~~~}~~~~~~~~~~~~~~~~+4\dfrac{bc}{a^{2}}\xi ^{2}\left( \eta
\right) +4\dfrac{b}{a}\xi ^{3}\left( \eta \right) \\ 
\\ 
\left\{ \xi _{2},\xi _{3}\right\} ^{D}\left( g,\eta \right) =2\xi _{1}\left(
\eta \right) -2\dfrac{b}{a}\xi _{3}\left( \eta \right) +4\dfrac{bc}{a^{2}}%
\xi ^{1}\left( \eta \right) \\ 
~~~~~~~~\text{~~}~~~~~~~~~~~~~~~~~~+2\left( 3\dfrac{b^{2}}{a^{2}}+\dfrac{%
c^{2}}{a^{2}}+\dfrac{1}{a^{4}}-1\right) \xi ^{2}\left( \eta \right) -4\dfrac{%
c}{a}\xi ^{3}\left( \eta \right)%
\end{array}
\label{fund Dirac bracket 2}
\end{equation}

Setting $\eta _{-}=0$, the only character in $\mathfrak{b}^{\ast }$, they
reduce to 
\begin{equation}
\begin{array}{l}
\left\{ \xi _{1},\mathrm{T}_{i}^{j}\right\} ^{D}\left( g,\eta \right) =-%
\mathrm{T}_{i}^{k}\left( g\right) \left[ T_{1}\right] _{k}^{j} \\ 
~~~~~~\text{~~~~~~~}~~~~~~~~~~~~~~~~~+\mathrm{T}_{i}^{k}\left( g\right) %
\left[ \left( 1-\dfrac{b^{2}}{a^{2}}-\dfrac{c^{2}}{a^{2}}-\dfrac{1}{a^{4}}%
\right) T^{2}+2\dfrac{c}{a}T^{3}\right] _{k}^{j} \\ 
\\ 
\left\{ \xi _{2},\mathrm{T}_{i}^{j}\right\} ^{D}\left( g,\eta \right) =-%
\mathrm{T}_{i}^{k}\left( g\right) \left[ T_{2}\right] _{k}^{j} \\ 
~~~~~~\text{~~~~~~~}~~~~~~~~~~~~~~~~~-\mathrm{T}_{i}^{k}\left( g\right) %
\left[ \left( 1-\dfrac{b^{2}}{a^{2}}-\dfrac{c^{2}}{a^{2}}-\dfrac{1}{a^{4}}%
\right) T^{1}+2\dfrac{b}{a}T^{3}\right] _{k}^{j} \\ 
\\ 
\left\{ \xi _{3},\mathrm{T}_{i}^{j}\right\} ^{D}\left( g,\eta \right) =-%
\mathrm{T}_{i}^{k}\left( g\right) \left[ T_{3}\right] _{k}^{j}-\mathrm{T}%
_{i}^{k}\left( g\right) \left[ 2\dfrac{c}{a}T^{1}+2\dfrac{b}{a}T^{2}\right]
_{k}^{j}%
\end{array}
\label{Fund Dirac brackets for a character}
\end{equation}%
\begin{equation}
\begin{array}{l}
\left\{ \xi _{1},\xi _{2}\right\} ^{D}\left( g,\eta \right) =2\xi _{3}\left(
\eta \right) +2\dfrac{b}{a}\xi _{1}\left( \eta \right) \\ 
~~~~~~\text{~~~~~~~}~~~~~~~~~~~~~~~-2\dfrac{c}{a}\xi _{2}\left( \eta \right)
-2\left( 1-\dfrac{b^{2}}{a^{2}}-\dfrac{c^{2}}{a^{2}}-\dfrac{1}{a^{4}}\right)
\xi _{3}\left( \eta \right) \\ 
\\ 
\left\{ \xi _{3},\xi _{1}\right\} ^{D}\left( g,\eta \right) =2\xi _{2}\left(
\eta \right) +2\dfrac{c}{a}\xi _{3}\left( \eta \right) \\ 
\\ 
\left\{ \xi _{2},\xi _{3}\right\} ^{D}\left( g,\eta \right) =2\xi _{1}\left(
\eta \right) -2\dfrac{b}{a}\xi _{3}\left( \eta \right)%
\end{array}
\label{Fund Dirac brackets for a character II}
\end{equation}

Let us now address some dynamical model on the phase space $\mathcal{N}%
\left( g_{-},0\right) $, taking a collective hamiltonian function%
\begin{equation*}
\mathcal{H}\left( g,\eta \right) =-\frac{1}{16}\kappa (\Pi _{\mathfrak{g}%
_{+}^{\circ }}Ad_{g}^{G}\bar{\psi}\left( \eta _{+}\right) ,\Pi _{\mathfrak{g}%
_{+}^{\circ }}Ad_{g}^{G}\bar{\psi}\left( \eta _{+}\right) )
\end{equation*}%
We may write then 
\begin{equation*}
\mathcal{H}\left( g,\eta \right) =\dfrac{1}{2}\sum_{a=1}^{3}\phi ^{a}(g,\eta
)\phi ^{a}(g,\eta )
\end{equation*}%
The Hamilton equation of motion are then%
\begin{eqnarray*}
\mathrm{\dot{T}}_{j}^{k} &=&\left\{ \mathrm{T}_{j}^{k},\mathcal{H}\right\}
^{D} \\
&& \\
\dot{\xi}_{a} &=&\left\{ \xi _{a},\mathcal{H}\right\} ^{D}
\end{eqnarray*}%
We may use the properties of the Poisson-Dirac bracket to get the Hamilton
equation of motion%
\begin{equation*}
\left\{ 
\begin{array}{l}
\mathrm{\dot{T}}_{u}^{v}=\sum_{a=1}^{3}\phi ^{a}\left\{ \mathrm{T}%
_{u}^{v},\phi ^{a}\right\} ^{D} \\ 
\\ 
\dot{\xi}_{c}=\sum_{a=1}^{3}\phi ^{a}\left\{ \xi _{c},\phi ^{a}\right\} ^{D}%
\end{array}%
\right.
\end{equation*}

A more explicit expression can be obtained in a compact matrix notation and
considering the Hamilton equations in the form $\left( \ref{Ham eq on N}%
\right) $. Therefore, for the Hamiltonian $\left( \ref{Ham example}\right) $%
, and introducing the identification $\hat{\kappa}_{\mathfrak{g}_{+}^{\circ
}}:\mathfrak{g}_{+}^{\circ }\longrightarrow \mathfrak{g}_{+}^{\ast }$, such
that $\left\langle \hat{\kappa}_{\mathfrak{g}_{+}^{\circ }}\left(
X_{+}\right) ,Y_{+}\right\rangle =\kappa (X_{+},Y_{+})$, with the relations $%
\hat{\kappa}_{\mathfrak{g}_{+}^{\circ }}(T_{i})=-8\mathbf{t}_{i}$, and using
the bijection $\left( \ref{su2->b basis}\right) $, $\bar{\psi}^{\ast }(\hat{%
\kappa}_{\mathfrak{g}_{+}^{\circ }}\left( T_{i}\right) )=-8T^{i}$, where $%
\Psi :\mathfrak{g}^{\ast }\longrightarrow \mathfrak{g}$ stands for the
identification provided by the non degenerate scalar product $\left(
,\right) _{\mathfrak{g}}$ which, when restricted to $\mathfrak{g}_{+}^{\circ
}$ coincides with $\psi $. So that differential is 
\begin{eqnarray*}
d\mathcal{H} &=&\left( \mathbf{d}\mathcal{H}\mathbf{,}\delta \mathcal{H}%
\right) \\
&=&\left( -\frac{1}{2}\lambda R_{g^{-1}}^{\ast }\left( ad_{Ad_{g}^{G}\bar{%
\Psi}\left( \eta \right) }^{\mathfrak{g}\ast }\hat{\kappa}_{\mathfrak{g}%
_{+}^{\circ }}\left( \Pi _{\mathfrak{g}_{+}^{\circ }}Ad_{g}^{G}\bar{\Psi}%
\left( \eta \right) \right) \right) \right. \\
&&~~~~~~~~~~~\left. ,\frac{1}{2}\lambda Ad_{g^{-1}}^{G}\bar{\Psi}^{\ast
}\left( \hat{\kappa}_{\mathfrak{g}_{+}^{\circ }}\left( \Pi _{\mathfrak{g}%
_{+}^{\circ }}Ad_{g}^{G}\bar{\Psi}\left( \eta \right) \right) \right) \right)
\end{eqnarray*}%
and remembering that $\eta _{-}=0$, the Hamilton equations reads, for $\eta
_{-}=0$, 
\begin{equation}
\begin{array}{l}
\left\{ 
\begin{array}{l}
g_{+}^{-1}\dot{g}_{+}=-\dfrac{1}{8}\Pi _{\mathfrak{g}_{+}^{\circ
}}Ad_{g_{+}^{-1}}^{G}\bar{\psi}^{\ast }\left( \hat{\kappa}_{\mathfrak{g}%
_{+}^{\circ }}\left( \Pi _{\mathfrak{g}_{+}^{\circ }}Ad_{g}^{G}\bar{\psi}%
\left( \eta _{+}\right) \right) \right) \\ 
\\ 
Ad_{g_{-}^{-1}}^{G\ast }\dot{\eta}_{+}=\dfrac{1}{8}\Pi _{\mathfrak{g}%
_{+}^{\ast }}ad_{\Pi _{\mathfrak{g}_{-}^{\circ }}Ad_{g_{+}^{-1}}^{G}\bar{\psi%
}^{\ast }\left( \hat{\kappa}_{\mathfrak{g}_{+}^{\circ }}\left( \Pi _{%
\mathfrak{g}_{+}^{\circ }}Ad_{g}^{G}\bar{\psi}\left( \eta _{+}\right)
\right) \right) }^{\mathfrak{g}\ast }Ad_{g_{-}^{-1}}^{G\ast }\eta _{+}%
\end{array}%
\right. \\ 
\\ 
\left\{ 
\begin{array}{l}
\dot{g}_{-}g_{-}^{-1}=0 \\ 
\\ 
\dot{\eta}_{-}=0%
\end{array}%
\right.%
\end{array}
\label{Example Ham eqs}
\end{equation}

To get some insight on the meaning physical content of this model, let us
built up the Lagrangian function associated with this hamiltonian system.
Since 
\begin{equation*}
\bar{\psi}^{\ast }\left( \eta \right) =\eta _{1}T^{1}+\eta _{2}T^{2}+\eta
_{3}T^{3}
\end{equation*}%
and 
\begin{equation*}
\begin{array}{l}
g_{+}^{-1}g_{+}^{T^{3}}=\dfrac{1}{2}i\left( \alpha \bar{\beta}-\bar{\alpha}%
\beta \right) T_{1}+\dfrac{1}{2}\left( \alpha \bar{\beta}+\bar{\alpha}\beta
\right) T_{2} \\ 
g_{+}^{-1}g_{+}^{T^{2}}=-\frac{1}{2}\left( \beta ^{2}+\bar{\beta}^{2}\right)
T_{1}-\frac{1}{2}i\left( \beta ^{2}-\bar{\beta}^{2}\right) T_{2}\mathbf{-}%
\frac{1}{2}\left( \bar{\alpha}\bar{\beta}+\alpha \beta \right) T_{3} \\ 
g_{+}^{-1}g_{+}^{T^{1}}=\frac{1}{2}i\left( \beta ^{2}-\bar{\beta}^{2}\right)
T_{1}-\frac{1}{2}\left( \beta ^{2}+\bar{\beta}^{2}\right) T_{2}+\frac{1}{2}%
i\left( \alpha \beta -\bar{\alpha}\bar{\beta}\right) T_{3}%
\end{array}%
\end{equation*}%
where it was used that 
\begin{equation}
\begin{array}{ccccc}
\bar{\psi}^{\ast }\left( \mathbf{t}_{1}\right) =T^{1} &  & \bar{\psi}^{\ast
}\left( \mathbf{t}_{2}\right) =T^{2} &  & \bar{\psi}^{\ast }\left( \mathbf{t}%
_{3}\right) =T^{3}%
\end{array}
\label{psi* barra}
\end{equation}%
we get%
\begin{equation*}
Ad_{g_{-}}^{G}\bar{\psi}\left( \eta _{+}\right) =a\left( a\eta _{+1}-b\eta
_{+3}\right) T^{1}+a\left( a\eta _{+2}+c\eta _{+3}\right) T^{2}+\eta
_{+3}T^{3}
\end{equation*}

Let us recall that $\phi _{X}(g,\eta )=\left\langle Ad_{g^{-1}}^{G^{\ast
}}\eta ,X\right\rangle $, so that $\phi ^{a}(g,\eta )=\left\langle
Ad_{g^{-1}}^{G^{\ast }}\eta _{+},T^{a}\right\rangle $ with%
\begin{eqnarray*}
\phi ^{1}(g,\eta ) &=&\frac{1}{2}ia^{2}\left( \beta ^{2}-\bar{\beta}%
^{2}\right) \eta _{+1}-\frac{1}{2}a^{2}\left( \beta ^{2}+\bar{\beta}%
^{2}\right) \eta _{+2} \\
&&+\frac{1}{2}i\left( \alpha \beta -\bar{\alpha}\bar{\beta}-a\bar{z}\beta
^{2}+az\bar{\beta}^{2}\right) \eta _{+3} \\
&& \\
\phi ^{2}(g,\eta ) &=&-\frac{1}{2}a^{2}\left( \beta ^{2}+\bar{\beta}%
^{2}\right) \eta _{+1}-\frac{1}{2}ia^{2}\left( \beta ^{2}-\bar{\beta}%
^{2}\right) \eta _{+2} \\
&&+\frac{1}{2}\left( a\bar{z}\beta ^{2}+az\bar{\beta}^{2}-\bar{\alpha}\bar{%
\beta}-\alpha \beta \right) \eta _{+3} \\
&& \\
\phi ^{3}(g,\eta ) &=&-\frac{1}{2}ia^{2}\left( \bar{\alpha}\beta -\alpha 
\bar{\beta}\right) \eta _{+1}+\frac{1}{2}a^{2}\left( \alpha \bar{\beta}+\bar{%
\alpha}\beta \right) \eta _{+2} \\
&&+\frac{1}{2}ia\left( \bar{z}\bar{\alpha}\beta -z\alpha \bar{\beta}\right)
\eta _{+3}
\end{eqnarray*}

Written in components, the first Hamilton equation in $\left( \ref{Example
Ham eqs}\right) $ gives rise to the differential equations 
\begin{eqnarray*}
\func{Im}\left( \bar{\alpha}\dot{\beta}-\beta \overset{\cdot }{\bar{\alpha}}%
\right) &=&-\dfrac{1}{2}\left( \left( \beta ^{2}+\bar{\beta}^{2}\right) \phi
^{2}-i\left( \beta ^{2}-\bar{\beta}^{2}\right) \phi ^{1}-i\left( \alpha \bar{%
\beta}-\bar{\alpha}\beta \right) \phi ^{3}\right) \\
&& \\
\func{Re}\left( \bar{\alpha}\dot{\beta}-\beta \overset{\cdot }{\bar{\alpha}}%
\right) &=&-\dfrac{1}{2}\left( \left( \beta ^{2}+\bar{\beta}^{2}\right) \phi
^{1}+i\left( \beta ^{2}-\bar{\beta}^{2}\right) \phi ^{2}-\left( \alpha \bar{%
\beta}+\bar{\alpha}\beta \right) \phi ^{3}\right) \\
&& \\
\left( \bar{\alpha}\dot{\alpha}+\beta \overset{\cdot }{\bar{\beta}}\right)
&=&-\dfrac{1}{2}\left( \left( \bar{\alpha}\bar{\beta}+\alpha \beta \right)
\phi ^{2}-i\left( \alpha \beta -\bar{\alpha}\bar{\beta}\right) \phi
^{1}\right)
\end{eqnarray*}

We use these equations to invert the Legendre transformation in order to
retrieve the Lagrange function. From these equations we write $\eta _{+1}$
and $\eta _{+2}$ in terms of the velocities $\bar{\alpha}\dot{\beta}-\beta 
\overset{\cdot }{\bar{\alpha}}$ and $\bar{\alpha}\dot{\alpha}+\beta \overset{%
\cdot }{\bar{\beta}}$. Thus, the expression of the momentum map associated
with the dressing action are 
\begin{eqnarray*}
\left[ \Pi _{\mathfrak{g}_{+}^{\circ }}Ad_{g}^{G}\bar{\psi}\left( \eta
_{+}\right) \right] _{1} &=&-\dfrac{1}{2\beta \bar{\alpha}}\left( \alpha
\beta +\gamma \delta \right) \left( \bar{\alpha}\dot{\beta}-\beta \overset{%
\cdot }{\bar{\alpha}}\right) \\
&&+\dfrac{1}{2\bar{\alpha}\left\vert \beta \right\vert ^{4}}\beta \left(
\left\vert \beta \right\vert ^{4}+\left\vert \beta \right\vert ^{2}-\gamma
^{2}\delta ^{2}\right) \left( \bar{\alpha}\dot{\alpha}+\beta \overset{\cdot }%
{\bar{\beta}}\right) \\
&& \\
\left[ \Pi _{\mathfrak{g}_{+}^{\circ }}Ad_{g}^{G}\bar{\psi}\left( \eta
_{+}\right) \right] _{2} &=&-i\dfrac{1}{2\beta \bar{\alpha}}\left( \alpha
\beta -\gamma \delta \right) \left( \bar{\alpha}\dot{\beta}-\beta \overset{%
\cdot }{\bar{\alpha}}\right) \\
&&+i\dfrac{1}{2\bar{\alpha}\left\vert \beta \right\vert ^{4}}\beta \left(
\gamma ^{2}\delta ^{2}+\left\vert \beta \right\vert ^{4}+\left\vert \beta
\right\vert ^{2}\right) \left( \bar{\alpha}\dot{\alpha}+\beta \overset{\cdot 
}{\bar{\beta}}\right) \\
&& \\
\left[ \Pi _{\mathfrak{g}_{+}^{\circ }}Ad_{g}^{G}\bar{\psi}\left( \eta
_{+}\right) \right] _{3} &=&\dfrac{\alpha }{\beta }\left( \bar{\alpha}\dot{%
\beta}-\beta \overset{\cdot }{\bar{\alpha}}\right) -\left( \bar{\alpha}\dot{%
\alpha}+\beta \overset{\cdot }{\bar{\beta}}\right)
\end{eqnarray*}%
where we may observe that there are no trace of $g_{-}$. So, the expression
for the Hamilton function in terms of the velocities is the same in each
phase space $\mathcal{N}\left( g_{-},0\right) $, 
\begin{eqnarray*}
&&\mathcal{H}(g,\dot{g}) \\
&=&\dfrac{1}{2\left\vert \beta \right\vert ^{2}}\left( \dfrac{\alpha \gamma 
}{\bar{\alpha}\beta }\left( \bar{\alpha}\dot{\beta}-\beta \overset{\cdot }{%
\bar{\alpha}}\right) ^{2}-2\dfrac{\gamma }{\delta }\left( \bar{\alpha}\dot{%
\alpha}+\beta \overset{\cdot }{\bar{\beta}}\right) \left( \bar{\alpha}\dot{%
\beta}-\beta \overset{\cdot }{\bar{\alpha}}\right) -\left( \bar{\alpha}\dot{%
\alpha}+\beta \overset{\cdot }{\bar{\beta}}\right) ^{2}\right)
\end{eqnarray*}

Let us then built up the Lagrangian function describing the dynamics of
these systems. It is obtained as usual%
\begin{equation*}
L_{\mathcal{N}\left( g_{-},0\right) }\left( g,\dot{g}\right) =\left\langle
\eta _{+},g_{+}^{-1}\dot{g}_{+}\right\rangle -\mathcal{H}(g_{+},\eta _{+})
\end{equation*}%
where 
\begin{equation*}
\left\langle \eta _{+},g_{+}^{-1}\dot{g}_{+}\right\rangle =\eta _{+1}\func{Im%
}\left( \bar{\alpha}\dot{\beta}-\beta \overset{\cdot }{\bar{\alpha}}\right)
+\eta _{+2}\func{Re}\left( \bar{\alpha}\dot{\beta}-\beta \overset{\cdot }{%
\bar{\alpha}}\right) -i\eta _{+3}\left( \bar{\alpha}\dot{\alpha}+\beta 
\overset{\cdot }{\bar{\beta}}\right)
\end{equation*}%
After some tedious calculations, we get

\begin{eqnarray*}
&&L_{\mathcal{N}\left( g_{-},0\right) }\left( g_{+},\dot{g}_{+}\right) \\
&=&\dfrac{1}{2a^{2}\left\vert \beta \right\vert ^{2}}\left\vert \bar{\alpha}%
\dot{\beta}-\beta \overset{\cdot }{\bar{\alpha}}\right\vert ^{2}+\dfrac{1}{%
2\left\vert \beta \right\vert ^{2}}\dfrac{\alpha \bar{\beta}}{\bar{\alpha}%
\beta }\left( \dfrac{1}{a^{2}}-1\right) \left( \bar{\alpha}\dot{\beta}-\beta 
\overset{\cdot }{\bar{\alpha}}\right) ^{2} \\
&&+\dfrac{1}{2a^{2}\left\vert \beta \right\vert ^{2}}\dfrac{\bar{\beta}}{%
\bar{\alpha}}\left( 2a^{2}-1-\dfrac{1}{\left\vert \beta \right\vert ^{2}}%
\right) \left( \bar{\alpha}\dot{\beta}-\beta \overset{\cdot }{\bar{\alpha}}%
\right) \left( \bar{\alpha}\dot{\alpha}+\beta \overset{\cdot }{\bar{\beta}}%
\right) \\
&&+\dfrac{1}{2a^{2}\left\vert \beta \right\vert ^{4}}\beta \bar{\alpha}%
\left( \bar{\alpha}\dot{\alpha}+\beta \overset{\cdot }{\bar{\beta}}\right) 
\overline{\left( \bar{\alpha}\dot{\beta}-\beta \overset{\cdot }{\bar{\alpha}}%
\right) }+\dfrac{1}{2\left\vert \beta \right\vert ^{2}}\left( \bar{\alpha}%
\dot{\alpha}+\beta \overset{\cdot }{\bar{\beta}}\right) ^{2} \\
&&+\eta _{+3}\dfrac{1}{2a^{2}\left\vert \beta \right\vert ^{2}}\left( 2\func{%
Im}\left( \bar{\beta}\left( a\bar{z}\beta -\alpha \right) \left( \bar{\alpha}%
\dot{\beta}-\beta \overset{\cdot }{\bar{\alpha}}\right) \right)
-2ia^{2}\left\vert \beta \right\vert ^{2}\left( \bar{\alpha}\dot{\alpha}%
+\beta \overset{\cdot }{\bar{\beta}}\right) \right)
\end{eqnarray*}%
Observe that the last term contains the Lagrange multiplier $\eta _{+3}$
realizing the constraint 
\begin{eqnarray*}
&&\Omega \left( g_{+},g_{-}\right) \\
&=&i\beta \left( az\bar{\beta}-\bar{\alpha}\right) \overline{\left( \bar{%
\alpha}\dot{\beta}-\beta \overset{\cdot }{\bar{\alpha}}\right) }-\bar{\beta}%
\left( a\bar{z}\beta -\alpha \right) \left( \bar{\alpha}\dot{\beta}-\beta 
\overset{\cdot }{\bar{\alpha}}\right) -2ia^{2}\left\vert \beta \right\vert
^{2}\left( \bar{\alpha}\dot{\alpha}+\beta \overset{\cdot }{\bar{\beta}}%
\right)
\end{eqnarray*}

Despite its rather complicated expression, this Lagrange function can be
written in the compact form%
\begin{equation*}
L_{\mathcal{N}\left( g_{-},0\right) }\left( g_{+},\dot{g}_{+}\right) =-\frac{%
1}{8}\kappa \left( g_{+}^{-1}\dot{g}_{+},\mathbb{K}\left( g_{+},g_{-}\right)
g_{+}^{-1}\dot{g}_{+}\right) -\lambda \Omega \left( g_{+},g_{-}\right)
\end{equation*}%
where $\lambda $ is a redefinition of the Lagrange multiplier, and $\mathbb{K%
}\left( g_{+},g_{-}\right) :\mathfrak{su}_{2}\longrightarrow \mathfrak{su}%
_{2}$ is a bijection that in the basis $\left\{ T_{1},T_{2},T_{3}\right\} $
is represented by the symmetric matrix, a metric tensor, 
\begin{equation*}
\mathbb{K}\left( g_{+}g_{-}\right) \mathbb{=}\dfrac{1}{2\left\vert \beta
\right\vert ^{2}}\left( 
\begin{array}{ccc}
1 & 0 & \mathrm{m} \\ 
0 & 1 & \mathrm{n} \\ 
\mathrm{m} & \mathrm{n} & 1%
\end{array}%
\right)
\end{equation*}%
where 
\begin{eqnarray*}
\mathrm{m} &=&\dfrac{\left( \left( a^{2}-1\right) \left( \beta \bar{\alpha}%
+\alpha \bar{\beta}\right) +a\left( \bar{z}+z\right) \beta \bar{\beta}%
\right) }{2a^{2}\beta \bar{\beta}} \\
&& \\
\mathrm{n} &=&i\dfrac{\left( \left( 1-a^{2}\right) \left( \alpha \bar{\beta}%
-\beta \bar{\alpha}\right) +a\left( z-\bar{z}\right) \beta \bar{\beta}%
\right) }{2a^{2}\beta \bar{\beta}}
\end{eqnarray*}%
It is worth to remark that $g_{-}=e$, it turns in 
\begin{equation*}
\mathbb{K=}\frac{1}{2\left\vert \beta \right\vert ^{2}}\left( 
\begin{array}{ccc}
1 & 0 & 0 \\ 
0 & 1 & 0 \\ 
0 & 0 & 1%
\end{array}%
\right)
\end{equation*}%
thus recovering the Lagrangian function described in ref \cite{CapMon}.

The constraint $\Omega $ can be written as%
\begin{equation*}
\Omega =\kappa \left( \mathbb{A}\left( g_{+}g_{-}\right) ,g_{+}^{-1}\dot{g}%
_{+}\right)
\end{equation*}%
for 
\begin{eqnarray}
\mathbb{A}\left( g_{+},g_{-}\right) &=&\frac{1}{8}i\left( \beta \left(
az\gamma -\bar{\alpha}\right) +\gamma \left( a\bar{z}\beta -\alpha \right)
\right) T_{1}  \label{n} \\
&&-\frac{1}{8}\left( \beta \left( az\gamma -\bar{\alpha}\right) -\gamma
\left( a\bar{z}\beta -\alpha \right) \right) T_{2}+\frac{1}{4}a^{2}\beta
\gamma T_{3}  \notag
\end{eqnarray}%
so that the velocities must be confined in the orthogonal complement of the
vector $\mathbb{A}\left( gb\right) $. The Lagrangian function is now written
as 
\begin{equation*}
L_{\mathcal{N}\left( g_{-},0\right) }\left( g_{+},\dot{g}_{+}\right) =-\frac{%
1}{8}\kappa \left( g_{+}^{-1}\dot{g}_{+},\mathbb{K}\left( g_{+}g_{-}\right)
g_{+}^{-1}\dot{g}_{+}\right) -\frac{1}{8}\lambda \kappa \left( \mathbb{A}%
\left( g_{+},g_{-}\right) ,g_{+}^{-1}\dot{g}_{+}\right) _{\mathfrak{g}%
_{+}^{\circ }}
\end{equation*}%
where $\mathbf{\mathbb{A}}\left( gb\right) \in \mathfrak{g}$ was given in $%
\left( \ref{n}\right) $. It is equivalent to%
\begin{equation*}
L_{\mathcal{N}\left( g_{-},0\right) }\left( g_{+},\dot{g}_{+}\right) =-\frac{%
1}{8}\kappa \left( g_{+}^{-1}\dot{g}_{+},\mathbb{K}\left( g_{+}g_{-}\right)
g_{+}^{-1}\dot{g}_{+}+\lambda \mathbb{A}\left( g_{+},g_{-}\right) \right)
\end{equation*}

\section{Conclusions}

Using the Dirac method we have studied a class of homogenous submanifolds in
the cotangent bundle of a factorizable Lie group. They appears as
generalizations of the cotangent bundles of the factors, in fact, they are
retrieved when the fibers at $\left( e,0\right) \in G\times g^{\ast }$ are
considered. These fibrations amounts to be symplectic: any horizontal curve
is a symplectomorphism between fibers.

By using the Dirac brackets, we described the restriction of the Lie algebra
of functions on $G\times g^{\ast }$ to the submanifolds $\mathcal{N}\left(
g_{-},\eta _{-}\right) $ and $\mathcal{M}\left( g_{+},\eta _{+}\right) $ in
a natural fashion, showing how symmetries and dynamics projects out from the
big phase space to the smaller ones. Dirac brackets of the momentum
functions associated with the let translations in $G\times g^{\ast }$ give
rise to an action of $G$ on $\mathcal{N}\left( g_{-},\eta _{-}\right) $ and $%
\mathcal{M}\left( g_{+},\eta _{+}\right) $ provided $\eta _{-}$ and $\eta
_{+}$ are characters of the corresponding coadjoint actions of $G_{-}$ and $%
G_{+}$. These actions play a central role in the connection with integrable
systems: collective dynamics built with these momentum functions on $G\times
g^{\ast }$, give rise to Dirac hamiltonian vector fields which are of the
AKS\ type, being integrable in this sense. Thus, Dirac brackets nicely
reduces a somehow trivial systems on $G\times g^{\ast }$ into a systems with
plenty of non trivial dynamics.

A deeper geometric insight, besides a description of AKS systems suitable
for our purposes, allowed us to explain this (apparently) unexpected
connection between Dirac constraints and integrable AKS system, as arising
from the description of the latter as reduced space of $T^{\ast }G$, and by
the existence of immersions of the cotangent bundle of some homogeneous
spaces of $G$ into $T^{\ast }G$.

\end{document}